\documentclass[xcolor=svgnames,11pt,reqno]{amsart}
\usepackage[T1]{fontenc}
\usepackage[round]{natbib}
\usepackage{amsmath}
\usepackage{amssymb}
\usepackage{amsthm}
\usepackage{fancyhdr}
\usepackage{enumitem}
\usepackage{comment}
\usepackage{relsize}
\usepackage{nicefrac}
\usepackage{bm}
\usepackage{mathrsfs}
\usepackage{graphicx}
\usepackage[utf8]{inputenc}
\usepackage{cancel}
\usepackage{dsfont}
\usepackage{mathtools}
\usepackage{tikz,xcolor,etoolbox}
\usepackage{dirtytalk}
\usetikzlibrary{cd,decorations.pathreplacing,matrix,arrows,positioning,automata,shapes,shadows,calc,fadings,decorations,through,intersections}
\usepackage{float} 
\usepackage{subcaption}
\usepackage{multirow}
\usepackage[left=2.5cm,top=2.8cm,right=2.5cm,bottom=2.5cm,head=3cm,headsep=1cm, foot=3cm]{geometry}
\usepackage{hyperref}
\usepackage[skip=0mm,indent=12pt]{parskip}
\definecolor{darkblue}{rgb}{0.1,0.1,0.9}
\definecolor{darkred}{rgb}{0.9,0.1,0.1}

\hypersetup{colorlinks, allcolors= blue}

\makeatletter
\pretocmd{\tagform@}{\color{blue}}{}{}
\makeatother

\newtheorem{theorem}{Theorem}[section]
\newtheorem{definition}[theorem]{Definition}
\newtheorem{proposition}[theorem]{Proposition}
\newtheorem{assumption}[theorem]{Assumption}
\newtheorem{corollary}[theorem]{Corollary}
\newtheorem{lemma}[theorem]{Lemma}
\newtheorem{remark}[theorem]{Remark}

\newcommand{\cF}{\mathcal{F}}

\newcommand{\cB}{\mathcal{B}}

\newcommand{\cC}{\mathcal{C}}

\newcommand{\cR}{\mathcal{R}}

\renewcommand{\P}{\mathbb{P}}

\newcommand{\R}{\mathbb{R}}

\newcommand{\N}{\mathbb{N}}
\newcommand{\e}{\enspace}

\usepackage{booktabs}
\usepackage{siunitx}
\usepackage[colorinlistoftodos,textsize=small]{todonotes}
\usepackage{cleveref}
\usepackage{caption}
\usepackage{anyfontsize}
\usepackage{yhmath}
\usepackage{comment}

\renewcommand{\l}{\left}
\renewcommand{\r}{\right}
\renewcommand{\tilde}{\widetilde}

\allowdisplaybreaks
\makeatletter

\newcommand{\Rmnum}[1]{\expandafter\@slowromancap\romannumeral #1@}

\title[Implementability In Insurance Markets With Adverse Selection]{Implementability In Insurance Markets\vspace{0.2cm}\\With Adverse Selection}

\author[Maria Andraos and Mario Ghossoub]{Maria Andraos\vspace{0.15cm}\\ University of Waterloo\vspace{0.8cm}\\
Mario Ghossoub\vspace{0.15cm}\\ University of Iowa\vspace{0.8cm}\\ \today}

\address{{\bf Maria Andraos}: University of Waterloo -- Department of Statistics and Actuarial Science -- 200 University Ave.\ W.\ -- Waterloo, ON, N2L 3G1 -- Canada\vspace{0.15cm}}\email{\href{mailto:mandraos@uwaterloo.ca}{mandraos@uwaterloo.ca}\vspace{0.3cm}}

\address{{\bf Mario Ghossoub}: University of Iowa -- Department of Statistics and Actuarial Science -- 20 East Washington Street -- Iowa City, IA, 52242 -- United States\vspace{0.15cm}}
\email{\href{mailto:mario-ghossoub@uiowa.edu}{mario-ghossoub@uiowa.edu}}

\thanks{ \vspace{0.2cm}}

\thanks{\vspace{0.2cm}}

\thanks{Mario Ghossoub acknowledges partial financial support from the Natural Sciences and Engineering Research Council of Canada (NSERC Grant No.\ 2024-03744).}

\begin{document}

\begin{abstract}
We consider an insurance market with hidden information, where the agent's type is private information and is drawn from an arbitrary type space. We study implementability of a collection of retention functions, namely, how to select premium schedules so that the resulting menu of contracts is incentive compatible, or truthful. Specifically, for general type spaces, implementability is equivalent to cyclical monotonicity of the collection of retention functions. For compact interval type spaces, we show that submodularity is a sufficient condition for implementability, under suitable type ordering assumptions. Moreover, for any implementable collection of retention functions, we characterize all corresponding premium schedules, up to a common additive constant. Finally, we apply our results to several standard classes of insurance contracts, for which the general implementability conditions admit simpler characterizations, and we provide several numerical illustrations.
\end{abstract}

\maketitle

\bigskip 
\section{Introduction}
Insurance markets play a fundamental role in the allocation and transfer of risk by allowing agents to insure against uncertain future losses.
In classical insurance markets, insurers are often assumed to possess complete information on the agent's risk characteristics, enabling contracts to be tailored directly to the agent's risk profile. 
However, in practice, this assumption is not always satisfied. In many insurance markets, the insurer has less information than the agent and cannot directly observe the agent's underlying risk characteristics, as noted by \cite{allais1953extension}. This information asymmetry gives rise to the problem of adverse selection. The seminal work of 
\cite{akerlof1970market} was the first to study the notion of asymmetric information through the analysis of the used cars market. Specifically, the seller will always possess superior information about the quality of the used car relative to potential buyers. Since then, adverse selection has become a central paradigm and has been extensively studied in insurance markets.

\medskip

In the presence of adverse selection, the insurer cannot provide a contract directly to the agent's privately observed type. Instead, the insurer must design a menu of contracts from which the agent self-selects a single contract. Consequently, a fundamental requirement is that the proposed menu must be incentive compatible to ensure that each type of the agent self-selects the contract intended for their own type over contracts designed for other types. 
In other words, incentive compatibility is a constraint that rules out profitable deviations by misreporting of the agent's type. 
Hence, the presence of adverse selection fundamentally changes the notion of efficiency in insurance markets to account for incentive compatibility in addition to participation constraints. 
A substantial body of the insurance literature studies the optimal design of contracts under adverse selection. In particular, early contributions, including \cite{prescott1984pareto}, \cite{jerez2003dual} and \cite{bisin2006efficient} analyze constrained Pareto efficiency in markets with asymmetric information and finitely many agent types. More recently, \cite{gershkov2023optimal} extend this framework to insurance markets with a continuum of types. \cite{ghossoub2025optimal} consider a monopolistic insurance market in which the agent's private type affects their level of risk aversion and discuss the notion of incentive efficiency, also called incentive Pareto optimality, providing a sufficient condition for a menu of contracts to be incentive efficient. \cite{andraos2026incentive} extend this framework by allowing the agent's private type to affect both the loss distribution and the level of risk aversion. Moreover, they provide a characterization of the optimal menu of contracts that maximizes the social welfare and study its properties. 
Hence, the aforementioned literature discusses that the meaningful notion of efficiency must be defined relative to the class of individually rational and incentive compatible menus of contracts, making incentive compatibility an essential prerequisite for efficiency under adverse selection. 

\medskip
From a practical perspective, insurers may begin with a prescribed collection of insurance contracts motivated by actuarial, regulatory, or commercial considerations, which is not guaranteed to satisfy incentive compatibility. Before addressing questions of optimality or efficiency, it is natural to ask whether the given collection of retention functions (or indemnity functions) can be supported by a premium schedule such that each type of agent prefers the contract intended for their type to every other contract in the menu. If such a premium schedule exists, the prescribed collection of retention functions is said to be implementable. 
The implementability problem has been studied primarily in the mechanism design literature. \cite{myerson1981optimal} showed that in a single dimensional environment, monotonicity of the allocation rule is sufficient for implementability. 
The subsequent literature identifies domains on which similar simple monotonicity conditions characterize implementability in multidimensional environments. For instance, \cite{gui2004dominant} establish such results for several restricted multidimensional domains, while \cite{bikhchandani2006weak} characterize deterministic implementability on a broad class of convex domains. \cite{saks2005weak} prove that weak monotonicity is sufficient for implementability on arbitrary convex domains, and \cite{ashlagi2010monotonicity} further characterize the class of domains on which monotonicity alone guarantees implementability.
On the other hand, \cite{rochet1987necessary} addresses the implementability problem from a different perspective by providing a complete characterization. Specifically, \cite{rochet1987necessary} shows that under quasi-linear preferences, an allocation rule is implementable if and only if it satisfies cyclical monotonicity.

\medskip

The present paper combines these two perspectives in the context of insurance markets under adverse selection, within a quasi-linear framework in which the agent's preferences are evaluated according to Yaari's dual utility. We provide a broad analysis of the implementability of insurance menus on arbitrary type spaces, in the spirit of \cite{rochet1987necessary}, and establish the equivalence between implementability and cyclical monotonicity in this insurance setting. We also study measurable implementability, which requires the existence of a measurable premium schedule supporting the given collection of retention functions. For compact metric type spaces, we show that implementability and measurable implementability are equivalent when the non-monetary utility function is jointly continuous.
Cyclical monotonicity is a global condition that requires checking an inequality for every finite cycle of types, which can be difficult to verify directly in applications. For compact interval type spaces and under suitable type ordering assumptions, we show that every submodular collection of retention functions is implementable, and we explicitly characterize the corresponding premium schedules. In a setting with ordered loss distributions and a common probability distortion, \citet[Proposition 1]{gershkov2023optimal} derive an envelope representation of supporting premiums and establish that submodularity of the retention function in loss and type is sufficient for incentive compatibility. In this paper, Corollary \ref{co:subimpliesration} and Proposition \ref{IC_characterization} extend this sufficient condition and premium representation, respectively, to a setting in which both the loss distribution and the probability distortion depend on the private type. Moreover, under suitable regularity and measurability assumptions, we show that these constructed premia are Borel measurable, yielding equivalence between implementability and measurable implementability without requiring the joint continuity condition of the non-monetary utility function.
We apply our results to several classes of insurance contracts, namely, deductible, proportional, policy-limit, and capped deductible insurance. 
For deductible, proportional and policy-limit insurance, implementability admits a complete characterization through a simple monotonicity condition on the relevant contract parameter. Specifically, implementability is equivalent to submodularity, which in turn is equivalent to the deductible being non-increasing in type, the coinsurance being non-decreasing in type, and the policy-limit being non-decreasing in type, respectively.
The capped deductible contract is determined by two type-dependent parameters: the deductible and the policy-limit. We first characterize submodularity and thereby obtain sufficient conditions on the two parameters for implementability. Finally, we illustrate the theoretical results obtained for each class of insurance contracts. 
 
\medskip
The rest of this paper is organized as follows. Section \ref{sec:insurancemarket} introduces the insurance market model and the preference framework. Section \ref{sec:implementability} studies implementability first on arbitrary type spaces and then specializes to interval type spaces. It provides a complete characterization of implementability in terms of cyclical monotonicity for arbitrary type spaces and establishes submodularity of the collection of retention functions as a sufficient condition for implementability on interval type spaces under suitable type ordering assumptions.
Section \ref{sec:examples} applies the theoretical results to several classes of insurance contracts, including deductible, proportional, policy-limit, and capped deductible insurance, and provides numerical illustrations. Section \ref{sec:conclusion} concludes. All proofs can be found in the \hyperref[appendix]{Appendix}. 

\bigskip
\section{The Insurance Market}
\label{sec:insurancemarket}

\subsection{Setting}
Let $(S, \Sigma, \P)$ be a probability space, and denote by $B(\Sigma)$ the space of bounded, real-valued, and $\Sigma$-measurable functions. We consider a monopolistic insurance market in which the agent has a privately observed type $\theta$ drawn from an arbitrary type space $\Theta$. Let $\cB(\Theta)$ be a given $\sigma$-algebra on the type space $\Theta$. 

\medskip
A type-$\theta$ agent faces a nonnegative type-dependent loss $L_\theta \in B(\Sigma)$. For each $\theta \in \Theta$, the random variable $L_\theta$ takes values in $[0, \bar L_\theta]$, where 
$0<\bar L_\theta< +\infty$. We assume that the collection of losses $\{L_\theta\}_{\theta \in \Theta}$ is uniformly bounded, with 
$$
0<\bar L:= \underset{\theta\in\Theta}{\sup}\, \bar L_\theta <+\infty. 
$$

\noindent Thus, $L_\theta$ takes values in $[0,\bar L]$ for every
$\theta\in\Theta$. An indemnity schedule is a measurable function $I :[0, \bar L]\to [0, \bar L]$, where $I(l)$ represents the coverage paid by the insurer to the agent when the realized loss is $l$. 
We restrict attention to indemnities belonging to the following set of 1-Lipschitz and non-decreasing functions:
$$
\cF_I:=\left\{ I:[0, \bar L]\to [0, \bar L]:\, I(0)=0, \, 0\leq I(l_2)-I(l_1) \leq l_2-l_1, \ \forall \, 0\leq l_1\leq l_2\leq\bar L\right\}. 
$$

\noindent Every indemnity schedule $I \in \cF_I$ is said to be feasible. This restriction ensures that the market only offers indemnities that satisfy the no-sabotage condition of \cite{carlier2003pareto}.

\medskip
\begin{definition}
A contract is a pair $(I, p)$ where $I \in \cF_I$ is the indemnity schedule and $p \in \R_+$ is the premium paid by the agent to the insurer in exchange for coverage.
\end{definition}

\medskip
Equivalently, an insurance contract can be represented in terms of its retention function 
$$R(l)=l-I(l) \geq 0,$$

\noindent which denotes the portion of a realized loss $l$ retained by the agent. An indemnity function $I$ is feasible if and only if the associated retention function $R$ belongs to the feasible set 
$$\cF_R= \left\{
R:[0,\bar L]\to[0,\bar L]:
R(0)=0,\; 0\le R(l_2)-R(l_1)\le l_2-l_1, \ \forall \, 0\leq l_1 \leq l_2 \leq \bar L\right\}.$$

\noindent Since every Lipschitz continuous function is absolutely continuous, every $R \in \cF_R$ is differentiable almost everywhere and satisfies 
$$
0 \leq R^\prime(l) \leq 1,  \ \text{ for a.e. $l \in [0, \bar L]$}. 
$$

\medskip

The insurer does not observe the agent's type and therefore cannot tailor a single contract directly to the agent's type. Instead, the insurer designs a menu of contracts from which the agent self-selects a single contract. 

\medskip
\begin{definition}
A menu of contracts is a collection of the form $(I_\theta, p_\theta)_{\theta \in \Theta}$. For each $\theta\in \Theta$, $(I_\theta, p_\theta) \in \cF_I\times \R_+$ is the contract intended for the type-$\theta$ agent, where $I_\theta$ is the indemnity schedule and $p_\theta$ is the corresponding premium. 
\end{definition}

\medskip
\begin{assumption}\label{ass:jointmeasurableR}
The map $(\theta, l)\mapsto R_\theta(l)$ is jointly measurable on $\Theta \times [0,\bar L]$. 
\end{assumption}

\medskip
\begin{lemma}\label{le:measurablederivative}
Suppose that Assumption \ref{ass:jointmeasurableR} holds, and that $R_\theta\in\cF_R$, for every $\theta\in\Theta$. Then there exists a $\cB(\Theta) \otimes \cB([0, \bar L])$-measurable map $r:\Theta \times [0, \bar L] \to [0,1]$, such that for every $\theta \in \Theta$,
$$
r(\theta, l)= \frac{\partial R_\theta(l)}{\partial l}, \ \text{ for a.e.\ $l \in [0, \bar L]$ }. 
$$
\end{lemma}

\medskip
\subsection{Dual Utility Framework}
Preferences in this market are represented by a utility functional
$$
U: \Theta \times \cF_R \times \R_+ \to \R,
$$

\medskip

\noindent where for a given triplet $(\theta, R, p) \in \Theta\times \cF_R \times\R_+$, the quantity $U(\theta, R, p)$ denotes the end-of-period utility of a type-$\theta$ agent after purchasing the contract $(R, p)$. For notational convenience, we write:
$$
U_\theta(R, p):= U(\theta, R, p).
$$

\medskip

\begin{definition}
For a given random variable $X \in B(\Sigma)$, the dual utility of $X$ (\cite{yaari1987dual}) is given by:
$$
DU(X) = \int X \, d(g\circ\P) = \int_{-\infty}^0 \left[ g(1-\P(X\leq x))-1\right] \,dx + \int_0^{+\infty} g(1-\P(X\leq x))\,dx, 
$$

\noindent where $g:[0,1]\to [0,1]$ is a distortion function, that is a non-decreasing function with $g(0)=0$ and $g(1) =1$. 
\end{definition}

\medskip
Hereafter, we assume that the agent's preferences are evaluated according to  Yaari's dual utility. Specifically, for each type-$\theta$ agent, 
$$
DU_\theta (\cdot) = \int \cdot \ d(g_\theta \circ \P), \ \text{ for each $\theta \in \Theta$},
$$

\noindent where $g_\theta$ denotes the type-$\theta$'s distortion function. The end-of-period wealth of a type-$\theta$ agent is 
$$-p_\theta - R_\theta(L_\theta), \ \text{ for $\theta \in \Theta$}.
$$

\medskip

\noindent By translation invariance of the dual utility, the end-of-period utility of the type-$\theta$ agent is given by:
$$
U_\theta(R_\theta, p_\theta) = -p_\theta + DU_\theta(-R_\theta(L_\theta)).
$$

\medskip

\noindent Moreover, since $-R_\theta(L_\theta) \leq 0$ for $\theta \in \Theta$, it follows that
\begin{align*}
U_\theta(R_\theta, p_\theta)
&= - p_\theta +\int_{-\infty}^0 \left[ g_\theta (1-\P(-R_\theta(L_\theta) \leq x)) -1\right] \,dx\\
&= - p_\theta - \int_0^{+\infty} \left[ 1 - g_\theta(\P(R_\theta(L_\theta)\leq l))\right] \, dl.
\end{align*}

\medskip

\noindent Let $F_\theta(l) := \P(L_\theta\leq l)$, for all $\theta \in \Theta$ and $l \in [0, \bar L]$, denote the cumulative distribution function of the type-dependent loss $L_\theta$, for all $\theta \in \Theta$. Then 
$$
U_\theta(R_\theta, p_\theta)
=  -p_\theta -\int_0^{\bar L} \left[ 1-g_\theta(F_\theta(l))\right] \frac{\partial R_\theta(l)}{\partial l} \,dl.
$$

\bigskip
\section{Implementability of Retention Functions}
\label{sec:implementability}
After observing the menu of contracts offered by the insurer, the agent selects a single contract. However, a type-$\theta$ agent is not forced to choose the contract $(R_\theta, p_\theta)$ designed for their type. Instead, the agent may select the contract $(R_{\hat\theta},p_{\hat\theta})$ intended for another type $\hat \theta \in \Theta$ if doing so yields a higher utility. Incentive compatibility rules out profitable misreporting and ensures that each type self-selects the contract designed for their own type. 

\medskip
\begin{definition} \label{def:IC}
A menu of contracts $(R_\theta, p_\theta)_{\theta\in \Theta}$ is said to be incentive compatible if no type $\theta$ can benefit from choosing the contract of another type $\hat\theta \in \Theta$. That is, 
$$
U_\theta(R_\theta, p_\theta)\geq U_\theta (R_{\hat \theta}, p_{\hat \theta}), \ \text{for all $\theta, \hat \theta \in \Theta$}. 
$$
\end{definition}

\medskip
Given a collection of feasible retention functions $\cR = \{R_\theta\}_{\theta \in \Theta}$, implementability is the problem of selecting the corresponding premium schedule that makes the resulting menu of contracts $(R_\theta, p_\theta)_{\theta \in\Theta}$ incentive compatible. We first study this problem on an arbitrary type space and subsequently specialize to interval type spaces.

\medskip
\subsection{Implementability On General Type Spaces}
\label{sec:implementability_general}
Suppose that the type space $\Theta$ is arbitrary, equipped with the given $\sigma$-algebra $\cB(\Theta)$. As established in Section \ref{sec:insurancemarket}, translation invariance of Yaari's dual utility implies that the agent's utility is quasi-linear. Specifically, the premium enters utility additively, while the remaining term depends on the retention function. Accordingly, for a given collection of retention functions $\cR=\{R_\theta\}_{\theta\in\Theta}$, let 
$$
u_\cR(\theta, \hat \theta):=-\int_0^{\bar L} \left[ 1-g_\theta(F_\theta(l))\right]\frac{\partial R_{\hat \theta}(l)}{\partial l}\,dl,
$$

\noindent denote the non-monetary utility of a type-$\theta$ agent from selecting the contract designed for type $\hat \theta \in \Theta$. For notational convenience, we write
$$
u_\cR(\theta):= u_\cR(\theta, \theta), \ \text{ for $\theta \in \Theta$},
$$

\noindent which represents the non-monetary utility of a type-$\theta$ agent under truthful reporting. For every $\theta, \hat \theta \in \Theta$, the end-of-period utility of the type-$\theta$ agent selecting the contract of $\hat \theta$, can be written as follows:
$$
U_\theta(R_{\hat \theta}, p_{\hat \theta})= -p_{\hat \theta} + u_\cR(\theta, \hat \theta). 
$$

\noindent In particular, under truthful reporting, 
$$
U_\theta(R_\theta, p_\theta) = -p_\theta + u_\cR(\theta)
$$

\medskip

\begin{definition}
\label{def:PointMeasImplem}
A premium schedule is a map
\begin{align*}
p: \Theta &\to \R_+\\
\theta &\mapsto p_\theta.
\end{align*}

\medskip

\noindent A collection $\cR := \{R_\theta\}_{\theta\in\Theta}
\subseteq \cF_R$ of retention functions is said to be:

\medskip

\begin{enumerate}
\item Implementable if there exists a premium schedule $p:\Theta \to \R_+$ such that the menu of contracts $(R_\theta,p_\theta)_{\theta\in\Theta}$
is incentive compatible; that is,
\begin{equation}
\label{eq:PointwiseIC}
u_\cR(\theta) - p_\theta
\geq
u_\cR(\theta,\hat\theta) - p_{\hat\theta},
\ \ \forall \, \theta, \hat\theta \in \Theta.
\end{equation}

\medskip

\item Measurably implementable if there exists a $\cB(\Theta)$-measurable premium schedule $p: \Theta \to \R_+$ that satisfies \eqref{eq:PointwiseIC}.
\end{enumerate}
\end{definition}

\medskip

Clearly, measurable implementability implies implementability. 

\medskip

\begin{definition}
A finite chain in $\Theta$ is a finite sequence of types from $\theta_0$ to $\theta$ given by 
$$(\theta_0,\theta_1,\ldots,\theta_m,\theta_{m+1}=\theta)$$ 

\smallskip

\noindent such that $\theta_k \in \Theta$, for all $k=0,1, \ldots, m+1$. 
A finite cycle in $\Theta$ is a finite chain whose terminal type coincides with its initial type. That is, a finite cycle is given by 
$$\cC :=(\theta_0, \theta_1, \ldots, \theta_n, \theta_{n+1})$$

\smallskip

\noindent such that $\theta_k \in \Theta$ for all $k \in \{0, \ldots, n, n+1\}$, and $\theta_{n+1}=\theta_0$. 
\end{definition}

\medskip
\begin{lemma}\label{le:finite}
Let $\cR= \{ R_\theta\}_{\theta \in \Theta}$ be a collection of feasible retention functions. Then,
$$
-\bar L \leq u_\cR(\theta,\hat \theta) - u_\cR(\hat \theta)\leq \bar L, \ \text{ for all $\theta, \hat \theta \in \Theta$}. 
$$
\end{lemma}

\medskip

\begin{theorem}
\label{th:ImplemIFF}
A collection of feasible retention functions $\cR$ is implementable if and only if, for every finite cycle $\cC$ in $\Theta$, we have
\smallskip
\begin{equation}
\label{eq:CycMono}
\sum_{k=0}^{n}\left[u_\cR(\theta_{k+1},\theta_k) - u_\cR(\theta_k) \right] \leq 0.
\end{equation}
\end{theorem}

\medskip

Theorem \ref{th:ImplemIFF} provides a complete characterization of implementability in the present insurance setting. Specifically, for an arbitrary type space $\Theta$, a collection of retention functions is implementable if and only if it satisfies the cyclical monotonicity condition, as in \cite{rochet1987necessary}. However, Theorem \ref{th:ImplemIFF} is a pointwise result. It does not necessarily imply that the premium schedule that implements $\cR$ is measurable. This is given in the following result.

\medskip
\begin{proposition}
\label{prop:MeasImplemCompTheta}
Suppose that $\Theta$ is a compact metric space equipped with its Borel
$\sigma$-algebra $\cB(\Theta)$, and suppose that the map $u_\cR:\Theta\times\Theta\to\R$ is continuous. If \eqref{eq:CycMono} holds for every finite cycle in $\Theta$, then $\cR$ is measurably implementable.
\end{proposition}

\medskip

It follows from Proposition \ref{prop:MeasImplemCompTheta} and Theorem \ref{th:ImplemIFF} that on a compact metric space, every implementable collection of retention functions, for which the map $u_\cR$ is jointly continuous, is also measurably implementable. Consequently, under these conditions, implementability and measurable implementability are equivalent. 

\medskip
\subsection{Implementability On Compact Interval Type Spaces}
\label{sec:implementability_interval}
We now specialize the model to compact interval type spaces of the form $\Theta=[\underline{\theta},\bar{\theta}]$, equipped with the Borel $\sigma$-algebra $\cB(\Theta)$.

\medskip
 
\begin{assumption} 
\label{ass:lipschitz1}
The following conditions hold. 

\smallskip
\begin{enumerate}
\item For every $l \in [0, \bar L]$, the map $\theta \mapsto F_\theta(l)$ is differentiable. 

\medskip

\item For every $t \in [0, 1]$, the map $\theta \mapsto g_\theta(t)$ is differentiable. 
Additionally, the map $(\theta,t) \to \frac{\partial g_\theta(t)}{\partial \theta}$ is jointly continuous. 

\medskip

\item For every $\theta \in \Theta$, the map $t \mapsto g_\theta(t)$ is differentiable.
\end{enumerate}
\end{assumption}

\medskip

It follows from Assumption \ref{ass:lipschitz1} that the partial derivatives 
$$
\frac{\partial F_\theta(l)}{\partial \theta}, \ \frac{\partial g_\theta(t)}{\partial \theta}, \ \text{ and } \ g^\prime_\theta(t):= \frac{\partial g_\theta(t)}{\partial t}
$$

\noindent exist on their respective domains. Moreover, Assumption \ref{ass:lipschitz1} ensures that the map $(\theta, t) \to g_\theta(t)$ is differentiable. Therefore, for every $l \in [0, \bar L]$, the chain rule gives
\begin{align*}
\frac{\partial }{ \partial \theta } \, g_{\theta} \big(  F_{\theta} (l)  \big) 
=\frac{\partial  g_{\theta} }{ \partial \theta } \big(  F_{\theta} (l)  \big) 
+ g^{\prime}_{\theta}\big(  F_{\theta} (l)  \big)  \frac{\partial  F_{\theta}(l) }{ \partial \theta}, \ \text{ for all $l \in [0, \bar L]$}. 
\end{align*} 

\medskip

\begin{lemma}
\label{le:jointlymeasurable}
The composite derivative $(\theta, l) \mapsto \frac{\partial }{ \partial \theta } \, g_{\theta} \big(  F_{\theta} (l)  \big)$ is jointly Borel measurable. 
\end{lemma}

\medskip

The following assumption ensures that the distorted cumulative distribution functions vary Lipschitz continuously with type, with a common Lipschitz constant across all loss levels. 

\medskip

\begin{assumption}\label{ass:lipschitz}
There exists $K< \infty$ such that 
$$
\left| \frac{\partial }{ \partial \theta } \,  g_{\theta} \big(  F_{\theta} (l)  \big) \right| \leq K, \ \text{ for all $\theta \in \Theta$ and $l \in [0, \bar L]$}.
$$
\end{assumption}

\medskip

In addition to the regularity assumption above, we impose the following type ordering assumptions throughout the remainder of the paper. These assumptions describe how types of the agent differ in terms of their loss distributions and risk attitudes. 

\medskip

\begin{assumption}
\label{Ass:cdf_family}
Let $L_{\theta}$ be the loss faced by a type-$\theta$ agent, with cumulative distribution function $F_\theta$. 
Type-dependent losses are ordered according to the first order stochastic dominance. Specifically, for $\theta_1 < \theta_2$, we have $L_{\theta_1} \preccurlyeq _{FOSD}   L_{\theta_2}$, that is, $F_{\theta_1}(l) \geq F_{\theta_2}(l)$, for all $l \in [0, \bar L]$. Equivalently, 
$$
\frac{\partial F_{\theta} (l) }{ \partial \theta} \leq 0, \ \text{for all $l\in [0, \bar L]$}.
$$
\end{assumption}

\medskip

Assumption \ref{Ass:cdf_family} states that higher types face stochastically larger losses in the sense of first order dominance.

\medskip

\begin{assumption}
\label{Ass:distortion_family}
The distortion family $\{g_\theta\}_{\theta \in \Theta}$ is ordered in the sense that 
$$
\frac{\partial g_{\theta} (t) }{\partial \theta} \leq 0, \ \text{  for all $t \in [0,1]$}. 
$$
\end{assumption}

\medskip

Assumption \ref{Ass:distortion_family} states that the distortion family is non-increasing in type. That is, if $\theta_1, \theta_2 \in \Theta$ are such that $\theta_1 \leq \theta_2$, then $g_{\theta_1}(t) \geq g_{\theta_2}(t)$ for all $t \in [0,1]$. This means that the type $\theta_2$-agent is weakly more risk averse than the type $\theta_1$-agent.  
Together, Assumptions \ref{Ass:cdf_family} and \ref{Ass:distortion_family} state that higher agent types (larger values of $\theta$) are more risk averse and face stochastically larger losses.

\medskip

\begin{remark} 
\label{Re:chain_rule}
Since $g^\prime_\theta (t)\geq0$ for all $t\in[0,1]$, it follows from Assumptions \ref{Ass:cdf_family} and \ref{Ass:distortion_family} that
$$
\frac{\partial }{ \partial \theta } \left [ g_{\theta} \big(  F_{\theta} (l)  \big)  \right ]  \leq 0, \ \text{ for all $l \in[0, \bar L]$}.
$$
\end{remark}

\medskip

Remark \ref{Re:chain_rule} shows that higher types, who are more risk averse, assign lower distorted cumulative distribution functions to the loss, reflecting the combined effects of stochastically larger losses and greater risk aversion. 

\medskip

\begin{definition}
A collection of retention functions $\cR=\{R_\theta\}_{\theta \in \Theta}$ is said to be submodular if the marginal retention $\frac{\partial R_\theta(l)}{\partial l}$ is non-increasing in type. That is, $\cR$ is submodular if for $\theta_1, \theta_2 \in \Theta$, $\theta_1 < \theta_2$, 
$$
\frac{\partial R_{\theta_2}(l)}{\partial l}\leq \frac{\partial R_{\theta_1}(l)}{\partial l}, \ \text{ for a.e. $l \in [0, \bar L]$}. 
$$
\end{definition}

\medskip

Submodularity of the collection of retention functions ensures that coverage becomes progressively more generous as risk aversion increases. Higher types receive weakly greater coverage at every loss level, transferring a larger portion of loss to the insurer and retaining less to themselves. 

\medskip

\begin{proposition}
\label{prop:increasingdifferences}
Suppose that the collection of retention functions $\cR$ is submodular. Then for any $\theta_a,\theta_b, \alpha, \beta \in \Theta$ such that $\theta_a<\theta_b$ and $\alpha<\beta$, we have 
$$
u_\cR(\theta_b, \beta) - u_\cR(\theta_b, \alpha) \geq u_\cR(\theta_a, \beta)  - u_\cR(\theta_a, \alpha).
$$
\end{proposition}

\medskip

\begin{proposition}
\label{prop:cyclicmonotone}
Suppose that the collection of retention functions $\cR$ is submodular. Then for every finite cycle $(\theta_0,\ldots,\theta_n,\theta_{n+1}=\theta_0)$ in
$\Theta =[\underline\theta,\bar\theta]$, the following holds
$$
\sum_{k=0}^{n} \left[ u_\cR(\theta_{k+1}, \theta_k) - u_\cR(\theta_k)\right]\leq 0.
$$
\end{proposition}

In a one-dimensional type space with scalar allocations, \cite{rochet1987necessary} shows that under the Spence-Mirrlees single-crossing condition \citep{mirrlees1971exploration,spence1973job}, cyclical monotonicity is equivalent to monotonicity of the allocation rule. In the present paper, under Yaari's dual utility representation and the type ordering assumptions,
Proposition \ref{prop:increasingdifferences} establishes the analogue of the Spence-Mirrlees condition by showing that submodularity of the collection of retention functions implies increasing differences in the non-monetary utility.
Consequently, Proposition \ref{prop:cyclicmonotone} implies that every submodular collection of retention functions satisfies cyclical monotonicity.

\medskip

\begin{corollary}
\label{co:subimpliesration}
Every submodular collection of retention functions $\cR$ is implementable. 
\end{corollary}

\medskip

The proof of Corollary \ref{co:subimpliesration} follows immediately from Proposition \ref{prop:cyclicmonotone} and Theorem \ref{th:ImplemIFF}. 
Indeed, Proposition \ref{prop:cyclicmonotone} shows that every submodular collection of retention functions satisfies cyclical monotonicity, while Theorem \ref{th:ImplemIFF} establishes that cyclical monotonicity is equivalent to implementability. Hence in the present insurance setting and under the maintained type ordering assumptions, Corollary \ref{co:subimpliesration} provides a sufficient condition for implementability.
In particular, by Definition \ref{def:PointMeasImplem}, there exists a premium schedule $p:\Theta \to \R_+$ such that the menu of contracts $(R_\theta,p_\theta)_{\theta\in\Theta}$ is incentive compatible.
A related sufficient condition for incentive compatibility is established in \citet[Proposition 1]{gershkov2023optimal} in a compact interval type-space setting with type-dependent loss distributions and a common probability distortion. Corollary \ref{co:subimpliesration} extends this result to a setting in which both the loss distribution and the probability distortion may depend on the agent's type.

\medskip

The following result provides an explicit representation of the premium schedules corresponding to an implementable collection of retention functions. \citet[Proposition 1]{gershkov2023optimal} and \citet[Proposition 3.2]{ghossoub2025optimal} obtain envelope based premium representations in settings with type-dependent loss distributions and type-dependent distortion functions, respectively. The representation below allows both components to vary with the agent's type. Moreover, this representation and its sufficiency for submodular menus, also appears in \citet[Propositions 4.10 and 4.12]{andraos2026incentive}. These results are used in the respective papers within different optimization problems subject to incentive constraints. In particular, \cite{gershkov2023optimal} study the insurer's profit maximization problem, \cite{andraos2026incentive} provide a welfare analysis of efficiency, while \cite{ghossoub2025optimal} study both the insurer's profit maximization problem and incentive efficiency. In contrast, the focus of the present paper is on the implementability problem itself, for a given collection of retention functions $\cR$. Proposition \ref{IC_characterization} below characterizes all premium schedules that support an implementable collection $\cR$ of retention functions, up to a common additive constant.

\medskip

\begin{proposition}
\label{IC_characterization}
Suppose that the collection of retention functions $\cR=\{R_\theta\}_{\theta \in \Theta}$ is implementable. Then a premium schedule $p: \Theta \to \R_+$ implements $\cR$ if and only if, for every $\theta \in \Theta$, 
\begin{align*}
p_\theta
&= p_{\underline\theta} + \int_0^{\bar L} \left[1-g_{\underline\theta} \left(F_{\underline\theta}(l)\right)\right]
\frac{\partial R_{\underline\theta}(l)}{\partial l}\,dl
\\& \ \ \  - \int_{\underline\theta}^{\theta} \int_0^{\bar L} \left[\frac{\partial g_s}{\partial s}(F_s(l)) + g_s'(F_s(l)) \frac{\partial F_s(l)}{\partial s} \right] \frac{\partial R_s(l)}{\partial l}\,dl\,ds
\\& \ \ \ - \int_0^{\bar L}\left[1-g_\theta\bigl(F_\theta(l)\bigr)\right]\frac{\partial R_\theta(l)}{\partial l}\,dl .
\end{align*}

\medskip

\noindent In particular, any two premium schedules implementing the same collection $\cR$ only differ by a common additive constant.
\end{proposition}

\medskip

The type space considered in this section, namely $\Theta=[\underline{\theta}, \bar \theta]$, is a compact metric space. Considering an implementable collection of retention functions $\cR$, it follows from Proposition \ref{prop:MeasImplemCompTheta} that if $u_\cR$ is continuous, then $\cR$ is measurably implementable. The following proposition drops the assumption of continuity of $u_\cR$ and establishes an equivalence between implementability and measurable implementability for the special case of compact interval type spaces, by verifying the measurability of the constructed premia in Proposition \ref{IC_characterization}, using Assumption \ref{ass:lipschitz}.

\medskip

\begin{proposition}
\label{prop:imp_measurably_interval}
Consider the compact interval type space $\Theta = [\underline{\theta}, \bar \theta]$. The collection of retention functions $\cR$ is measurably implementable if and only if it is implementable. 
\end{proposition}

\bigskip
\section{Applications to Standard Insurance Contracts}
\label{sec:examples}

In this section, we study implementability for several standard classes of insurance contracts: deductible contracts, proportional insurance contracts, contracts with policy limits and capped deductible contracts. While deductible and policy-limit insurance arise in \cite{gershkov2023optimal} as solutions to the insurer's optimal problem under suitable conditions, our objective in this section is different. We characterize implementability for deductible, proportional, and policy-limit insurance. For capped deductible insurance, we characterize submodularity and obtain sufficient conditions for implementability. Finally, we illustrate the theoretical results through numerical examples.

\medskip

The weak type ordering assumptions imposed in Section \ref{sec:implementability_interval} are sufficient to establish implementability of the submodular collection of retention functions. However, to obtain a complete characterization for the specific classes of insurance contracts, we require a strict separation of types. 
Hence in addition to Assumption \ref{Ass:cdf_family}, we impose the following strict assumption in the remainder of this paper. 

\medskip

\begin{assumption}
\label{ass:strictorder}
For each $\theta \in \Theta$, we have 
$$
0<F_\theta(l) <1, \ \text{ for all $l \in (0, \bar L)$}.
$$

\noindent Moreover, the distortion family is strictly ordered in the following sense:
$$
\frac{\partial g_\theta(t)}{\partial \theta} <0, \ \text{ for all $t\in(0,1)$}. 
$$
\end{assumption}

Assumption \ref{ass:strictorder} ensures that, for every loss level in $(0, \bar L)$, the corresponding cumulative function $F_\theta(l)$ lies strictly between $0$ and $1$. Consequently, every interior loss threshold corresponds to an event that is neither impossible nor certain.
Moreover, Assumption \ref{ass:strictorder}  strengthens Assumption \ref{Ass:distortion_family} by ruling out any two distinct types sharing identical risk attitudes. In particular, higher agent types have
strictly smaller distortion functions and are therefore strictly more risk averse. 

\medskip
\begin{remark}\label{re:strict_chain}
Indeed, every distortion function satisfies  $g_\theta(0) =0$ and $g_\theta(1)=1$ for all $\theta \in \Theta$. Consequently, the distortion family cannot be strictly ordered at the boundary points of its domain, since
$$
\frac{\partial g_\theta (t)}{\partial \theta} =0,  \ \text{ if $t\in \{0,1\}$}. 
$$

\noindent Thus the strict ordering of Assumption
\ref{ass:strictorder} is imposed only on the interior domain $(0,1)$. Hence under Assumption \ref{ass:strictorder}, it follows that:
$$
\frac{\partial }{ \partial \theta } \left [ g_{\theta} \big(  F_{\theta} (l)  \big)  \right ] < 0, \ \text{ for all $l \in (0, \bar L)$}.
$$
\end{remark}

\medskip
\subsection{Deductible Insurance}
For each $\theta \in \Theta$, let $d(\theta) \in [0, \bar L]$ denote the deductible assigned to the type-$\theta$ agent. 
We assume that the deductible
$$d:\Theta \to [0, \bar L]$$
is Borel measurable. 
The corresponding retention function is given by
$$
R_\theta(l) = \min \{ l, d(\theta) \},  \ \text{ for $\theta \in \Theta$ and $l \in [0, \bar L]$}. 
$$

\medskip
\subsubsection{Implementability under Deductible Insurance}
We begin by characterizing implementable collections of deductible retention functions. 

\medskip
\begin{proposition}\label{prop:rationiffsub_deductible}
The following statements are equivalent.
\smallskip
\begin{enumerate}
\item The map $\theta \mapsto d(\theta)$ is non-increasing.

\medskip
\item The collection of deductible retention functions $\{R_\theta\}_{\theta \in \Theta}$ is submodular.

\medskip
\item The collection of deductible retention functions $\{R_\theta\}_{\theta \in \Theta}$ is implementable. 
\end{enumerate}
\end{proposition}

\medskip

Under the maintained 
type ordering assumptions, Proposition \ref{prop:rationiffsub_deductible} shows that a collection of deductible retention
functions is implementable if and only if deductible is non-increasing in type.
In other words, Proposition \ref{prop:rationiffsub_deductible} reduces the abstract cyclical monotonicity
condition of Theorem \ref{th:ImplemIFF} to a simple monotonicity condition on the deductible function when the type space is given by $[\underline{\theta}, \bar \theta]$.

\medskip

\begin{lemma}\label{prop:deductible_premium}
Suppose that the collection of deductible retention functions $\{R_\theta\}_{\theta \in \Theta}$ is implementable. Then the corresponding premium schedule
$\theta \mapsto p_\theta$ is non-decreasing. Moreover, for each $\theta \in \Theta$, 
\begin{align*}
p_\theta
&=p_{\underline\theta}
+\int_0^{d(\underline\theta)}
\left[
1-g_{\underline\theta}(F_{\underline\theta}(l))
\right]\,dl-
\int_{\underline\theta}^{\theta}
\int_0^{d(s)}\left[\frac{\partial g_s}{\partial s}(F_s(l))
+g_s'(F_s(l))\frac{\partial F_s(l)}{\partial s}\right]\,dl\,ds\\
&\quad -
\int_0^{d(\theta)}\left[1-g_\theta(F_\theta(l))\right]\,dl.
\end{align*}
\end{lemma}

\medskip
\subsubsection{Numerical Illustration of Deductible Insurance}
\label{sec:ded_ex}
Consider the type space $\Theta = [0,1]$, and let $\bar L =100$. Suppose that 
$$
\frac{L_\theta}{100} \sim \operatorname{Beta}(1+\theta, 1), \ \text{ for $\theta \in [0,1]$}.
$$  

\noindent Equivalently, the cumulative distribution function of the type-$\theta$ loss $L_\theta$ is given by 
$$
F_\theta(l) = \left( \frac{l}{100} \right) ^{1+\theta}, \ \text{ for all $(\theta,l) \in [0,1] \times [0, 100]$}. 
$$

\noindent Suppose further that the distortion function of the type-$\theta$ agent is given by 
$$
g_\theta(t) =  t^{1+\theta}, \ \text{ for all $(\theta,t) \in [0,1]\times [0,1]$}.
$$

\medskip
\begin{lemma}\label{re:numericalexampleassumption}
The loss distribution family $\{F_\theta\}_{\theta\in \Theta}$ and the distortion family $\{g_\theta\}_{\theta \in \Theta}$ satisfy Assumptions \ref{ass:lipschitz1}, \ref{ass:lipschitz}, \ref{Ass:cdf_family}, and \ref{ass:strictorder}.
\end{lemma}

\medskip
We define the deductible by
$$d(\theta) = 80 - 50 \,\theta,\ \text{ for  $\theta \in [0,1]$}.$$ 

\noindent The corresponding retention function is given by
$$
R_\theta(l) = \min \{l, 80-50 \theta\}, \ \text{for $\theta \in [0,1]$ and $l \in [0, 100]$}.
$$

\medskip
\begin{figure}[H]
\centering
\begin{subfigure}[t]{0.46\linewidth}
\centering
\includegraphics[width=\linewidth]{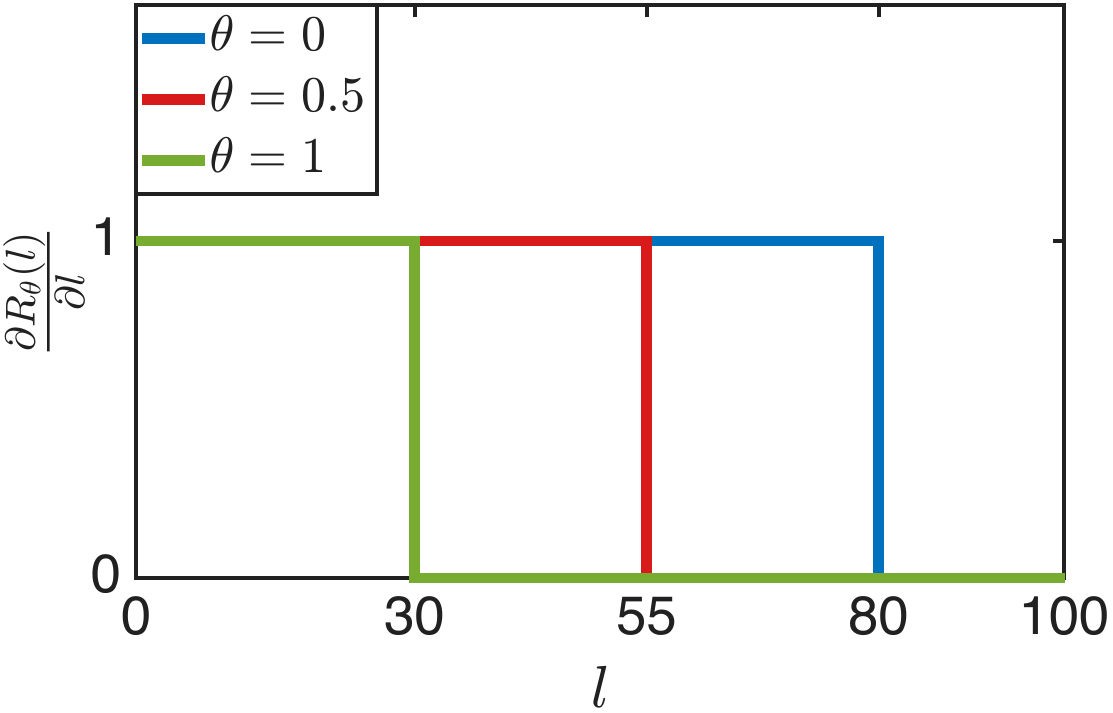}
\caption{Marginal retention functions}
\label{fig:deductibles_submodularity}
\end{subfigure}
\begin{subfigure}[t]{0.47\linewidth}
\centering
\includegraphics[width=\linewidth]{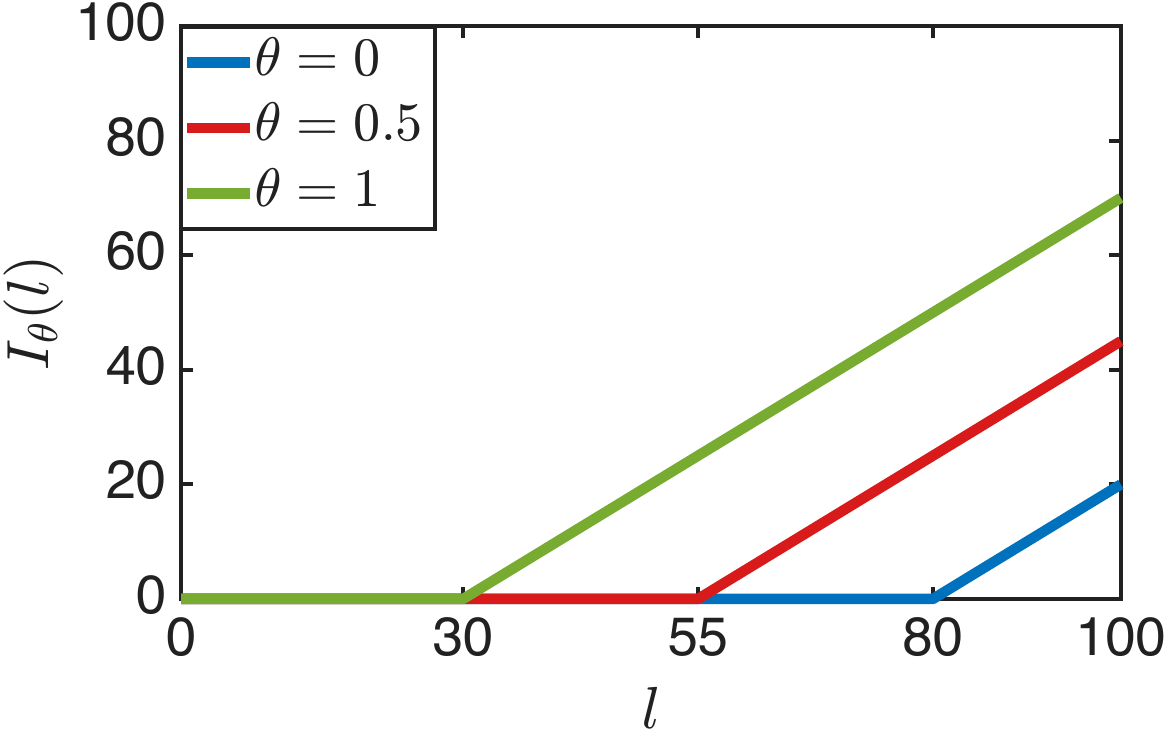}
\caption{Indemnity functions}
\label{fig:deductible_indemnity}
\end{subfigure}
\caption{Coverage and marginal retentions under deductible insurance}
\label{fig:deductible_coverage}
\end{figure}

\medskip
Since the deductible map $\theta \mapsto d(\theta)$ is non-increasing, Proposition \ref{prop:rationiffsub_deductible} implies that the collection of deductible retention functions $\{R_\theta\}_{\theta \in \Theta}$ is submodular and therefore implementable.
Figure \ref{fig:deductibles_submodularity} illustrates this submodularity for the three representative types $\theta=0, \ \theta=0.5,$ and $\theta=1$. Specifically, for every fixed loss level $l$, the marginal retention weakly decreases as the type increases.

\medskip
As the type increases, the lower deductible shifts the loss threshold at which coverage begins to the left. 
Figure \ref{fig:deductible_indemnity} illustrates the corresponding indemnity functions for the three representative types and shows that coverage begins at a lower loss level for higher types. 
Consequently, for any given realized loss, higher types of the agent, who are more risk averse and face stochastically larger losses, receive weakly greater coverage. 

\medskip
Let $\{p_\theta\}_{\theta\in\Theta}$ denote the corresponding collection of premia to the implementable collection of deductible retention functions $\{R_\theta\}_{\theta \in \Theta}$. It follows from Lemma \ref{prop:deductible_premium} that for each $\theta \in [0,1]$, 
\begin{align*}
p_\theta
&= p_0 +48 - \int_0^\theta\int_0^{80-50s}2(1+s)
\left(\frac{l}{100}\right)^{(1+s)^2}
\ln\left(\frac{l}{100}\right)\,dl\,ds
- \int_0^{80-50 \, \theta}\left[1-\left(\frac {l}{100}\right)^{(1+\theta)^2}\right]dl.
\end{align*}

\medskip
\begin{assumption}
\label{ass:p0}
We normalize the premium of the lowest type $\underline{\theta}=0$ by setting $p_0=0$. 
\end{assumption}

\medskip

\begin{figure}[H]
\centering
\begin{subfigure}[t]{0.44\linewidth}
\centering
\includegraphics[width=\linewidth]{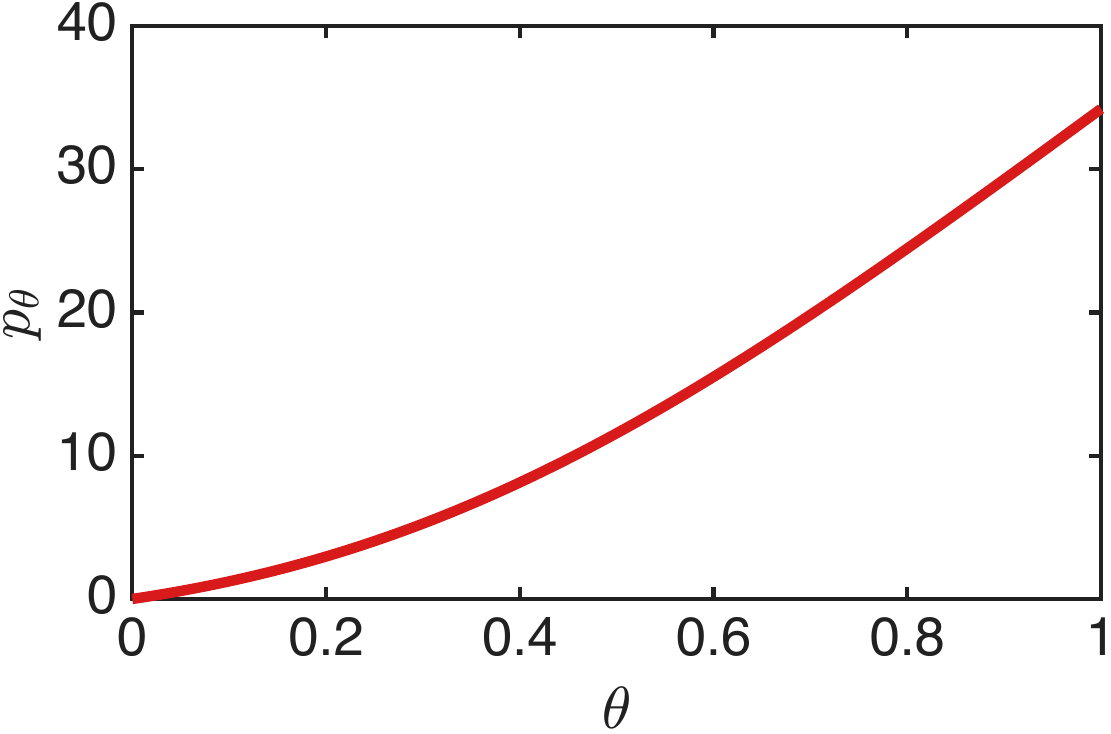}
\caption{Corresponding Constructed Premia}
\label{fig:deductible_premium}
\end{subfigure}
\begin{subfigure}[t]{0.45\linewidth}
\centering
\includegraphics[width=\linewidth]{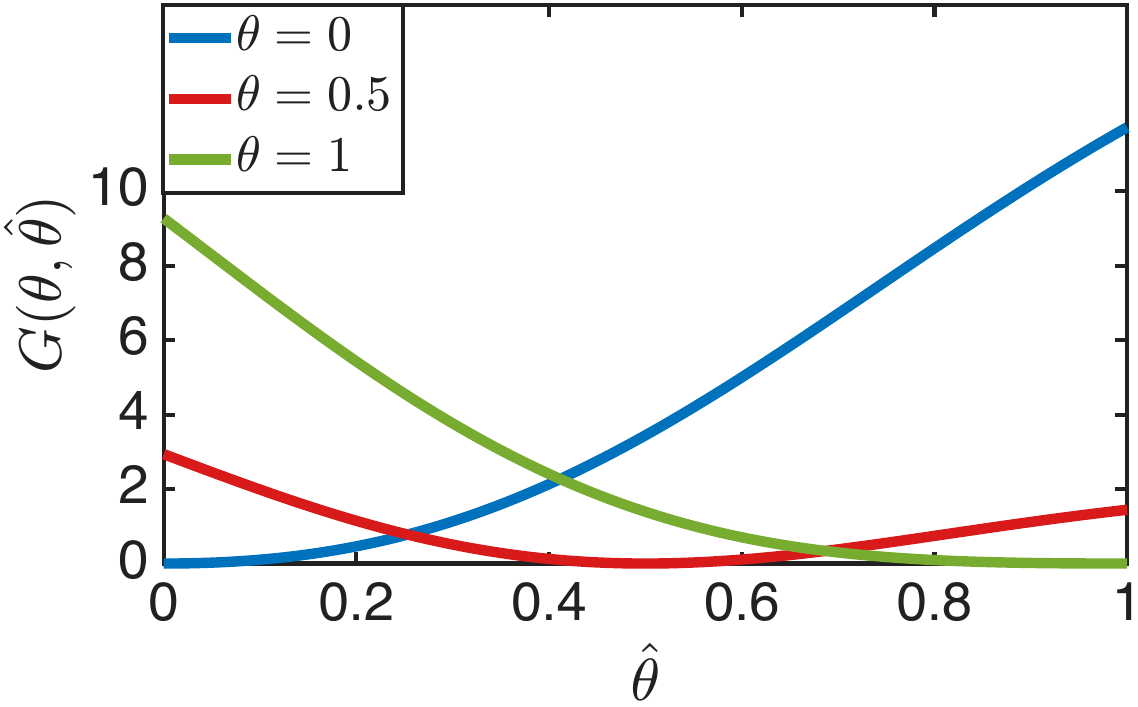}
\caption{Incentive Compatibility}
\label{fig:deductible_IC}
\end{subfigure}
\caption{Numerical illustration under deductible insurance}
\label{fig:deductible_premium_IC}
\end{figure}

The expression of the premium paid by the type-$\theta$ agent to the insurer depends on the value of $p_0 \in \R_+$ for all $\theta \in \Theta$. 
For any constant $c\in\R$ such that $p_\theta + c \geq 0$ for all $\theta \in \Theta$, the translated collection
$\{p_\theta+c\}_{\theta\in\Theta}$ generates the same shape of the
premium schedule, shifted vertically by $c$. Thus, changing the value
of $p_0$ changes the level of the entire premium curve but not its
monotonicity, convexity, or differences across types. Moreover, we know that the constructed premium schedule is non-decreasing in $\theta$ as illustrated in Figure \ref{fig:deductible_premium} consistently with Lemma \ref{prop:deductible_premium}. Then, we have that 
$$
p_\theta \geq p_0 \geq 0,\  \forall \theta \in \Theta. 
$$

\noindent Hence, we can choose $p_0=0$, which justifies Assumption \ref{ass:p0}. Figure \ref{fig:deductible_premium} illustrates that higher types, who are more risk averse and face stochastically larger losses, receive lower deductibles and are willing to pay higher premia in exchange for greater coverage. Moreover, the premium is convex, indicating that the marginal increase in the premium becomes larger as the agent's type increases. In other words, the willingness to pay for additional coverage increases with the agent's type. 

\medskip

\medskip

For a true type $\theta$ and a reported type $\hat\theta$, define the
incentive compatibility gap by 
$$
G(\theta, \hat \theta) = \left[u_\cR(\theta)-p_\theta \right]- \left[ u_\cR (\theta, \hat \theta) - p_{\hat \theta}\right],  
$$

\noindent that is the difference between the utility of the type-$\theta$ agent from truthful selection and the utility of the type-$\theta$ agent from misreporting.
Note that the incentive compatibility gap remains unchanged under a common translation of the premia $\{p_\theta+c\}_{\theta \in \Theta}$ for all $c \in \R$, since
\begin{align*}
G_c(\theta, \hat \theta) 
&= \left[u_\cR(\theta)-(p_\theta+c) \right]- \left[ u_\cR (\theta, \hat \theta) - (p_{\hat \theta}+c)\right] \\
&= \left[u_\cR(\theta)-p_\theta \right]- \left[  u_\cR (\theta, \hat \theta) - p_{\hat \theta}\right]\\
&= G(\theta, \hat \theta).
\end{align*}

\noindent Hence, the normalization adopted in Assumption \ref{ass:p0} does not affect incentive compatibility.
Figure \ref{fig:deductible_IC} illustrates the incentive compatibility gap $G(\theta, \hat \theta)$ as a function of the reported type for the representative true types $\theta=0$, $\theta=0.5$, and $\theta=1$. For each true type, the gap attains its minimum value of zero at the truthful report $\hat\theta=\theta$, that is, $G(\theta, \theta)=0$, and remains non-negative for all alternative reports, that is, $G(\theta, \hat \theta) \geq0$ for all $\hat \theta \neq \theta$. Consequently, truthful reporting weakly dominates every possible misreport, confirming the incentive compatibility of the constructed deductible menu $(R_\theta, p_\theta)_{\theta \in \Theta}$.

\medskip
\subsection{Proportional Insurance}
For each $\theta\in \Theta=[\underline{\theta}, \bar \theta ]$, let $a(\theta) \in [0, 1]$ denote the coinsurance rate, that is, the proportion of the loss of the type-$\theta$ agent covered by the insurer. We assume that the coinsurance map
$$a:\Theta \to [0, 1]$$ 
is Borel measurable. The corresponding retention function is given by
$$
R_\theta(l) = \left[ 1-a(\theta) \right]\cdot l, \ \text{ for $\theta \in \Theta$ and $l \in [0, \bar L]$}. 
$$

\medskip
\subsubsection{Implementability under Proportional Insurance} 
We begin by characterizing implementability of the collection of retention functions under proportional insurance. 

\medskip
\begin{proposition}\label{prop:proportional_rationiffsubmod}
The following statements are equivalent. 
\smallskip
\begin{enumerate}
\item The map $\theta \mapsto a(\theta)$ is non-decreasing.

\medskip
\item The collection of retention functions $\{R_\theta\}_{\theta \in \Theta}$ is submodular. 

\medskip
\item The collection of retention functions $\{R_\theta\}_{\theta \in \Theta}$ is implementable. 
\end{enumerate}
\end{proposition}

\medskip

Under the maintained type ordering assumptions, Proposition \ref{prop:proportional_rationiffsubmod} reduces the cyclical monotonicity condition of Theorem \ref{th:ImplemIFF} to a simple monotonicity condition on the coinsurance map $\theta \mapsto a(\theta)$. In particular, a collection of proportional retention functions is implementable if and only if the map $\theta\mapsto a(\theta)$ is non-decreasing. 

\medskip
\begin{lemma}\label{prop:proportional_premia}
Suppose that the collection of proportional retention functions
$\{R_\theta\}_{\theta\in\Theta}$ is implementable, then the corresponding premium schedule 
$\theta \mapsto p_\theta$
is non-decreasing. Moreover, if $a:\Theta \to [0,1]$ is absolutely continuous, then for each $\theta \in\Theta$, 
\begin{align*}
p_\theta=p_{\underline\theta}
+\int_{\underline\theta}^{\theta}
\int_0^{\bar L}a^\prime (s)\left[1-g_s(F_s(l))\right]dl\,ds.
\end{align*}
\end{lemma}

\medskip
\subsubsection{Numerical Illustration of Proportional Insurance}
Consider the type space $\Theta = [0,1]$ and let $\bar L =100$. We adopt the same loss distribution and distortion family as in the numerical example of Section \ref{sec:ded_ex}. Particularly, 
$$
F_\theta(l)= \left( \frac{l}{100} \right) ^{1+\theta}, \ \text{ for all $(\theta,l) \in [0,1]\times [0,100]$},  
$$
and,
$$
g_\theta(t) =  t^{1+\theta}, \ \text{ for all $(\theta,t) \in [0,1] \times [0,1]$}.
$$

\medskip

\noindent It follows from Lemma \ref{re:numericalexampleassumption} that Assumptions \ref{ass:lipschitz}, \ref{Ass:cdf_family} and \ref{ass:strictorder} are satisfied. 
We define the proportion of the loss of the type-$\theta$ agent covered by the insurer by
$$
a(\theta)= 0.2 + 0.5 \, \theta , \ \text{for each $\theta \in [0,1]$}. 
$$

\noindent The corresponding retention function is given by
$$
R_\theta(l) = [0.8 - 0.5 \, \theta] \,l, \ \text{ for $(\theta, l) \in [0,1] \times  [0,100]$}. 
$$

\medskip
\begin{figure}[H]
\centering
\begin{subfigure}[t]{0.45\linewidth}
\centering
\includegraphics[width=\linewidth]{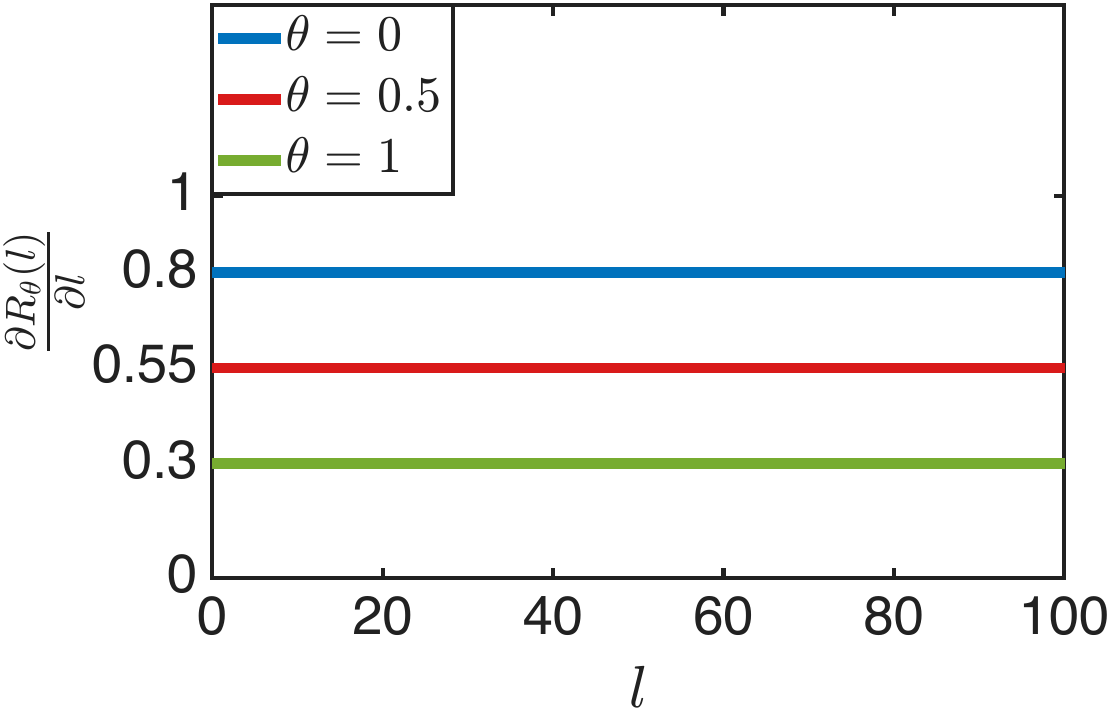}
\caption{Marginal Retention Functions}
\label{fig:proportional_retentions}
\end{subfigure}
\begin{subfigure}[t]{0.46\linewidth}
\centering
\includegraphics[width=\linewidth]{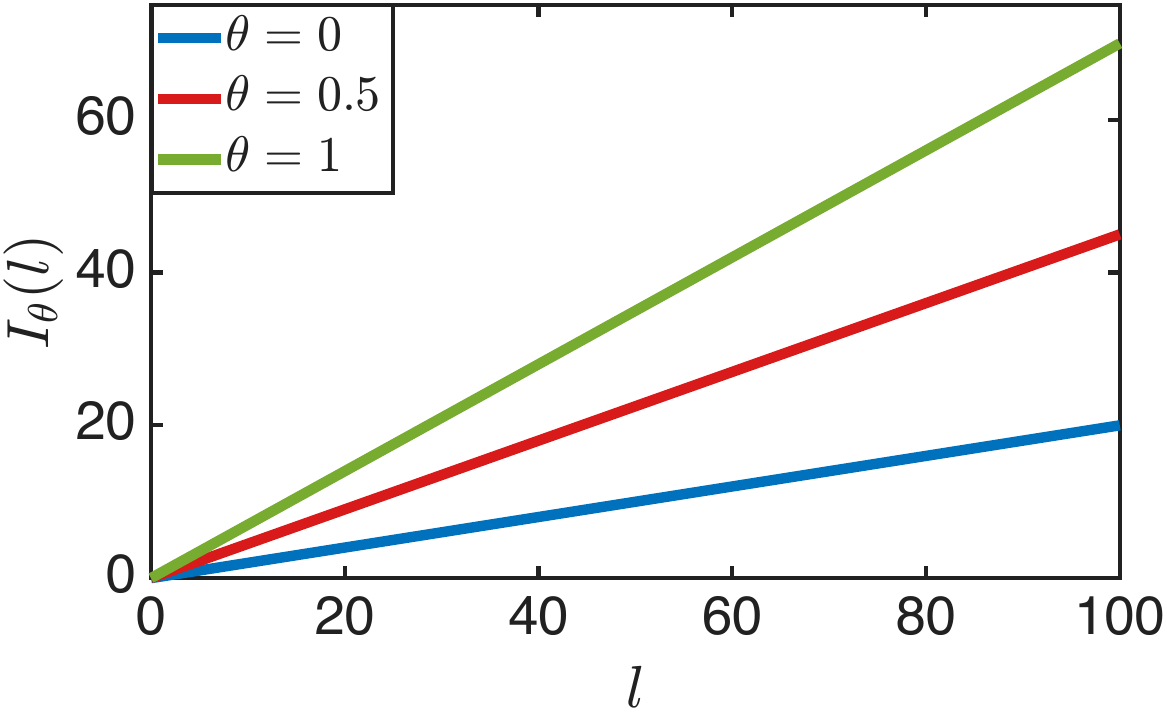}
\caption{Indemnity Functions}
\label{fig:proportional_indemnities}
\end{subfigure}
\caption{Coverage and marginal retentions under proportional insurance.}
\label{fig:proportional_coverage}
\end{figure}

\medskip
Since $\theta \mapsto a(\theta)$ is non-decreasing, it follows from Proposition \ref{prop:proportional_rationiffsubmod} that the collection of retention functions $\{R_\theta\}_{\theta \in \Theta}$ is submodular under proportional insurance and therefore implementable.
Figure \ref{fig:proportional_retentions} illustrates the submodularity of the marginal retention functions for the three representative types $\theta=0$, $\theta=0.5$, and $\theta=1$. The figure shows that the marginal retention under proportional insurance decreases as the type $\theta$ increases.

\medskip 
Figure \ref{fig:proportional_indemnities} displays the corresponding indemnity functions for the three representative types and shows that higher types, who are more risk averse and face stochastically larger losses, receive more coverage for every realized loss.

\medskip
Using the normalization of Assumption \ref{ass:p0}, i.e. $p_0=0$, we know from Lemma \ref{prop:proportional_premia}, that the premium paid by the type-$\theta$ agent is given by
\begin{align*}
p_\theta 
&=50\left[\theta-\arctan(1+\theta)+\frac{\pi}{4}\right], \ \text{ for all $\theta \in [0,1]$.}
\end{align*}

\medskip
\begin{figure}[H]
\centering
\begin{subfigure}{0.45\linewidth}
\centering
\includegraphics[width=\linewidth]{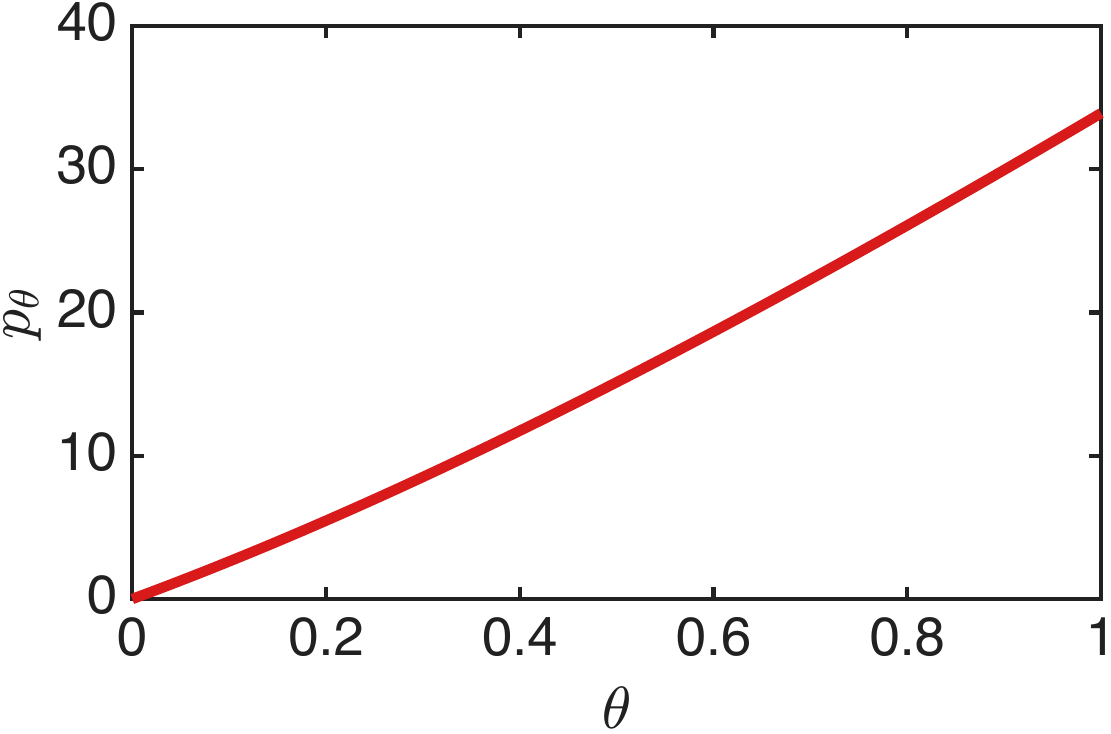}
\caption{Corresponding Constructed Premia}
\label{fig:proportional_premium}
\end{subfigure}
\begin{subfigure}{0.46\linewidth}
\centering
\includegraphics[width=\linewidth]{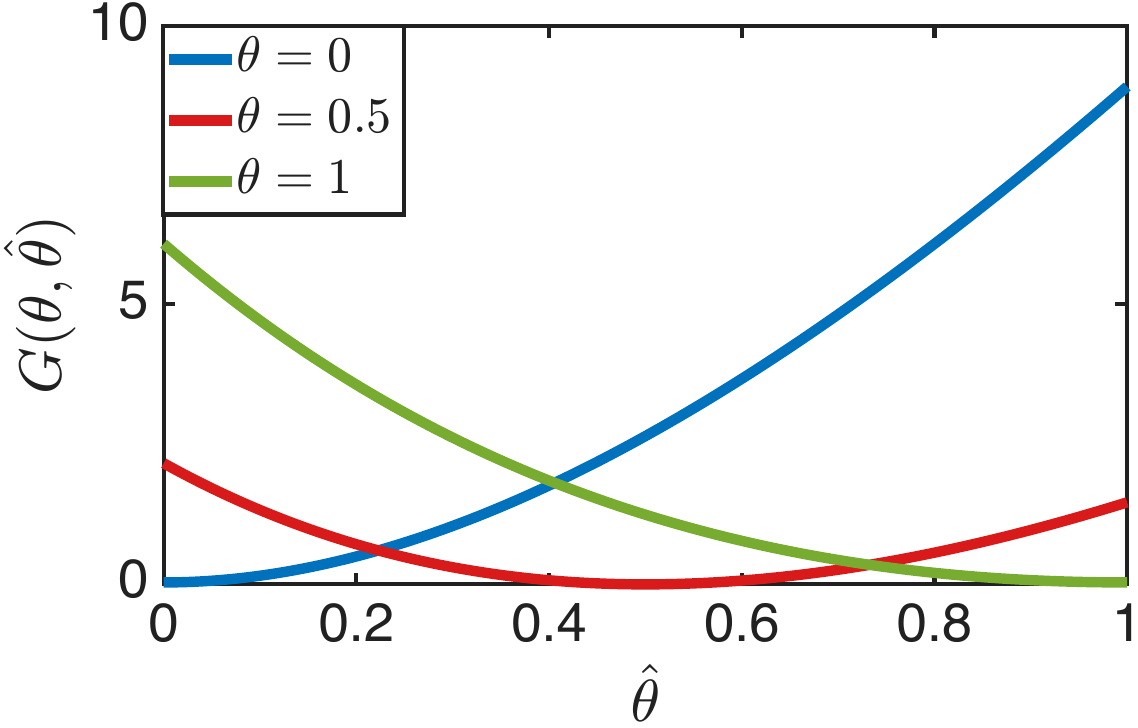}
\caption{Incentive Compatibility}
\label{fig:proportional_IC}
\end{subfigure}
\caption{Numerical Illustration of Proportional Insurance}
\label{fig:proportional_numerical}
\end{figure}

\medskip

Figure \ref{fig:proportional_premium} shows that, consistently with
Lemma \ref{prop:proportional_premia}, the premium is non-decreasing in type. Higher types, who are more risk averse and face stochastically larger losses, receive a larger proportional coverage rate and pay a weakly higher premium. Moreover, the premium is convex, indicating that the premium required to support additional coverage increases at an increasing rate. 

\medskip

For a true type-$\theta$ agent and reported type $\hat\theta$, consider the incentive compatibility gap 
$$
G(\theta,\hat\theta)=\bigl[u_\cR(\theta)-p_\theta\bigr]-\bigl[ u_\cR (\theta, \hat \theta)-p_{\hat\theta}\bigr], \ \text{ for $\theta, \hat \theta \in [0,1]$}.
$$

\noindent Figure \ref{fig:proportional_IC} plots this gap for the representative true types. In each case, the gap attains its minimum value of zero at the truthful report $\hat\theta=\theta$, that is, $G(\theta, \theta)=0$, and remains non-negative for all alternative reports, that is, $G(\theta, \hat \theta)\geq0$. This illustrates the incentive compatibility
of the constructed proportional menu $(R_\theta, p_\theta
)_{\theta \in \Theta}$.

\medskip
\subsection{Policy-Limit Insurance}
For each $\theta \in \Theta$, let $m(\theta) \in [0, \bar L]$ denote the maximum amount paid by the insurer to the type-$\theta$ agent. We assume that the policy-limit map 
$$m:\Theta \to [0, \bar L]$$
is Borel measurable. The corresponding retention function is given by
$$
R_\theta(l) = l - \min \{l, m(\theta)\} = ( l - m(\theta))^+ , \ \text{ for $l \in [0, \bar L]$}. 
$$

\medskip
\subsubsection{Implementability under Policy-Limit Insurance}
We begin by characterizing implementability for collections of policy-limit retention functions.

\medskip
\begin{proposition}\label{prop:policylimit_submodulariffimplementable}
The following statements are equivalent. 
\smallskip
\begin{enumerate}
\item The map $\theta \mapsto m(\theta)$ is non-decreasing. 

\medskip
\item The collection of retention functions $\{R_\theta\}_{\theta \in \Theta}$ is submodular.

\medskip
\item The collection of retention functions $\{R_\theta\}_{\theta \in \Theta}$ is implementable. 
\end{enumerate}
\end{proposition}

\medskip

Under the maintained type ordering assumptions, Proposition \ref{prop:policylimit_submodulariffimplementable} reduces the cyclic monotonicity condition of Theorem \ref{th:ImplemIFF} and shows that implementability of the collection of retention functions is equivalent to the policy-limit map being non-decreasing. 

\medskip
\begin{lemma}
\label{prop:policylimit_premia}
Suppose that the collection of retention functions $\{R_\theta\}_{\theta \in \Theta}$ is implementable, then the corresponding premium schedule $\theta \mapsto p_\theta$ is non-decreasing. Moreover, if $m:\Theta \to [0, \bar L]$ is absolutely continuous, then for each $\theta \in \Theta$, 
\begin{align*}
p_\theta 
&= p_{\underline{\theta}} +\int_{\underline{\theta}}^\theta m^\prime(s) \left[ 1-g_s\big(F_s(m(s))\big) \right] \,ds. 
\end{align*}
\end{lemma}

\medskip
\subsubsection{Numerical Illustration of Policy-Limit Insurance}
\label{sec:pl_ex}
Consider the type space $\Theta = [0,1]$ and let $\bar L =100$. We adopt the same loss distribution and distortion family as in the numerical example of Section \ref{sec:ded_ex}. Particularly, 
$$
F_\theta(l)= \left( \frac{l}{100} \right) ^{1+\theta}, \ \text{ for all $(\theta,l) \in [0,1]\times [0,100]$},  
$$
and,
$$
g_\theta(t) =  t^{1+\theta}, \ \text{ for all $(\theta,t) \in [0,1] \times [0,1]$}.
$$

\medskip

\noindent It follows from Lemma \ref{re:numericalexampleassumption} that Assumptions \ref{ass:lipschitz}, \ref{Ass:cdf_family} and \ref{ass:strictorder} are satisfied. We define the policy-limit by 
$$
m(\theta) = 20 + 50 \, \theta, \ \text{for all $\theta \in [0,1]$}. 
$$

\noindent The corresponding retention function is given by
$$
R_\theta(l)=(l-20-50\theta)^+,  \ \text{ for $(\theta,l) \in [0,1] \times [0,100]$}. 
$$

\medskip
\begin{figure}[H]
\centering
\begin{subfigure}{0.45\linewidth}
\centering
\includegraphics[width=\linewidth]{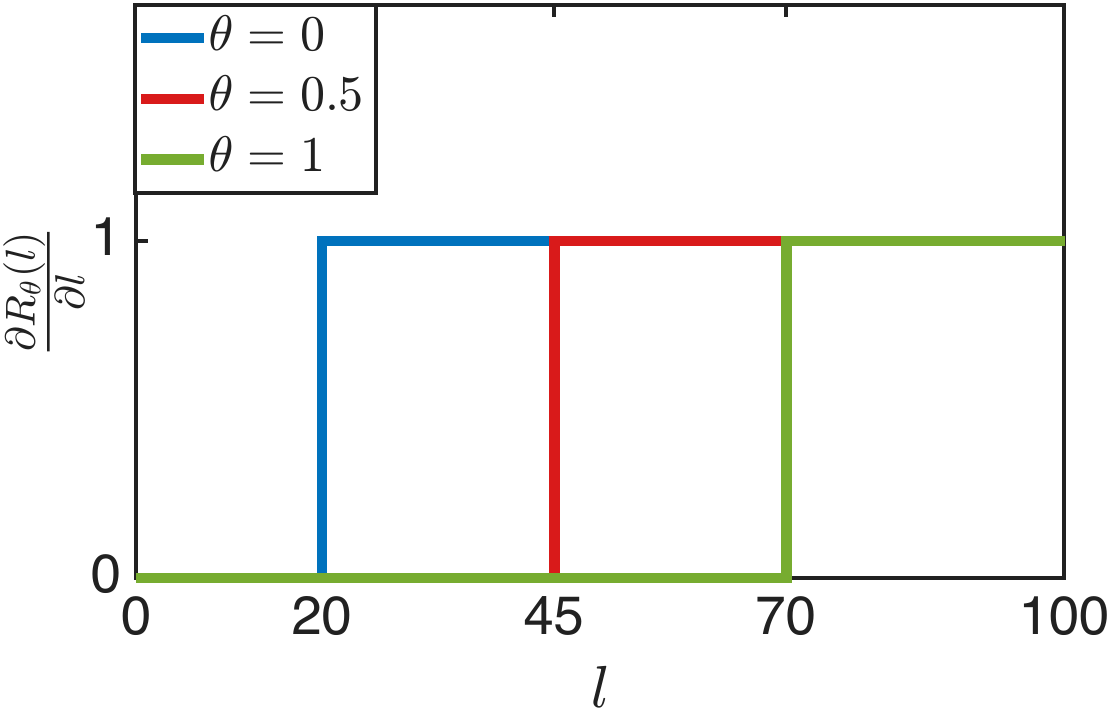}
\caption{Marginal Retention Functions}
\label{fig:policylimit_R}
\end{subfigure}
\begin{subfigure}{0.46\linewidth}
\centering
\includegraphics[width=\linewidth]{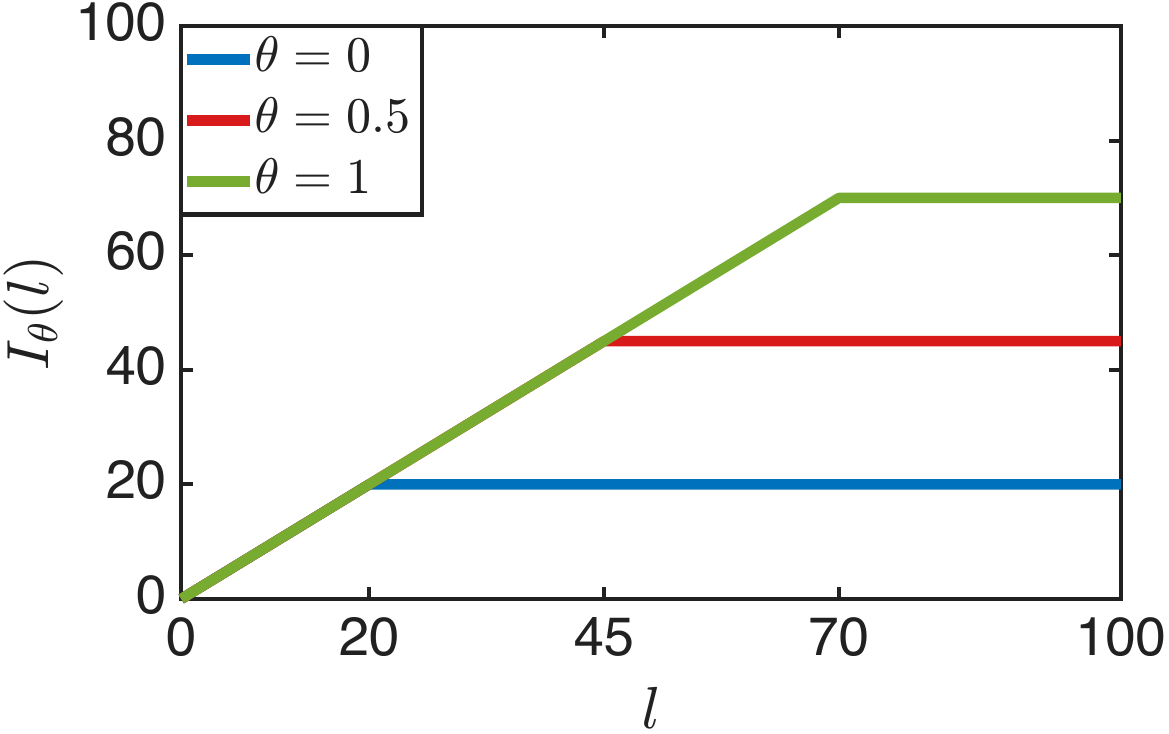}
\caption{Indemnity Functions}
\label{fig:policylimit_I}
\end{subfigure}
\caption{Coverage and marginal retentions under policy-limit insurance.}
\label{fig:policylimitfig1}
\end{figure}

\medskip
Since the policy-limit map $\theta \mapsto m(\theta)$ is non-decreasing, it follows from Proposition \ref{prop:policylimit_submodulariffimplementable} that the collection of retention functions $\{R_\theta\}_{\theta \in \Theta}$ is submodular and therefore implementable. Figure \ref{fig:policylimit_R} illustrates the submodularity of the collection of retention functions using three representative types $\theta=0$, $\theta=0.5$ and $\theta=1$. Moreover, the figure shows that the jump in the marginal retention occurs at progressively larger loss levels as the type increases, reflecting the increase in the policy-limit.

\medskip

Figure \ref{fig:policylimit_I} presents the corresponding indemnity functions and shows that higher types, who are more risk averse and face stochastically larger losses, receive full coverage over a larger range of losses and are assigned a higher maximum indemnity. Consequently, indemnity functions are ordered by type, and higher types receive weakly greater coverage at every loss level.  

\medskip

Using the normalization of Assumption \ref{ass:p0}, $p_0=0$, it follows from Lemma \ref{prop:policylimit_premia} that the premium paid by the type-$\theta$ agent satisfies
$$
p_\theta
=50\int_0^\theta
\left[1-\left(0.2+0.5s\right)^{(1+s)^2}
\right]ds, \ \text{ for each $\theta \in [0,1]$}. 
$$

\medskip
\begin{figure}[H]
\centering
\begin{subfigure}{0.45\linewidth}
\centering
\includegraphics[width=\linewidth]{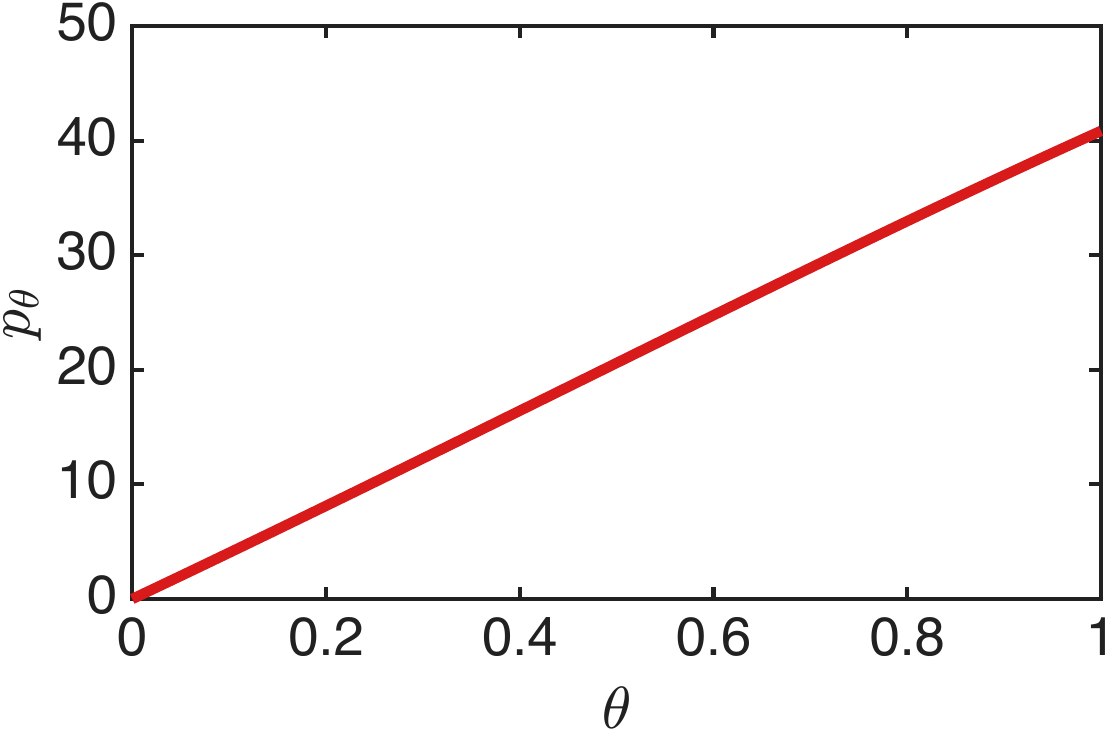}
\caption{Corresponding Constructed Premia}
\label{fig:policylimit_p}
\end{subfigure}
\begin{subfigure}{0.45\linewidth}
\centering
\includegraphics[width=\linewidth]{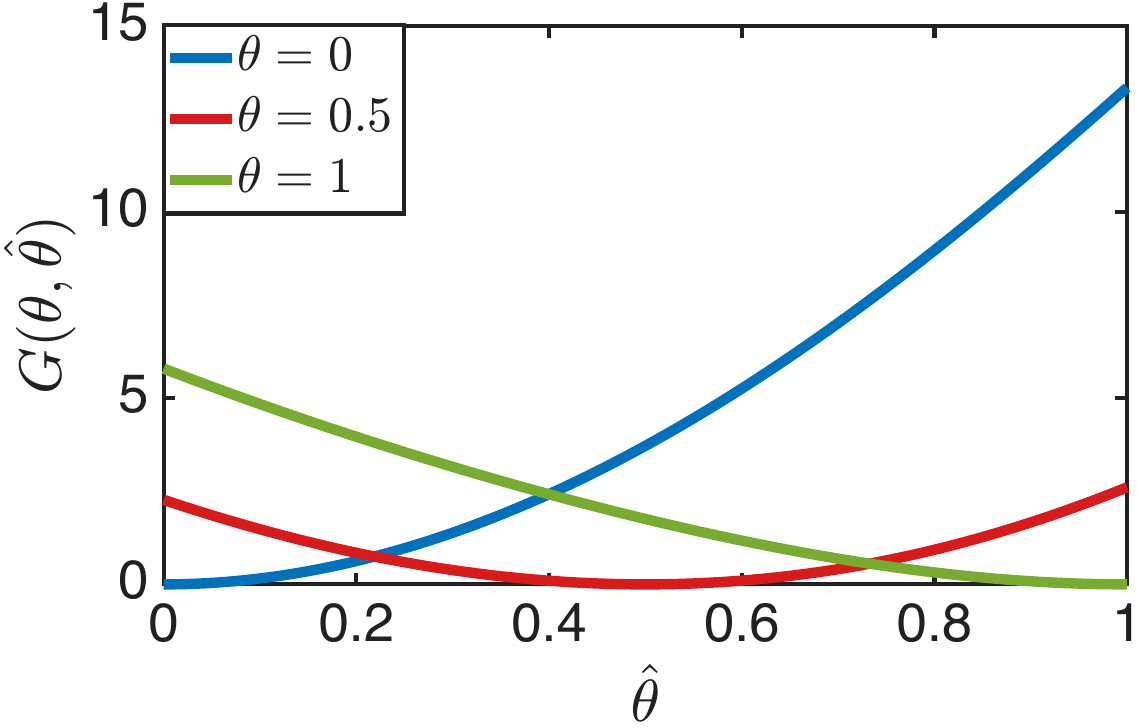}
\caption{Incentive Compatibility}
\label{fig:policylimit_IC}
\end{subfigure}
\caption{Numerical illustration under policy-limit insurance.}
\label{fig:policylimitfig2}
\end{figure}

\medskip
Figure \ref{fig:policylimit_p} displays the corresponding premia. Consistently with Lemma \ref{prop:policylimit_premia}, the premium is non-decreasing in type, that is, contracts providing larger policy limits yield more coverage to the agent and are supported by weakly higher premia. Unlike deductible and proportional insurance, the premium is not convex on the whole interval $[0,1]$. Specifically, the premium is convex for lower types, however beyond a certain point, the premium becomes concave. 
Although higher types continue to pay higher premia as they receive larger policy limits, the marginal increase in the premium gradually declines. In other words, for sufficiently high policy limits, each additional unit of coverage requires a smaller increase in the corresponding premium than at lower policy limits. 

\medskip

For a true type $\theta$ and a reported type $\hat\theta$, define the
incentive compatibility gap by 
$$
G(\theta, \hat \theta) = \left[u_\cR(\theta)-p_\theta \right]- \left[ u_\cR (\theta, \hat \theta) - p_{\hat \theta}\right].  
$$

\noindent Figure \ref{fig:policylimit_IC} plots the gap $G(\theta, \hat \theta)$ as a function of the reported type $\hat \theta \in \Theta$ for the representative true types $\theta=0$, $\theta=0.5$, and $\theta=1$. For each true type, the gap attains its minimum value of zero at the truthful report $\hat\theta=\theta$, that is, $G(\theta, \theta)=0$, and remains non-negative for all alternative reports, that is, $G(\theta, \hat \theta) \geq0$ for all $\hat \theta \neq \theta$, confirming the incentive compatibility of the constructed menu $(R_\theta, p_\theta)_{\theta \in \Theta}$.

\medskip
\subsection{Capped Deductible Insurance} 
For each $\theta \in \Theta$, let $d(\theta)\in[0,\bar L]$ denote the deductible assigned to the type-$\theta$ agent and let $m(\theta) \in (0, \bar L]$ denote the maximum indemnity paid by the insurer to the type-$\theta$ agent. We assume that the deductible map and policy-limit map
$$
d:\Theta\to[0,\bar L], \ \text{ and } \ m:\Theta \to (0, \bar L]
$$ 
are Borel measurable and satisfy
$$
d(\theta)+ m(\theta) \leq \bar L , \ \text{for all $\theta \in \Theta$}. 
$$

\noindent The corresponding retention function is given by $$R_\theta(l) = l- \min \{(l-d(\theta))^+, m(\theta) \}\ \text{ for all $(\theta,l) \in \Theta \times [0, \bar L]$}.
$$

\noindent Or equivalently, 
$$
R_\theta(l)= 
\begin{cases}
l & 0\leq l\leq d(\theta), \\
d(\theta) & d(\theta) \leq l \leq d(\theta)+m(\theta), \\
l-m(\theta) &l>d(\theta) +m(\theta), 
\end{cases}
\ \text{ for all $(\theta,l) \in \Theta \times [0, \bar L]$}.
$$

\medskip
\subsubsection{Implementability under Capped Deductible Insurance}
We first characterize submodularity for collections of capped deductible retention functions and then use this characterization to obtain sufficient conditions for implementability. 

\medskip
\begin{proposition}\label{prop:rationiffsub_capped}
The collection of capped deductible retention functions $\{R_\theta\}_{\theta \in \Theta}$ is submodular if and only if both of the following conditions hold:
\smallskip
\begin{enumerate}
\item The deductible map $\theta \mapsto d(\theta)$ is non-increasing.

\medskip
\item The map $\theta \mapsto d(\theta) + m(\theta)$ is non-decreasing. 
\end{enumerate}
\end{proposition}

\medskip

Proposition \ref{prop:rationiffsub_capped} reduces submodularity of the collection of capped deductible retention functions to two monotonicity conditions on the contract parameters. The first requires the deductible to be non-increasing in type. The second requires the map $\theta \mapsto d(\theta)+ m(\theta)$ to be non-decreasing. Assuming that the deductible and policy-limit maps are absolutely continuous, these conditions can be equivalently written as
$$
d^\prime(\theta) \leq 0 
\ \ \text{ and } \ \ 
m^\prime(\theta) \geq - d^\prime(\theta), \ \text{for a.e. $\theta \in \Theta$}.
$$

\medskip

\noindent Thus, the policy limit is non-decreasing in type, and its increase must be at least as large as the decrease in the deductible. In other words, the policy-limit is more sensitive to increases in type than the deductible.

\medskip

The following result follows immediately from Proposition \ref{prop:cyclicmonotone} and Theorem \ref{th:ImplemIFF}.

\medskip
\begin{corollary}\label{prop:capped_implementable}
Every submodular collection of capped deductible retention functions $\{R_\theta\}_{\theta \in \Theta}$ is implementable. 
\end{corollary}

\medskip
\begin{lemma}\label{prop:capped_premium}
Suppose that $\theta\mapsto d(\theta)$ is non-increasing and that
$\theta\mapsto d(\theta)+m(\theta)$ is non-decreasing. Then the corresponding premium schedule $\theta \mapsto p_\theta$ is non-decreasing. Moreover, if $d:\Theta \to [0, \bar L]$ and $m:\Theta \to (0, \bar L]$ are absolutely continuous, then for each $\theta \in \Theta$, the premium is given by:
\begin{align*}
p_\theta
&= p_{\underline\theta}
-\int_{\underline\theta}^{\theta}
d^\prime(s)
\left[1-g_s(F_s(d(s)))\right]\,ds
+
\int_{\underline\theta}^{\theta}
\left[d^\prime(s)+m^\prime(s)\right]
\left[1-g_s\big( F_s(d(s)+m(s)) \big)\right]\,ds.
\end{align*}
\end{lemma}

\medskip
\subsubsection{Numerical Illustration of Capped Deductible Insurance}
Consider the type space $\Theta = [0,1]$ and let $\bar L =100$. We adopt the same loss distribution and distortion family as in the numerical example of Section \ref{sec:ded_ex}. Particularly, 
$$
F_\theta(l)= \left( \frac{l}{100} \right) ^{1+\theta}, \ \text{ for all $(\theta,l) \in [0,1]\times [0,100]$}, 
$$
and
$$
g_\theta(t) =  t^{1+\theta}, \ \text{ for all $(\theta,t) \in [0,1] \times [0,1]$}.
$$

\noindent It follows from Lemma \ref{re:numericalexampleassumption} that Assumptions \ref{ass:lipschitz}, \ref{Ass:cdf_family} and \ref{ass:strictorder} are satisfied. We define the deductible as 
$$
d(\theta) = 70-40 \, \theta, \ \text{ for $\theta \in [0,1]$},
$$
and the policy-limit as
$$
m(\theta) = 20 + 50 \, \theta, \ \text{ for $\theta \in [0,1]$}.
$$

\noindent Then, 
$$
d(\theta) +m(\theta) = 90 + 10 \,\theta \leq 100, \ \text{ for all $\theta \in [0,1]$}. 
$$

\noindent The corresponding capped deductible retention function is given by
$$
R_\theta(l) = 
\begin{cases}
l & 0\leq l \leq  70-40 \, \theta,\\
70-40 \, \theta &  70-40 \, \theta<l\leq 90 + 10 \,\theta, \\
l- (20 +50\,\theta) &  90 + 10 \,\theta< l \leq 100, 
\end{cases}
\ \text{ for all $(\theta, l) \in [0,1]\times[0,100]$}.
$$

\medskip
\begin{figure}[H]
\centering
\begin{subfigure}[t]{0.45\linewidth}
\centering
\includegraphics[width=\linewidth]{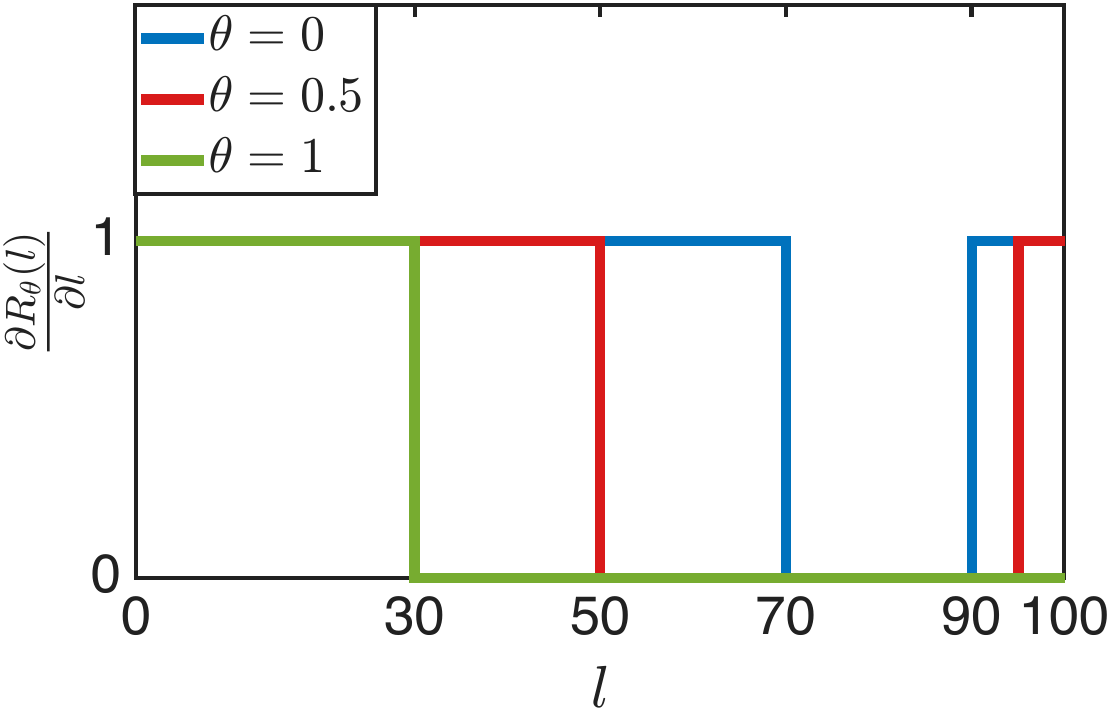}
\caption{Marginal retention functions}
\label{fig:capped_submodularity}
\end{subfigure}
\begin{subfigure}[t]{0.46\linewidth}
\centering
\includegraphics[width=\linewidth]{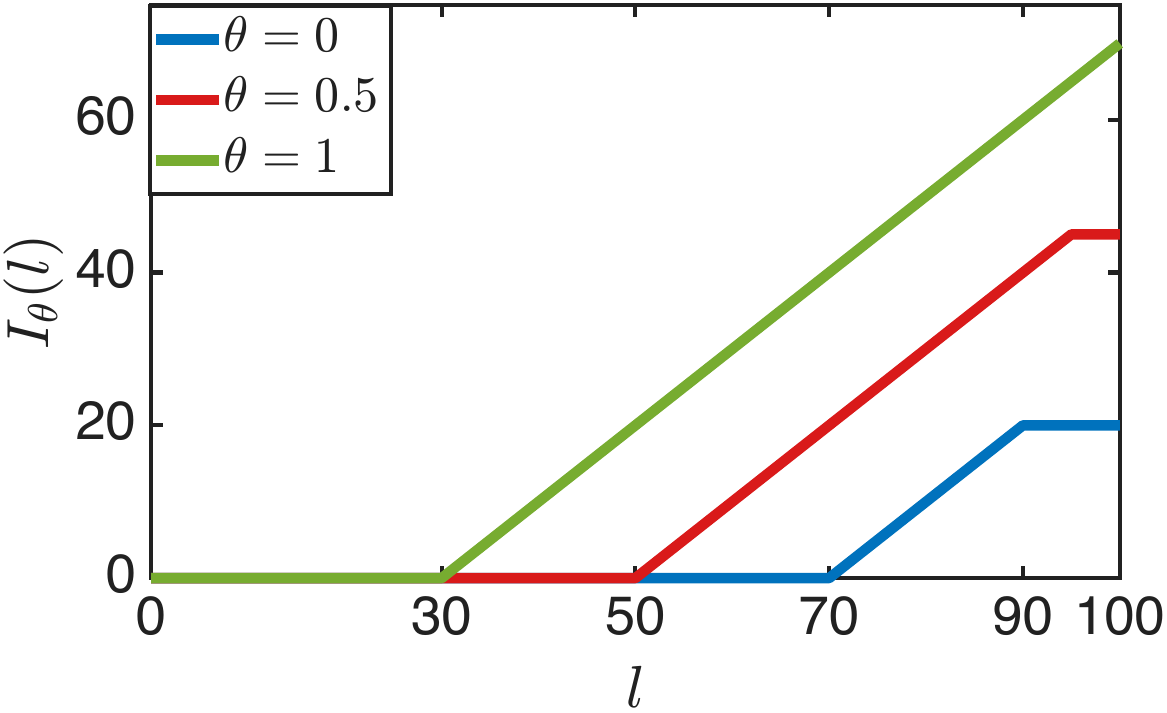}
\caption{Indemnity functions}
\label{fig:capped_indemnity}
\end{subfigure}
\caption{Coverage and marginal retentions under capped deductible insurance}
\label{fig:cappeddeductible_coverage}
\end{figure}

\medskip

Since the deductible map $\theta \mapsto d(\theta)$ is non-increasing, and the map $\theta \mapsto d(\theta) + m(\theta)$ is non-decreasing, it follows from Proposition \ref{prop:rationiffsub_capped} that the collection of capped deductible retention functions is submodular and therefore implementable by Corollary \ref{prop:capped_implementable}. Figure \ref{fig:capped_submodularity} illustrates this submodularity for the three representative types $\theta=0$, $\theta=0.5$ and $\theta=1$.

\medskip

Figure \ref{fig:capped_indemnity} presents the corresponding indemnity functions. As the agent's type increases, coverage begins at a lower loss level and the indemnity reaches its maximum at higher loss levels. This combines the two coverage patterns considered previously: the decrease in the deductible allows coverage to begin earlier as in Section \ref{sec:ded_ex}, while the increase in the policy-limit extends coverage to larger losses as in Section \ref{sec:pl_ex}. Consequently, the indemnity functions are ordered by type. Higher types, who are more risk averse and face stochastically larger losses, receive weakly more coverage at every loss level.

\medskip

Using the normalization of Assumption \ref{ass:p0}, i.e. $p_0=0$, it follows from Lemma \ref{prop:capped_premium} that the premium paid by the type-$\theta$ agent satisfies
$$
p_\theta
= 40\int_0^\theta
\left[1-\left(\frac{70-40s}{100}\right)^{(1+s)^2}\right]ds
+
10\int_0^\theta
\left[1-\left(\frac{90+10s}{100}\right)^{(1+s)^2}
\right]ds, \ \text{ for all $\theta \in [0,1]$}. 
$$

\medskip

\begin{figure}[H]
\centering
\begin{subfigure}{0.45\linewidth}
\centering
\includegraphics[width=\linewidth]{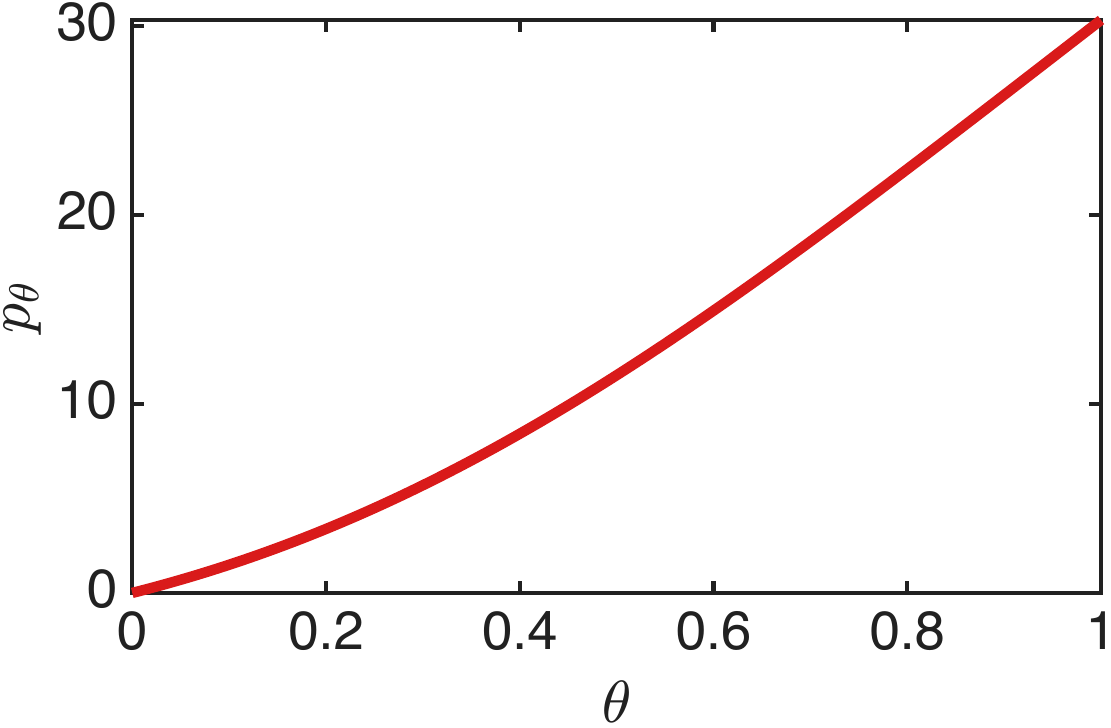}
\caption{Corresponding Constructed Premia}
\label{fig:capped3}
\end{subfigure}
\begin{subfigure}{0.46\linewidth}
\centering
\includegraphics[width=\linewidth]{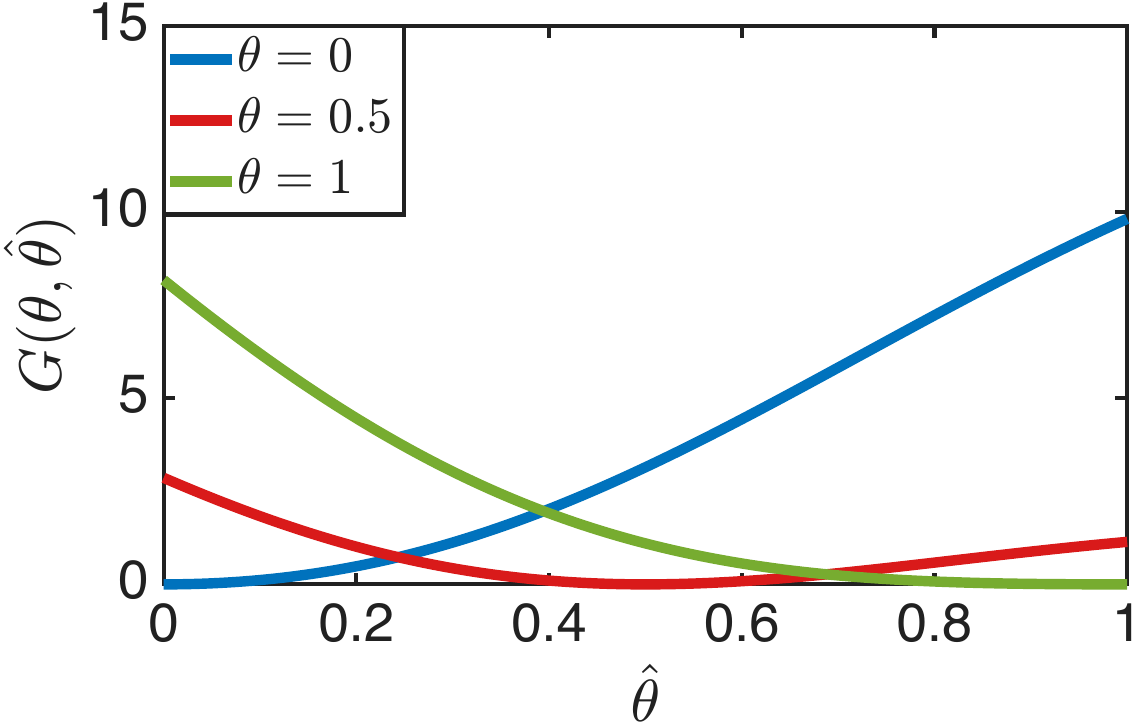}
\caption{Incentive Compatibility}
\label{fig:capped4}
\end{subfigure}
\caption{Numerical illustration under capped deductible insurance.}
\label{fig:cappeddeductiblefig2}
\end{figure}

\medskip

Figure \ref{fig:capped3} displays the corresponding constructed premia. Consistently with Lemma \ref{prop:capped_premium}, the premium is non-decreasing in type. Higher types are assigned lower deductibles and larger policy limits, providing more coverage supported by higher premia. The premium is not globally convex on $[0,1]$. Specifically, for types sufficiently close to the upper bound, the premium becomes concave. For a true type $\theta$ and a reported type $\hat\theta$, define the
incentive compatibility gap by 
$$
G(\theta, \hat \theta) = \left[u_\cR(\theta)-p_\theta \right]- \left[ u_\cR (\theta, \hat \theta) - p_{\hat \theta}\right].  
$$

\noindent Figure \ref{fig:capped4} plots the gap $G(\theta, \hat \theta)$ as a function of the reported type $\hat \theta$, using the three representative true types $\theta=0$, $\theta=0.5$ and $\theta=1$. For each type, the gap is minimized at the truthful report $\hat\theta=\theta$, where $G(\theta,\theta)=0$, and remains non-negative for every alternative report which confirms incentive compatibility of the given capped deductible insurance.

\bigskip
\section{Conclusion}
\label{sec:conclusion}
This paper studies whether a given collection of type-dependent retention functions can be supported by a premium schedule that induces the agent to self-select the contract intended for their own private type. If such a premium schedule exists, then the collection of retention functions is said to be implementable. The agent's preferences are evaluated according to Yaari's dual utility, and the agent's private type affects both their loss distribution and their level of risk aversion.

\bigskip

For arbitrary type spaces, we show that implementability is equivalent to cyclical monotonicity. Additionally, we show that for compact metric spaces, if the agent's utility functional is continuous, then implementability is equivalent to measurable implementability. 
For compact interval type spaces, under suitable type ordering assumptions, submodularity of the collection of retention functions is sufficient for implementability, and we also characterize the corresponding premium schedule. Moreover, we show that under our measurability assumptions, this premium schedule is Borel measurable. Thus, implementability and measurable implementability are equivalent without requiring continuity of the utility functional.

\bigskip

For deductible, proportional, and policy-limit insurance indemnity schedules, and under the strict ordering assumptions, implementability is satisfied if and only if the deductible is non-increasing in type, the proportional coverage rate is non-decreasing in type, and the policy limit is non-decreasing in type, respectively. Within each class of indemnity schedules, these conditions are equivalent to submodularity of the collection of retention functions. Therefore, higher types, who are more risk averse and face stochastically larger losses, receive weakly greater coverage at any given loss. 
For capped deductible insurance, submodularity remains sufficient, although it is not shown to be necessary for implementability. In this case, submodularity is satisfied if and only if the deductible is non-increasing in types and the policy limit is non-decreasing in types, and the increase in the policy limit is at least as large as the decrease in the deductible. 

\bigskip

In summary, the results of this paper provide criteria for the implementability of a given collection of retention functions. For the standard classes of insurance contracts that we consider, we obtain characterizations or sufficient conditions expressed as monotonicity restrictions on the contract parameters with respect to the agent's types.


\newpage
\appendix
\phantomsection
\label{appendix}
\numberwithin{equation}{subsection}

\section{Proofs of Main Results}

\subsection{Proof of Lemma \ref{le:measurablederivative}}
Define the continuous projection
\begin{align*}
\pi: \mathbb R &\to [0,\bar L]\\
l &\mapsto \pi(l):=\min\{\max\{l,0\},\bar L\},
\end{align*}

\noindent and extend each retention function such that  $\widetilde R_\theta(l):=R_\theta(\pi(l))$, for all $\theta \in \Theta$ and $l \in \R$. Since the map $(\theta,l)\mapsto(\theta,\pi(l))$ is measurable and $(\theta,l)\mapsto R_\theta(l)$ is $\cB(\Theta) \otimes \cB([0, \bar L])$-measurable by Assumption \ref{ass:jointmeasurableR}, the composition
$(\theta,l)\mapsto \widetilde R_\theta(l)$ is $\cB(\Theta)\otimes\cB(\mathbb R)$-measurable. 

\medskip

For every $n \geq 1$, let 
$$r_n(\theta, l) 
:= n \left[ \widetilde R_\theta\left(l+\frac1n\right) - \widetilde R_\theta(l)\right], 
\ \ \forall \, \theta \in \Theta, \ \forall 
\, l \in [0, \bar L].$$

\noindent The map $(\theta, l)\mapsto (\theta, l +\frac{1}{n})$ is measurable since both coordinate maps are measurable. Consequently, each $r_n$ is $\cB(\Theta)\otimes\cB([0,\bar L])$-measurable.

\medskip

Now, let $r(\theta, l) := \underset{n \to \infty}{\limsup} \, r_n(\theta, l)$, for all $\theta \in \Theta$ and $l \in [0, \bar L]$. Since a pointwise $\limsup$ of a sequence of measurable functions is measurable, the map $r$ is $\cB(\Theta) \otimes \cB([0, \bar L])$-measurable.
For every $\theta\in\Theta$, the condition $R_\theta\in\cF_R$ implies
that $R_\theta$ is non-decreasing and $1$-Lipschitz. Since the projection
$\pi$ is also non-decreasing and $1$-Lipschitz, the extension
$\widetilde R_\theta=R_\theta\circ\pi$ is non-decreasing and
$1$-Lipschitz on $\mathbb R$. Consequently, for all $\theta \in \Theta$, $l \in [0, \bar L]$, and $n \in \N$, we have
$$
0 \leq
\widetilde R_\theta\left(l+\frac1n\right)
-\widetilde R_\theta(l)
\leq \frac{1}{n}.
$$

\noindent Thus, $0\leq r_n(\theta,l)\leq 1$, for every $(\theta,l) \in \Theta \times [0, \bar L]$ and every $n \in \N$. Consequently, it follows that $0\leq r(\theta,l)\leq 1$, for every $(\theta,l) \in \Theta \times [0, \bar L]$.

\medskip

For each $\theta \in \Theta$, since $R_\theta \in \cF_R$, it is Lipschitz continuous and hence differentiable, for a.e.\ $l \in (0, \bar L)$. Let $l\in(0,\bar L)$ be a point at which
$R_\theta$ is differentiable. For all sufficiently large $n$, we have $l+\frac{1}{n}<\bar L$, and therefore for all $\theta \in \Theta$, 
$$
r_n(\theta,l) =\frac{R_\theta(l+\frac{1}{n})-R_\theta(l)}{\frac{1}{n}}.
$$

\noindent Hence, the differentiability of $R_\theta$ at $l$ gives $\underset{n\to+\infty} \lim \, r_n(\theta,l) = \frac{\partial R_\theta(l)}{\partial l}$. Consequently, $r(\theta,l)=\frac{\partial R_\theta(l)}{\partial l}$, for a.\e.\ $l\in[0,\bar L]$.\qed

\bigskip
\subsection{Proof of Lemma \ref{le:finite}}
\label{App:le:finite}
Let $\cR= \{ R_\theta\}_{\theta \in \Theta}$ be a collection of feasible retention functions. We know that 
$$
u_\cR(\hat \theta) = - \int_0^{\bar L} \left[ 1- g_{\hat \theta}(F_{\hat \theta}(l)) \right]\frac{\partial R_{\hat \theta}(l)}{\partial l} \, dl . 
$$

\noindent Since $g_{\hat \theta}:[0,1]\to[0,1]$ is a distortion function, we have $0\leq 1-g_{\hat \theta}(F_{\hat \theta}(l))\leq 1$, for all $l \in [0, \bar L]$. Moreover, since $R_{\hat \theta}\in \cF_R$, we have that $\frac{\partial R_{\hat \theta}(l)}{\partial l} \in [0,1]$ for a.e.\ $l \in [0, \bar L]$. Therefore,
$$0 \leq \left[ 1- g_{\hat \theta}(F_{\hat \theta}(l)) \right]\frac{\partial R_{\hat \theta}(l)}{\partial l}\leq 1, \ \text{ for a.e. $l \in [0, \bar L]$},$$

\noindent and so
$0 \leq \displaystyle\int_0^{\bar L} \left[ 1- g_{\hat \theta}(F_{\hat \theta}(l)) \right]\frac{\partial R_{\hat \theta}(l)}{\partial l} \, dl\leq \bar L,$
which implies that $-\bar L\leq u_\cR(\hat \theta) \leq 0$, where $\bar L < +\infty$. Using the same argument for $\theta \in \Theta$, we can show that $-\bar L\leq u_\cR(\theta,\hat \theta)\leq 0$. Therefore both quantities are finite. Moreover, their difference is also bounded and satisfies 
$$-\infty < -\bar L \leq
u_\cR(\theta,\hat \theta)-u_\cR(\hat \theta)
\leq \bar L < +\infty.$$\qed

\bigskip
\subsection{Proof of Theorem \ref{th:ImplemIFF}}
\label{App:th:ImplemIFF}
Suppose that the collection of retention functions $\cR=\{R_\theta\}_{\theta \in \Theta}$ is implementable. It follows from Definition \ref{def:PointMeasImplem} that there exists a premium schedule $p:\Theta \to \R_+$ such that
$u_\cR(\theta) - p_\theta 
\geq 
u_\cR (\theta, \hat \theta) - p_{\hat \theta}$, for all $\theta, \hat \theta \in \Theta$. 
For a finite cycle $\cC= (\theta_0, \ldots, \theta_{n+1})$ in $\Theta$, implementability of $\cR$ implies that $u_\cR(\theta_{k+1}, \theta_k) -u_\cR(\theta_{k+1}) \leq p_{\theta_k} - p_{\theta_{k+1}}$, for all $k \in \{0, \ldots, n\}$. 
Summing these inequalities over $k$ yields
\begin{equation}
\label{eq:ratio_leq0}
\sum_{k=0}^n \left[u_\cR(\theta_{k+1}, \theta_k)-u_\cR(\theta_{k+1})\right] 
\leq  \sum_{k=0}^n \left[ p_{\theta_k} - p_{\theta_{k+1}}\right]
= p_{\theta_0} - p_{\theta_{n+1}}
=0,
\end{equation}

\medskip

\noindent where the last equality follows from the fact that $\theta_{n+1}=\theta_0$. Moreover, 
\begin{align*}
\sum_{k=0}^{n}u_\cR(\theta_{k+1})
&=u_\cR(\theta_{1}) + u_\cR(\theta_2)+ \cdots +u_\cR(\theta_{n-1+1}) +u_\cR(\theta_{n+1}) \\
&=u_\cR(\theta_{1}) + u_\cR(\theta_2)+ \cdots +u_\cR(\theta_{n}) +u_\cR(\theta_{0}) \ \text{ \ (since $\cC$ is a finite cycle)}\\
&=\sum_{k=0}^n u_\cR(\theta_k).
\end{align*}

\noindent Hence the inequality in \eqref{eq:ratio_leq0} can be rewritten as
$\sum_{k=0}^{n}
\left[u_\cR(\theta_{k+1}, \theta_k)-u_\cR(\theta_k)
\right]\leq 0,$
where
\begin{align*}
\sum_{k=0}^{n}
\left[ u_\cR(\theta_{k+1}, \theta_k)-u_\cR(\theta_k) \right] 
&= \sum_{k=0}^{n} 
\left[-\int_0^{\bar L}\left[1-g_{\theta_{k+1}}(F_{\theta_{k+1}}(l))\right]\frac{\partial R_{\theta_k}(l)}{\partial l} \, dl
+ \int_0^{\bar L}\left[1-g_{\theta_{k}}(F_{\theta_{k}}(l))\right]\frac{\partial R_{\theta_k}(l)}{\partial l} \, dl
\right]\\
&= \sum_{k=0}^n \int_0^{\bar L}\left[ g_{\theta_{k+1}}(F_{\theta_{k+1}}(l))-g_{\theta_{k}}(F_{\theta_{k}}(l)) \right]\frac{\partial R_{\theta_k}(l)}{\partial l}  \, dl . 
\end{align*}

\bigskip

Conversely, suppose that for every finite cycle $\cC= (\theta_0, \ldots, \theta_{n+1})$ in $\Theta$, we have
\begin{equation}
\label{eq:ration_leq0_2}
\sum_{k=0}^{n}
\left[u_\cR(\theta_{k+1}, \theta_k)-u_\cR(\theta_k)
\right]\leq 0.
\end{equation}

\noindent For each $\theta \in \Theta$, consider the chains from $\theta_0$ to $\theta$ given by $\theta_0, \theta_1, \ldots, \theta_m, \theta_{m+1}=\theta,$ and let
$$\Phi(\theta)
:=
\sup_{\{ \text{chains}\ \theta_0,\ldots,\theta_m, \theta_{m+1}=\theta \}} \ \sum_{k=0}^{m}\left[u_\cR(\theta_{k+1}, \theta_k)-u_\cR(\theta_k)\right].$$

\medskip

\noindent We first show that $\Phi$ is well defined. When $\theta=\theta_0$,  every nonempty chain from $\theta_0$ to $\theta_0$ is a finite cycle, and it then follows by assumption that \eqref{eq:ration_leq0_2} holds. Moreover, the empty chain gives $\Phi(\theta_0)=0$. Consider any type $\theta \in \Theta$ and let 
$\theta_0, \theta_1, \dots, \theta_{m+1}=\theta,$
be any chain from $\theta_0$ to $\theta$. Appending the link from $\theta$ to $\theta_0$, we obtain the finite cycle $$\theta_0,\theta_1,\ldots,\theta_m,\theta_{m+1}=\theta, \theta_0.$$ 

\medskip

\noindent Applying \eqref{eq:ration_leq0_2} to this cycle gives 
$\sum_{k=0}^{m} \left[u_\cR(\theta_{k+1}, \theta_k)-u_\cR(\theta_k)\right] 
\leq u_\cR(\theta) - u_\cR(\theta_0, \theta),$
which is finite and independent of the chain. Taking the supremum over all chains from $\theta_0$ to $\theta$ and using Lemma \ref{le:finite}, we obtain 
$$\Phi(\theta) 
\leq u_\cR(\theta) - u_\cR(\theta_0, \theta) 
\leq \bar L<+\infty.$$

\medskip

\noindent To show that $\Phi(\theta)> - \infty$, consider the direct chain from $\theta_0$ to $\theta$ (i.e., $m=0$). The value of this chain is given by $u_\cR(\theta, \theta_0) - u_\cR(\theta_0)$. Since this quantity is finite by Lemma \ref{le:finite}, we obtain 
$$\Phi(\theta) 
\geq  
u_\cR(\theta, \theta_0) - u_\cR(\theta_0) 
\geq - \bar L
>-\infty.$$

\medskip

\noindent Hence $\Phi$ is well defined and finite. Now let
$$\tilde p_\theta:= u_\cR(\theta)-\Phi(\theta), 
\ \ \forall \, \theta \in \Theta.$$

\medskip

\noindent Since $-\bar L\leq u_\cR(\theta) \leq 0$, and since $-\bar L\leq \Phi(\theta)\leq \bar L$ by construction, we obtain
$$\tilde p_\theta 
= u_\cR(\theta) - \Phi(\theta)
\geq - \bar L -\bar L 
= -2 \bar L, 
\ \ \forall \, \theta \in \Theta. $$

\noindent Choosing $p_\theta = \tilde p_\theta + 2 \bar L \geq 0$, 
we aim to show that the menu $(R_\theta,p_\theta)_{\theta\in\Theta}$ is incentive compatible. We start by fixing two types $\theta, \hat \theta \in \Theta$. Let   $\theta_0, \theta_1, \ldots, \theta_m,\theta_{m+1}=\hat\theta
$ be an arbitrary chain from $\theta_0$ to $\hat \theta$. We denote the value of this chain by 
$$S(\theta_0,\ldots,\theta_m,\hat\theta)
:=\sum_{k=0}^{m}
\left[ u_\cR(\theta_{k+1}, \theta_k)- u_\cR(\theta_k)\right].$$

\noindent We now extend this chain by an additional link from $\hat \theta$ to $\theta$ to obtain the chain 
$$\theta_0,\theta_1, \ldots, \theta_m,\theta_{m+1}=\hat\theta, \theta_{m+2}=\theta.$$

\noindent The value of this extended chain is the value of the original chain plus the term $\left[ u_\cR(\theta, \hat \theta) - u_\cR(\hat \theta)\right]$. That is, 
\begin{align*}
S(\theta_0,\ldots,\theta_m, \hat \theta, \theta) 
&=\sum_{k=0}^{m}
\left[ u_\cR(\theta_{k+1}, \theta_k) - u_\cR(\theta_k)\right]+\left[u_\cR(\theta_{m+2} ,\theta_{m+1} ) -u_\cR(\theta_{m+1} )\right] \\
&=S(\theta_0,\ldots,\theta_m,\hat\theta) + \left[ 
u_\cR(\theta, \hat \theta) - u_\cR(\hat \theta)\right].
\end{align*}

\medskip

\noindent Taking the supremum over all chains ending at $\theta$, we obtain
$$\Phi(\theta) 
\geq S(\theta_0,\ldots,\theta_m,\hat\theta) + u_\cR(\theta, \hat \theta) - u_\cR(\hat \theta),$$ 

\noindent which means that $\Phi(\theta)$ is an upper bound for the set 
$$\l\{S(\theta_0,\ldots,\theta_m,\hat\theta) + u_\cR(\theta, \hat \theta) - u_\cR(\hat \theta): \ S(\theta_0,\ldots,\theta_m,\hat\theta) \ \text{ is any chain from $\theta_0$ to $\hat \theta$} \r\}.$$ 

\medskip

\noindent Consequently, 
\begin{align*}
\Phi(\theta) 
&\geq \underset{\text{all chains ending at $\hat \theta$}} \sup \l\{ S(\theta_0,\ldots,\theta_m,\hat\theta) + u_\cR(\theta, \hat \theta) - u_\cR(\hat \theta)\r\} \\
&= \underset{\text{all chains ending at $\hat \theta$}} \sup  \
\left\{S(\theta_0,\ldots,\theta_m,\hat\theta)\right\} + u_\cR(\theta, \hat \theta) - u_\cR(\hat \theta). 
\end{align*}

\medskip

\noindent Hence we obtain
\begin{align} 
\label{app:eq:supinequality}
\Phi(\theta)
\geq 
\Phi(\hat\theta) + u_\cR(\theta, \hat \theta) - u_\cR(\hat \theta).
\end{align}

\noindent By definition, we know that
$\Phi(\theta) 
= u_\cR(\theta) - \tilde p_\theta 
= u_\cR(\theta) - p_\theta + 2\bar L ,$
and 
$\Phi(\hat \theta) 
= u_\cR(\hat \theta) - \tilde p_{\hat \theta} 
= u_\cR(\hat \theta) -  p_{\hat \theta} + 2\bar L.$
Substituting these identities into \eqref{app:eq:supinequality} yields
$$u_\cR(\theta) - p_\theta + 2\bar L
\geq
u_\cR(\hat \theta) - p_{\hat\theta} + 2\bar L + u_\cR(\theta, \hat \theta) - u_\cR(\hat \theta).$$

\noindent Thus, 
$u_\cR( \theta) - p_\theta
\geq u_\cR(\theta, \hat \theta) - p_{\hat\theta},$ and hence the menu of contracts $(R_\theta, p_\theta)_{\theta \in \Theta}$ is incentive compatible, and hence the collection of retention functions $\cR$ is implementable.\qed

\bigskip
\subsection{Proof of Proposition \ref{prop:MeasImplemCompTheta}}
\label{App:prop:MeasImplemCompTheta}

Fix some $\theta_0 \in \Theta$. As in the proof of Theorem
\ref{th:ImplemIFF}, for each $\theta \in \Theta$ let
$$\Phi(\theta) :=
\sup\l\{
\sum_{k=0}^{m}
\l[u_\cR(\theta_{k+1},\theta_k) - u_\cR(\theta_k)\r]:
\, m\in \N \cup \{0\}, \, 
\theta_1,\ldots,\theta_m\in\Theta, \, \theta_{m+1}=\theta
\r\}.$$

\noindent Then $\Phi(\theta)$ is finite for every $\theta \in \Theta$. Moreover, the function
\begin{equation}
\label{eq:MeasPrePrem2}
\tilde p_\theta
:= u_\cR(\theta) - \Phi(\theta)
\end{equation}
is bounded below and satisfies
\begin{equation}
\label{eq:PrePremiumIC}
u_\cR(\theta) - \tilde p_\theta
\geq
u_\cR(\theta,\hat\theta) - \tilde p_{\hat\theta},
\ \ \forall \, \theta, \hat\theta \in \Theta.
\end{equation}

\noindent We show below that the map $\theta \mapsto \tilde p_\theta$ is Borel measurable. For each $m \in \N \cup \{0\}$, define the map
\smallskip
\begin{equation}
\begin{split}
\label{eq:FixedLengthPath}
\Phi_m: \Theta &\to \R\\[-0.8em]
\theta &\mapsto \Phi_m(\theta)
:= \underset{(\theta_1,\ldots,\theta_m)\in\Theta^m} \max \, 
\sum_{k=0}^{m}
\l[u_\cR(\theta_{k+1},\theta_k) - u_\cR(\theta_k)\r],
\end{split}
\end{equation}

\noindent where $\theta_0$ is the fixed reference type and $\theta_{m+1}:=\theta$. When $m=0$, the set $\Theta^0$ is a singleton, so
that $\Phi_0(\theta) = u_\cR(\theta,\theta_0) -  u_\cR(\theta_0)$. 

\medskip

Fix $m \in \N$. For every $k \in \l\{0,\ldots,m\r\}$, the maps
$\l(\theta,\theta_1,\ldots,\theta_m\r) \mapsto \theta_k$ 
and 
$\l(\theta,\theta_1,\ldots,\theta_m\r) \mapsto \theta_{k+1}$
are continuous. Consequently, the map
$\l(\theta,\theta_1,\ldots,\theta_m\r) \mapsto \l(\theta_{k+1},\theta_k\r)$
is also continuous. Hence, since the function $u_\cR: \Theta \times \Theta \to \R$ is assumed to be continuous, the map
\begin{align*}
\Gamma_m: \Theta \times \Theta^m &\to \R\\
(\theta,\theta_1,\ldots,\theta_m) &\mapsto  \Gamma_m(\theta,\theta_1,\ldots,\theta_m)
:= \sum_{k=0}^{m}
\l[u_\cR(\theta_{k+1},\theta_k) - u_\cR(\theta_k)\r]
\end{align*}
\noindent is continuous on $\Theta\times\Theta^m$. Since $\Theta$ is assumed to be compact, the product space $\Theta^m$ is also compact. Therefore, the maximum in \eqref{eq:FixedLengthPath} is attained. Moreover, by Berge's maximum theorem \cite[Theorem 17.31]{AliprantisBorder2006}, the value function
$\Phi_m:\Theta\to\R$ is continuous. Additionally, the function $\Phi_0$ is continuous by definition.

\medskip

Now, for a fixed $\theta\in\Theta$, and for each $m \in \N\cup\{0\}$, let
$$\mathfrak C_m(\theta)
:=\l\{
(\theta_0,\theta_1,\ldots,\theta_m,\theta_{m+1}):
\theta_1,\ldots,\theta_m\in\Theta,\ \theta_{m+1}=\theta
\r\}$$

\medskip

\noindent denote the collection of all chains from $\theta_0$ to $\theta$ that have exactly $m$ types in between. When $m=0$, the chain is simply $
(\theta_0,\theta)$. Let
$\mathfrak C(\theta)
:=\bigcup_{m\,\in\,\N\cup\{0\}} \mathfrak C_m(\theta)$
be the collection of all finite chains from $\theta_0$ to $\theta$. Every element of $\mathfrak C(\theta)$ is in some $\mathfrak C_m(\theta)$, for some $m \in \N \cup \{0\}$, namely, the number of intermediate types in that
chain. 

\medskip

For a chain $\gamma := \l(\theta_0,\theta_1,\ldots,\theta_m,\theta_{m+1}=\theta\r) \in \mathfrak C_m(\theta)$, let
$J(\gamma)
:= \sum_{k=0}^{m}
\l[u_\cR(\theta_{k+1},\theta_k) - u_\cR(\theta_k)\r],$
so that
$\Phi(\theta)
=\sup_{\gamma\in\mathfrak C(\theta)}J(\gamma).$
Moreover, for each fixed $m$, we have
$\Phi_m(\theta)
= \max_{\gamma\in\mathfrak C_m(\theta)}J(\gamma),$
since the maximum is attained because of compactness of $\Theta^m$ and continuity of $J$.
Next, we now show that
$\Phi(\theta)
= \underset{m\,\in\,\N\cup\{0\}} \sup \Phi_m(\theta).$
First, for every $m\in\N\cup\{0\}$, we have
$\mathfrak C_m(\theta) \subseteq \mathfrak C(\theta),$
and so
$$\Phi_m(\theta)
= \underset{\gamma\,\in\,\mathfrak C_m(\theta)} \sup J(\gamma)
\leq \underset{\gamma\,\in\,\mathfrak C(\theta)} \sup J(\gamma)
= \Phi(\theta),$$

\noindent implying that 
$\underset{m\,\in\,\N\cup\{0\}}\sup\Phi_m(\theta)
\leq \Phi(\theta).$
Conversely, fix some $\gamma\in\mathfrak C(\theta)$. Since $\gamma$ is a finite chain, it has some
finite number $m$ of intermediate types. Hence $\gamma\in\mathfrak C_m(\theta)$, for some $m\in\N\cup\{0\}$. Therefore,
$$J(\gamma)
\leq
\underset{\eta\,\in\,\mathfrak C_m(\theta)} \sup J(\eta)
= \Phi_m(\theta)
\leq \underset{r\,\in\,\N\cup\{0\}}\sup\Phi_r(\theta).$$

\noindent Since this holds for every $\gamma\in\mathfrak C(\theta)$, taking the supremum over all finite chains gives
$$\Phi(\theta)
= \underset{\gamma\,\in\,\mathfrak C(\theta)} \sup J(\gamma)
\leq \underset{r\,\in\,\N\cup\{0\}} \sup \Phi_r(\theta),$$
and so
$\Phi(\theta)
= \underset{m\,\in\,\N\cup\{0\}} \sup \Phi_m(\theta)$, for all $\theta \in \Theta$. In particular, being the pointwise supremum of a family of continuous
real-valued functions, $\Phi$ is lower semicontinuous, and therefore it is Borel measurable.

\medskip

Since the diagonal map
\begin{align*}
\Delta: \Theta &\to \Theta \times \Theta \\
\theta &\mapsto \Delta(\theta):= (\theta,\theta)
\end{align*}

\noindent is continuous, and since $u_\cR$ is assumed to be continuous on $\Theta \times \Theta$, the map
$\theta \mapsto
u_\cR(\theta)
= u_\cR(\theta,\theta)
= u_\cR\l(\Delta(\theta)\r)$
is continuous and hence Borel measurable. Therefore, it follows from \eqref{eq:MeasPrePrem2} that the map
$\theta \mapsto \tilde p_\theta$
is Borel measurable.

\medskip

Now, as in the proof of Theorem \ref{th:ImplemIFF}, $\tilde p$ is bounded below, and so $\underline p := \underset{\theta\,\in\,\Theta} \inf \, \tilde p_\theta > - \infty$. Let
\begin{equation}
\label{eq:NonnegMeasPrem}
c := \max\{0,-\underline p\}
\ \ \hbox{ and } \ \ 
p_\theta
:= \tilde p_\theta + c,
\ \ \forall \, \theta \in \Theta.
\end{equation}

\noindent Then $p_\theta \geq \underline p + c \geq 0$, for all $\theta \in \Theta$, and hence
$p: \Theta \to \R_+$.
Moreover, $p$ is Borel measurable since $\tilde p$ is Borel measurable. Finally, for all $\theta, \hat\theta \in \Theta$, we have
\begin{align*}
\l[u_\cR(\theta) - p_\theta\r]
- \l[u_\cR(\theta,\hat\theta) - p_{\hat\theta}\r]
&=\l[u_\cR(\theta) - \tilde p_\theta - c\r]
-
\l[u_\cR(\theta,\hat\theta) - \tilde p_{\hat\theta} - c\r]\\
&=\l[u_\cR(\theta) - \tilde p_\theta\r]
- \l[u_\cR(\theta,\hat\theta) - \tilde p_{\hat\theta}\r]\\
&\geq 0,
\end{align*}
where the inequality follows from \eqref{eq:PrePremiumIC}. Therefore,
$$u_\cR(\theta) - p_\theta
\geq u_\cR(\theta,\hat\theta) - p_{\hat\theta},
\ \ \forall \, \theta,\hat\theta \in \Theta.$$

\noindent Hence $p: \Theta \to \R_+$ is a Borel-measurable premium schedule that implements $\cR$. \qed

\bigskip
\subsection{Proof of Lemma \ref{le:jointlymeasurable}}
First, the chain rule gives 
$$
\frac{\partial }{ \partial \theta } \, g_{\theta} \big(  F_{\theta} (l)  \big) 
=\frac{\partial  g_{\theta} }{ \partial \theta } \big(  F_{\theta} (l)  \big) 
+ g^{\prime}_{\theta}\big(  F_{\theta} (l)  \big)  \frac{\partial  F_{\theta}(l) }{ \partial \theta}, \ \text{ for all $l \in [0, \bar L]$}. 
$$

\medskip

The map $\theta\mapsto F_\theta(l)$ is continuous for every $l$ by Assumption \ref{ass:lipschitz1}. Additionally, for every $\theta \in \Theta$, the map $l\mapsto F_\theta(l)$ is non-decreasing since it is a cumulative distribution function, and hence Borel measurable. Thus, $(l, \theta)\mapsto F_\theta(l)$ is a Carath\'eodory function. Since $\Theta = [\underline{\theta}, \bar \theta]$ is a separable metric space, the map $(\theta,l)\mapsto F_\theta(l)$ is jointly Borel measurable \cite[Lemma 4.51]{AliprantisBorder2006}.
Similarly, $(\theta,t)\mapsto g_\theta(t)$ is jointly Borel measurable because it is separately continuous.

\medskip

For each $n\geq 1$, define
$$
H^F_n(\theta, l):= 
\begin{cases}
n \bigl[ F_{\theta+\frac{1}{n}} (l) -F_\theta(l)\bigr] & \theta+\frac{1}{n} \leq \bar \theta, \\
0 & \text{otherwise,}
\end{cases}
$$
and
$$
H^g_n (\theta, t):= 
\begin{cases}
n \bigl[ g_\theta (t+\frac{1}{n}) -g_\theta(t)\bigr] & t+\frac{1}{n} \leq 1, \\
0 & \text{otherwise,}
\end{cases}
$$

\noindent Each $H^F_n$ and $H^g_n$ is jointly Borel measurable. For every $\theta < \bar \theta$ and $t<1$, the corresponding forward difference quotients are defined for all sufficiently large $n$. Therefore, by differentiability, 
$$
\frac{\partial F_\theta(l)}{\partial\theta}=
\underset{n\to\infty}{\lim}H^F_n(\theta, l), \ \text{ and } \  g_\theta'(t)= \underset{n\to\infty}{\lim}H^g_n(\theta, t).
$$

\noindent Thus the maps $(\theta, l)\mapsto \frac{\partial F_\theta(l)}{\partial\theta}$ and $(\theta, t) \mapsto g^\prime_\theta(t)$ are jointly Borel measurable on $[\underline{\theta}, \bar \theta) \times [0,\bar L]$ and $[\underline{\theta}, \bar \theta] \times [0,1)$ respectively.  
At $\theta = \bar \theta$, the derivative is obtained from the backward difference quotients
$$
\frac{\partial F_\theta(l)}{\partial\theta}\bigg|_{\theta=\bar \theta}=
\underset{n\to\infty}{\lim}n \bigl[ F_{\bar \theta} (l) -F_{\bar\theta-\frac{1}{n}}(l)\bigr],
$$

\noindent for all sufficiently large $n$, which is Borel measurable in $l$. 
Similarly, at $t=1$, the derivative is obtained from the backward difference quotients
$$ g_\theta'(t)\big|_{t=1}= \underset{n\to\infty}{\lim}n \left[ g_\theta (1) -g_\theta\left(1-\frac{1}{n}\right)\right] ,$$

\noindent for all sufficiently large $n$, which is Borel measurable in $\theta$. Since pointwise limits of sequences of measurable functions are measurable, and the endpoint sets, $\{\bar \theta\}$ and $\{1\}$, are Borel, both derivative maps $(\theta, l)\mapsto \frac{\partial F_\theta(l)}{\partial\theta}$ and $(\theta, t) \mapsto g^\prime_\theta(t)$ are jointly Borel measurable. 

\medskip

Finally, the map $(\theta, t) \mapsto \frac{\partial g _\theta(t)}{\partial \theta}$ is jointly Borel measurable since it is jointly continuous by Assumption \ref{ass:lipschitz1}. Hence, the chain rule identity implies that the map $(\theta, l) \mapsto \frac{\partial }{ \partial \theta } \, g_{\theta} \big(  F_{\theta} (l)  \big) $ is jointly Borel measurable. \qed

\bigskip
\subsection{Proof of Proposition \ref{prop:increasingdifferences}}
\label{App:prop:increasingdifferences}

Let $\theta_a, \theta_b, \alpha, \beta \in \Theta$ with $\theta_a<\theta_b$ and $\alpha<\beta$. We aim to show that
$$
u_\cR(\theta_b, \beta) - u_\cR(\theta_b, \alpha)
\geq
u_\cR(\theta_a, \beta)  - u_\cR(\theta_a, \alpha).
$$

\noindent Firstly, we know that 
$$
u_\cR(\theta, \alpha) = -\int_0^{\bar L} \left[ 1-g_\theta(F_\theta(l))\right] \frac{\partial R_\alpha(l)}{\partial l} \,dl.
$$

\noindent Then, 
\begin{align*}
u_\cR(\theta_b, \beta) - u_\cR(\theta_b, \alpha)
&= -\int_0^{\bar L} \left[ 1-g_{\theta_b}(F_{\theta_b}(l))\right] \frac{\partial R_{\beta}(l)}{\partial l} \,dl +  \int_0^{\bar L} \left[ 1-g_{\theta_b}(F_{\theta_b}(l))\right] \frac{\partial R_{\alpha}(l)}{\partial l}\,dl\\
&= \int_0^{\bar L} \left[ 1-g_{\theta_b}(F_{\theta_b}(l))\right] \left[\frac{\partial R_{\alpha}(l)}{\partial l} - \frac{\partial R_{\beta}(l)}{\partial l} \right]\,dl.
\end{align*}

\noindent On the other hand, 
\begin{align*}
u_\cR(\theta_a, \beta) - u_\cR(\theta_a, \alpha)
&= -\int_0^{\bar L} \left[ 1-g_{\theta_a}(F_{\theta_a}(l))\right] \frac{\partial R_{\beta}(l)}{\partial l} \,dl +  \int_0^{\bar L} \left[ 1-g_{\theta_a}(F_{\theta_a}(l))\right] \frac{\partial R_{\alpha}(l)}{\partial l}\,dl\\
&= \int_0^{\bar L} \left[ 1-g_{\theta_a}(F_{\theta_a}(l))\right] \left[\frac{\partial R_{\alpha}(l)}{\partial l} - \frac{\partial R_{\beta}(l)}{\partial l} \right]\,dl.
\end{align*}

\noindent Hence, 
\begin{align*}
u_\cR(\theta_b, \beta) - &u_\cR(\theta_b, \alpha) -\left[ u_\cR(\theta_a, \beta) - u_\cR(\theta_a, \alpha) \right]\\
&= \int_0^{\bar L} \left[ 1-g_{\theta_b}(F_{\theta_b}(l))\right] \left[  \frac{\partial R_{\alpha}(l)}{\partial l} - \frac{\partial R_{\beta}(l)}{\partial l}\right] \, dl - \int_0^{\bar L} \left[ 1-g_{\theta_a}(F_{\theta_a}(l))\right] \left[\frac{\partial R_{\alpha}(l)}{\partial l} - \frac{\partial R_{\beta}(l)}{\partial l} \right]\,dl \\
&= \int_0^{\bar L} \left[g_{\theta_a}(F_{\theta_a}(l)) - g_{\theta_b}(F_{\theta_b}(l)) \right] \left[\frac{\partial R_{\alpha}(l)}{\partial l} - \frac{\partial R_{\beta}(l)}{\partial l} \right]\,dl. 
\end{align*}

\noindent The above expression is non-negative since $g_\theta(F_\theta(l))$ is non-increasing in $\theta$ by Remark \ref{Re:chain_rule} and $\{R_\theta\}_{\theta\in \Theta}$ is submodular. Hence,
$$
u_\cR(\theta_b, \beta) - u_\cR(\theta_b, \alpha)
\geq
u_\cR(\theta_a, \beta)  - u_\cR(\theta_a, \alpha).
$$
\qed

\bigskip
\subsection{Proof of Proposition \ref{prop:cyclicmonotone}}
\label{App:prop:cyclicmonotone}
Consider a collection of submodular retention functions $\cR$. We aim to prove that the following inequality holds
$$
\sum_{k=0}^{n} \left[ u_\cR(\theta_{k+1}, \theta_k) - u_\cR(\theta_k)\right]\leq 0.
$$

\noindent Firstly, consider a two-cycle given by $\cC_2=(\theta_0, \theta_1, \theta_2=\theta_0)$. Then, 
\begin{align*}
S(\cC_2):=\sum_{k=0}^{1}
\left[ u_\cR(\theta_{k+1}, \theta_k) - u_\cR(\theta_k) \right]
&= \left[ u_\cR(\theta_1, \theta_0)- u_\cR(\theta_0)\right] + \left[u_\cR(\theta_2, \theta_1) - u_\cR(\theta_1) \right]\\
&= \left[ u_\cR(\theta_1, \theta_0)- u_\cR(\theta_0)\right]+ \left[u_\cR(\theta_0, \theta_1) - u_\cR(\theta_1) \right],
\end{align*}

\noindent which could be rearranged as follows:
\begin{align*}
S(\cC_2)
&=u_\cR(\theta_0, \theta_1) - u_\cR(\theta_0) - u_\cR(\theta_1) +  u_\cR(\theta_1, \theta_0). 
\end{align*}

\noindent By Proposition \ref{prop:increasingdifferences}, for $\theta_0<\theta_1$, $S(\cC_2)\leq0$. 

\medskip

We now prove the result for arbitrary finite cycles by induction on the order of the cycle. Suppose that the result holds for all cycles $\cC_N$ of order $N$. That is, 
$$
S(\cC_{N}):=\sum_{k=0}^{N-1}\left[ u_\cR(\theta_{k+1}, \theta_k) - u_\cR(\theta_k)\right]\leq 0. 
$$

\noindent Consider a cycle of order $N+1$ given by
$
\cC_{N+1} = (\theta_0, \theta_1, \ldots,\theta_{N-1},\theta_N, \theta_{N+1}=\theta_0)$, and let us prove that 
$$
S(\cC_{N+1}):=\sum_{k=0}^{N}\left[u_\cR(\theta_{k+1}, \theta_k) - u_\cR(\theta_k)\right] \leq 0.
$$

\noindent Since the cyclic sum is independent of the starting point, we can choose the starting point so that 
$$\theta_N = \underset{0\leq k \leq N}{\max} \theta_k.$$
Define the cycle 
$\cC= (\theta_0, \theta_1, \ldots, \theta_{N-1}, \theta_0)$
obtained by removing $\theta_N$. By the induction hypothesis, we know that $S(\cC) \leq0$. Moreover, 
$S(\cC_{N+1}) - S(\cC)$ 
is exactly the contribution of inserting $\theta_N$ between $\theta_{N-1}$ and $\theta_0$. Therefore, 
\begin{align*}
S(\cC_{N+1}) - S(\cC)
&=\left[ u_\cR(\theta_N, \theta_{N-1}) - u_\cR(\theta_{N-1})\right]  
+\left[ u_\cR(\theta_0, \theta_N) - u_\cR(\theta_N)\right]  
-\left[ u_\cR(\theta_0, \theta_{N-1}) - u_\cR(\theta_{N-1})\right]\\
&= \left[ u_\cR(\theta_0, \theta_N) -  u_\cR(\theta_0, \theta_{N-1}) \right] - \left[u_\cR(\theta_N) - u_\cR(\theta_N, \theta_{N-1}) \right]. 
\end{align*}

\medskip

\noindent We know that $\theta_0 \leq \theta_N$ and $\theta_{N-1}\leq \theta_N$. Moreover, by submodularity of the collection of retention functions, we have
$\frac{\partial R_{\theta_N}(l)}{\partial l} \leq \frac{\partial R_{\theta_{N-1}}(l)}{\partial l}$, for a.e.\ $l \in [0, \bar L]$. It follows from Proposition \ref{prop:increasingdifferences} that 
$$
u_\cR(\theta_N) - u_\cR(\theta_N, \theta_{N-1}) 
\geq
u_\cR(\theta_0, \theta_N) -  u_\cR(\theta_0, \theta_{N-1}).
$$

\noindent Thus, $S(\cC_{N+1}) - S(\cC)\leq 0.$ Since $S(\cC)\leq 0$, we obtain 
$$
S(\cC_{N+1}) = S(\cC) + \left[S(\cC_{N+1}) - S(\cC) \right] \leq 0,
$$

\noindent as desired, and hence the cyclical monotonicity condition holds for every finite cycle. \qed

\bigskip
\subsection{Proof of Proposition \ref{IC_characterization}}
Consider an implementable collection of retention functions $\cR$, and let $p:\Theta \to \R_+$ be any premium schedule that implements $\cR$. Such a premium schedule exists by Definition \ref{def:PointMeasImplem}. Then the menu of contracts $(R_{\theta}, p_{\theta})_{ \theta \in \Theta}$ is incentive compatible. We aim to find an expression of the corresponding premium schedule. First, we know that for $\theta \in \Theta$,
\begin{equation}
\label{eqStar}
U_{\theta} ( R_{\theta} , p_{\theta} )
= - p_{\theta} - \int_0^{\bar L}\left[1 - g_{\theta} ( F_{\theta}(l))\right]\, \frac{ \partial R_{\theta}(l) }{ \partial l } \,dl .
\end{equation}

\noindent We can also express the utility of a type-$\theta$ agent using $(R_{\hat \theta }, p_{\hat \theta})$ where $\hat \theta  \in \Theta$, as follows:
\begin{align*}
U_{\theta} ( R_{\hat \theta} , p_{\hat \theta} )
&= - p_{\hat \theta} - \int_0^{ \bar L }  \left[ 1 - g_{\theta} \big( F_{\theta}(l)  \big)   \right] \,\, \frac{ \partial R_{\hat \theta}(l) }{ \partial l } \, dl  .
\end{align*}

\medskip

Define the supremum of utilities associated with the menu of contracts $(R_\theta, p_\theta)_{\theta \in \Theta}$ as follows: 
$$
V(\theta):=\underset{\hat \theta \in \Theta}{\sup} \,U_\theta(R_{\hat{\theta}}, p_{\hat \theta}), \ \ \forall \, \theta \in \Theta. 
$$
Since the menu $(R_\theta, p_\theta)_{\theta \in \Theta}$ is incentive compatible, truthful reporting maximizes the type-$\theta$ agent's utility. That is,
$$V(\theta) = U_\theta(R_\theta, p_\theta), \ \ \forall \, \theta \in \Theta.$$

\noindent Moreover, 
\begin{align*}
\left| \frac{\partial U_{\theta} (R_{\hat \theta}, p_{\hat \theta})}{\partial \theta} \right| 
&= \left| \int_0^{\bar L} 
\frac{\partial }{\partial \theta} \, g_\theta(F_\theta(l)) \, \frac{ \partial R_{\hat \theta}(l) }{ \partial l } \, dl\right| \leq
\int_0^{\bar L } 
\left| \frac{\partial }{\partial \theta} \, g_\theta(F_\theta(l))\frac{ \partial R_{\hat \theta}(l) }{ \partial l } \right|  dl. 
\end{align*}

\medskip

\noindent Since $0 \leq \frac{\partial R_{\hat \theta}(l)}{\partial l}\leq 1$, and by Assumption \ref{ass:lipschitz}, we obtain
$$
\left| \frac{\partial U_{\theta} (R_{\hat \theta}, p_{\hat \theta})}{\partial \theta} \right| \leq  K\bar L < +\infty.
$$

\noindent Thus, for every fixed $\hat \theta \in \Theta$, the map $\theta \mapsto U_{\theta}(R_{\hat \theta} , p_{\hat \theta})$ is Lipschitz continuous with common Lipschitz constant $K \, \bar L$, and therefore absolutely continuous.  Hence, $V$ is absolutely continuous and by the envelope theorem (e.g.,  \cite[Theorem 2]{milgrom2002envelope}), for every $\theta \in \Theta$, we have:
\begin{equation}
\label{eqStarStar}
\begin{split}
U_{\theta}(R_{\theta}, p_{\theta}) 
&= U_{ \underline{ \theta }} (R_{ \underline{ \theta} }, p_{ \underline{ \theta } })  + \int_{\underline{\theta} } ^ {\theta} \frac{\partial U_{s'}(R_s, p_s) }{ \partial s'}\bigg|_{s'=s} \, ds \\
&= - p_{\underline{ \theta }} - \int_0^{ \bar L}  \left[ 1 - g_{ \underline{ \theta }}  \big( F_{\underline{ \theta }}(l)  \big)   \right]  \frac{ \partial R_{\underline{ \theta } }(l) }{ \partial l } \, dl 
+
\int_{\underline{\theta} } ^ {\theta} \int_0 ^{\bar L}  \left[ \frac{\partial g_s}{ \partial s} (F_s(l)) + g'_s(F_s(l)) \frac{\partial F_s(l)}{\partial s} \right] \frac{\partial R_s(l)}{\partial l } \, dl \, ds.
\end{split}
\end{equation}

\noindent Equating \eqref{eqStar} and \eqref{eqStarStar} yields
\begin{equation}
\label{eq:PremRep}
\begin{split}
p_\theta
&= p_{\underline\theta} + \int_0^{\bar L} \left[1-g_{\underline\theta} \left(F_{\underline\theta}(l)\right)\right]
\frac{\partial R_{\underline\theta}(l)}{\partial l}\,dl
\\& \ \ \  - \int_{\underline\theta}^{\theta} \int_0^{\bar L} \left[\frac{\partial g_s}{\partial s}(F_s(l)) + g_s'(F_s(l)) \frac{\partial F_s(l)}{\partial s} \right] \frac{\partial R_s(l)}{\partial l}\,dl\,ds
\\& \ \ \ - \int_0^{\bar L}\left[1-g_\theta\bigl(F_\theta(l)\bigr)\right]\frac{\partial R_\theta(l)}{\partial l}\,dl .
\end{split}
\end{equation}

\bigskip

To show the converse, suppose that $\tilde p: \Theta \to \R_+$ is a premium schedule that satisfies \eqref{eq:PremRep}, with $p_\theta$ and $p_{\underline\theta}$ replaced by $\tilde p_\theta$ and $\tilde p_{\underline\theta}$, respectively. We show that the premium schedule $\tilde p$ implements $\cR$. First, since $\cR$ is implementable, fix one premium schedule $p^0: \Theta \to \R_+$ that implements $\cR$. By the first part of the proof above, $p^0$ satisfies \eqref{eq:PremRep}. The terms on the right-hand side of \eqref{eq:PremRep} other than the premium of the lowest type depend only on the collection of retention functions and on the agents' type-dependent preferences. Therefore, subtracting the representations of $\tilde p$ and $p^0$ gives
$$\tilde p_\theta-p^0_\theta
= \tilde p_{\underline\theta} - p^0_{\underline\theta},
\ \ \forall \, \theta\in\Theta.$$

\noindent Hence, if $c:= \tilde p_{\underline\theta} - p^0_{\underline\theta}$, then $\tilde p_\theta = p^0_\theta + c$, for all $\theta\in\Theta$, and therefore it follows that for every $\theta, \hat\theta \in \Theta$, we have
\begin{align*}
\left[u_\cR(\theta)-\tilde p_\theta\right]
- \left[u_\cR(\theta,\hat\theta)-\tilde p_{\hat\theta}\right]        &=\left[u_\cR(\theta)-(p^0_\theta+c)\right]
-\left[u_\cR(\theta,\hat\theta)-(p^0_{\hat\theta}+c)\right]\\
&=\left[u_\cR(\theta)-p^0_\theta\right]
-\left[u_\cR(\theta,\hat\theta)-p^0_{\hat\theta}\right]\\
&\geq0,
\end{align*}

\noindent where the inequality follows since $p^0$ implements $\cR$. Therefore,
$$u_\cR(\theta)-\tilde p_\theta
\geq
u_\cR(\theta,\hat\theta)-\tilde p_{\hat\theta},
\ \ \forall \, \theta,\hat\theta\in\Theta,$$

\medskip

\noindent and so the menu $(R_\theta,\tilde p_\theta)_{\theta\in\Theta}$ is incentive compatible. Since $\tilde p$ takes values in $\R_+$ by assumption, it is admissible and therefore it implements $\cR$.

\bigskip

Finally, the previous argument shows that any two premium schedules
that implement the same collection of retention functions must differ only by a common additive constant. Conversely, translating any premium schedule that implements $\cR$ by a common additive constant that gives a nonnegative premium schedule makes that premium schedule also an implementing premium schedule for the collection $\cR$.\qed

\bigskip
\subsection{Proof of Proposition \ref{prop:imp_measurably_interval}}
Consider the interval type space $\Theta = [\underline{\theta}, \bar \theta]$. Suppose that the collection of retention functions $\cR$ is measurably implementable, then it is trivially implementable. 

\medskip

Conversely, suppose that the collection of retention functions $\cR$ is implementable. It follows from Proposition \ref{IC_characterization} that the corresponding premium schedule $\theta \mapsto p_\theta$ satisfies the following form:
\begin{align*}
p_{\theta} = 
p_{\underline {\theta} } \,\, &+\int_0^{ \bar L}  \left[ 1 - g_{ \underline{ \theta }} \big( F_{\underline{ \theta }}(l)  \big)   \right] \,\, \frac{ \partial R_{\underline{\theta}}(l) }{ \partial l } \, dl 
-
\int_{ \underline{\theta}} ^ { \theta} \int_0 ^{\bar L}  \left[ \frac{\partial g_s}{ \partial s} (F_s(l)) + g'_s(F_s(l)) \frac{\partial F_s(l)}{\partial s} \right] 
\frac{\partial R_s(l)}{\partial l } \, dl \, ds \\
&\quad  -
\int_0^{ \bar L }  \left[ 1 - g_{\theta} \big( F_{\theta}(l)  \big)   \right] \,\frac{ \partial R_{\theta}(l) }{ \partial l } \,dl.  
\end{align*} 

\medskip

We show that the above expression of the premium is $\cB(\Theta)$-measurable. Firstly, we know that 
$$
k:= p_{\underline {\theta} } +\int_0^{ \bar L}  \left[ 1 - g_{ \underline{ \theta }} \big( F_{\underline{ \theta }}(l)  \big)   \right] \,\, \frac{ \partial R_{\underline{\theta}}(l) }{ \partial l } \, dl 
$$

\noindent is a constant in $\theta$ and hence is $\cB(\Theta)$-measurable. Thus the premium can be rewritten as 
$$
p_\theta = k - \int_{\underline{\theta}}^\theta B(s) \,ds - A(\theta), 
$$
where
$$
A(\theta) := \int_0^{\bar L}
\left[1-g_\theta(F_\theta(l))\right] \frac{\partial R_\theta(l)}{\partial l}\,dl, \ \text{ for $\theta
\in \Theta$},
$$
and
$$
B(s) := \int_0^{\bar L}\left[
\frac{\partial g_s}{\partial s}(F_s(l))+g_s'(F_s(l))
\frac{\partial F_s(l)}{\partial s}\right]\frac{\partial R_s(l)}{\partial l}\,dl, \ \text{ for $s \in \Theta$}. 
$$

It remains to show that $\theta \mapsto A(\theta)$ and $\theta \mapsto \displaystyle\int_{\underline{\theta}}^\theta B(s) \, ds$ are Borel measurable.
Using Lemma \ref{le:measurablederivative}, there exists a jointly Borel
measurable function $r:\Theta\times[0,\bar L]\to[0,1]$ such that, for
every $\theta\in\Theta$,
$r(\theta,l)=\frac{\partial R_\theta(l)}{\partial l}$, for a.e.\ $l\in[0,\bar L]$.
Thus, the functions $A$ and $B$ can be rewritten as follows:
$$
A(\theta) = \int_0^{\bar L} \left[1-g_\theta(F_\theta(l))\right]r(\theta,l)\,dl, \ \text{ for $\theta \in \Theta$}, 
$$
and
$$
B(\theta)=\int_0^{\bar L}
\left[\frac{\partial g_\theta}{\partial\theta}(F_\theta(l))+g_\theta'(F_\theta(l))
\frac{\partial F_\theta(l)}{\partial\theta} \right]
r(\theta,l)\,dl, \ \text{ for $\theta \in \Theta$}.   
$$

\medskip

\noindent It follows from Assumption \ref{ass:lipschitz1} and Lemma \ref{le:jointlymeasurable} that the maps
$(\theta, l) \mapsto 1-g_\theta(F_\theta(l))$ and
$(\theta,l)\mapsto
\frac{\partial }{\partial\theta}g_\theta(F_\theta(l)) $ are jointly Borel measurable. Thus the integrands defining $A$ and $B$, namely, 
$$
(\theta, l) \mapsto \left[ 1- g_\theta(F_\theta(l)) \right] r(\theta, l) 
$$
and
$$
(\theta,l)\mapsto \left[
\frac{\partial g_\theta}{\partial\theta}(F_\theta(l)) +g_\theta'(F_\theta(l))\frac{\partial F_\theta(l)}{\partial\theta}\right] r(\theta, l) 
$$

\medskip

\noindent are jointly Borel measurable. Moreover, we know that for all $\theta \in \Theta$ and $l \in [0, \bar L]$, $0\leq 1-g_\theta(F_\theta(l))\leq 1,$
and by Assumption \ref{ass:lipschitz},
$$
\left|
\frac{\partial g_\theta}{\partial\theta}(F_\theta(l))
+g_\theta'(F_\theta(l))
\frac{\partial F_\theta(l)}{\partial\theta}
\right| \leq K.
$$

\medskip

\noindent Since $0\leq r(\theta,l)\leq1$ and $[0,\bar L]$ has finite Lebesgue measure, the integrands defining $A$ and $B$ are integrable with respect to $l$ for all $\theta \in \Theta$.
Thus it follows that the maps $\theta \mapsto  A(\theta)$ and $\theta \mapsto B(\theta)$ are Borel measurable \cite[Lemma 1.28]{kallenberg2021foundations}. Moreover, since $B$ is bounded on the compact interval $\Theta$, it belongs to $L^1(\Theta)$. Consequently, the map $\theta \mapsto \int_{\underline\theta}^{\theta}B(s)\,ds$
is absolutely continuous, and hence continuous and Borel measurable. Thus, the implementing premium schedule $p:\Theta \to \R_+$ is Borel measurable, and the collection $\{R_\theta\}_{\theta\in\Theta}$ is measurably implementable. \qed 

\bigskip
\subsection{Proof of Proposition \ref{prop:rationiffsub_deductible}}
We know that for deductible insurance, the retention function is given by $R_\theta(l) = \min \{ l, d(\theta) \}$ for $l \in [0, \bar L]$, then 
$$
\frac{\partial R_\theta(l)}{\partial l} = \mathbf 1_{\{l<d(\theta)\}} = 
\begin{cases}
1 & l <d(\theta), \\
0 & l\geq d(\theta), 
\end{cases} 
\ \text{ for a.e. $l \in [0, \bar L]$}. 
$$

\bigskip

\noindent\underline{(1)$\implies$(2):} Suppose that the deductible map $\theta \mapsto d(\theta)$ is non-increasing, that is, for any $\theta_1, \theta_2\in \Theta$, $\theta_1<\theta_2$, $d(\theta_2)\leq d(\theta_1)$. Thus
$\{l\in[0,\bar L]:l<d(\theta_2)\}
\subseteq
\{l\in[0,\bar L]:l<d(\theta_1)\},$
and so $\mathbf 1_{\{l<d(\theta_2)\}} \leq \mathbf 1_{\{l<d(\theta_1)\}}$ for a.e. $l \in [0, \bar L]$. Hence, the collection of retention functions $\{R_\theta\}_{\theta \in \Theta}$ is submodular since 
$\frac{\partial R_{\theta_2}(l)}{\partial l} \leq \frac{\partial R_{\theta_1}(l)}{\partial l}$, for a.e., $l \in [0, \bar L]$.

\bigskip

\noindent\underline{(2)$\implies$(1):} Assume that the collection of retention functions $\{R_\theta\}_{\theta \in \Theta}$ is submodular and suppose for the sake of contradiction that $\theta \mapsto d(\theta)$ is not non-increasing. Then there exists $\theta_1, \theta_2 \in \Theta$, $\theta_1<\theta_2$ such that 
$d(\theta_2)>d(\theta_1)$. For every loss level $l$ between the two deductibles $d(\theta_1)<l<d(\theta_2)$, we have
$\frac{\partial R_{\theta_1}(l)}{\partial l} =0$ and $\frac{\partial R_{\theta_2}(l)}{\partial l} =1.$
That is, on the interval $\left(d(\theta_1), d(\theta_2)\right)$ that has a positive Lebesgue measure, we have
$$
\frac{\partial R_{\theta_1}(l)}{\partial l} < \frac{\partial R_{\theta_2}(l)}{\partial l},
$$ 

\noindent contradicting the submodularity of $\{R_\theta\}_{\theta \in \Theta}$. Hence $\theta \mapsto d(\theta)$ is non-increasing. 

\bigskip

\noindent\underline{(2)$\implies$(3):} Suppose that $\{R_\theta\}_{\theta\in\Theta}$ is submodular, then it follows from Proposition \ref{prop:cyclicmonotone} that the cyclical monotonicity condition is satisfied. Moreover, by Theorem \ref{th:ImplemIFF}, we conclude that the collection of deductible retention functions is implementable. 

\bigskip

\noindent\underline{(3)$\implies$(2):} Suppose that $\{R_\theta\}_{\theta\in\Theta}$ is implementable. To show submodularity of $\{R_\theta\}_{\theta\in\Theta}$, it suffices to show that the map $\theta \mapsto d(\theta)$ is non-increasing since (1)$\iff$(2). For the sake of contradiction, suppose that $\theta \mapsto d(\theta)$ is not non-increasing. Then there exists $\theta_1, \theta_2 \in \Theta$, 
$\theta_1<\theta_2$ such that $d(\theta_2) > d(\theta_1)$. We know that  $\frac{\partial R_\theta(l)}{\partial l} = \mathbf{1}_{\{l<d(\theta)\}}. $
By Theorem \ref{th:ImplemIFF}, since $\{R_\theta\}_{\theta\in\Theta}$ is implementable, the
cyclical monotonicity is satisfied. That is,
$$u_\cR(\theta_2, \theta_1) - u_\cR(\theta_1) + u_\cR(\theta_1, \theta_2) -u_\cR(\theta_2) \leq 0, 
$$
where 
\begin{align*}
u_\cR(\theta_i, \theta_j)
&= -  \int_0^{\bar L} \left[ 1- g_{\theta_i}(F_{\theta_i}(l)) \right] \mathbf{1}_{\{l<d(\theta_j)\}} \,dl \\
&=-  \int_0^{d(\theta_j)}\left[1-g_{\theta_i}(F_{\theta_i}(l))\right]\,dl\\
&= - d(\theta_j) + \int_0^{d(\theta_j)}g_{\theta_i}(F_{\theta_i}(l))\,dl. 
\end{align*}

\noindent Then, 
\begin{align*}
\int_0^{d(\theta_1)}
\left[
g_{\theta_2}(F_{\theta_2}(l))-
g_{\theta_1}(F_{\theta_1}(l))\right]dl
+
\int_0^{d(\theta_2)}
\left[g_{\theta_1}(F_{\theta_1}(l))
-g_{\theta_2}(F_{\theta_2}(l))\right]dl \leq 0.  
\end{align*}

\medskip

\noindent Since $d(\theta_1)<d(\theta_2)$ by assumption, splitting the second integral of the above expression at $d(\theta_1)$ yields $\displaystyle\int_{d(\theta_1)}^{d(\theta_2)}
\left[g_{\theta_1}(F_{\theta_1}(l))-g_{\theta_2}(F_{\theta_2}(l))\right]\,dl \leq 0$.
On the other hand, for $\theta_1<\theta_2$, Remark \ref{re:strict_chain} implies that $g_{\theta_1}(F_{\theta_1}(l)) - g_{\theta_2}(F_{\theta_2}(l)) > 0$ for all $l \in (0, \bar L)$. Moreover, since $(d(\theta_1), d(\theta_2))$ has positive Lebesgue measure, it follows that
$$
\int_{d(\theta_1)}^{d(\theta_2)}
\left[g_{\theta_1}(F_{\theta_1}(l))
-g_{\theta_2}(F_{\theta_2}(l))\right]dl > 0. 
$$

\noindent This contradicts the preceding inequality. Hence, the map $\theta \mapsto d(\theta)$ is non-increasing, and therefore $\{R_\theta\}_{\theta \in \Theta}$ is submodular by (1)$\iff$(2).  \qed

\bigskip
\subsection{Proof of Lemma \ref{prop:deductible_premium}}
Consider an implementable collection of retention functions $\{R_\theta\}_{\theta\in\Theta}$, then by Definition \ref{def:PointMeasImplem}, there exists $p:\Theta \to \R_+$ such that the menu $(R_\theta, p_\theta)_{\theta \in \Theta}$ is incentive compatible. That is, for $\theta_1, \theta_2 \in \Theta$, $\theta_1<\theta_2$, we have $p_{\theta_2} -p_{\theta_1} \geq u_\cR(\theta_1, \theta_2) - u_\cR(\theta_1),$
where
\begin{align*}
u_\cR(\theta_1, \theta_2) - u_\cR(\theta_1)
&= - \int_0^{\bar L}\left[ 1- g_{\theta_1}(F_{\theta_1}(l)) \right]\, \frac{\partial R_{\theta_2}(l)}{\partial l } \, dl +  \int_0^{\bar L}\left[ 1- g_{\theta_1}(F_{\theta_1}(l)) \right]\, \frac{\partial R_{\theta_1}(l)}{\partial l } \, dl \\
&= \int_0^{\bar L}\left[ 1- g_{\theta_1}(F_{\theta_1}(l)) \right]\, \left[\frac{\partial R_{\theta_1}(l)}{\partial l } -\frac{\partial R_{\theta_2}(l)}{\partial l } \right] \, dl. 
\end{align*}

\medskip

\noindent Since $g_{\theta_1}$ is a distortion function, we have $ 1- g_{\theta_1}(F_{\theta_1}(l)) \geq 0$ for all $l \in [0, \bar L]$. Moreover, it follows from Proposition \ref{prop:rationiffsub_deductible} that the collection of deductible retention functions $\{R_\theta\}_{\theta\in\Theta}$ is submodular. That is, 
$\frac{\partial R_{\theta_1}(l)}{\partial l } -\frac{\partial R_{\theta_2}(l)}{\partial l }  \geq 0$, for a.e. $l \in [0, \bar L]$. Hence, we obtain
$$
p_{\theta_2} -p_{\theta_1} \geq u_\cR(\theta_1, \theta_2) - u_\cR(\theta_1) \geq 0,
$$

\noindent which means that $\theta \mapsto p_\theta$ is non-decreasing. Additionally, it follows from Proposition \ref{IC_characterization} that for each $\theta \in \Theta$, the corresponding premium satisfies
\begin{align*}
p_{\theta}
&=p_{\underline{\theta}}
+\int_0^{\bar L}
\left[1-g_{\underline{\theta}}\bigl(F_{\underline{\theta}}(l)\bigr)
\right]\frac{\partial R_{\underline{\theta}}(l)}{\partial l}\,dl -
\int_{\underline{\theta}}^{\theta}
\int_0^{\bar L}
\left[\frac{\partial g_s}{\partial s}(F_s(l))
+g'_s(F_s(l))\frac{\partial F_s(l)}{\partial s}
\right]\frac{\partial R_s(l)}{\partial l}\,dl\,ds\\
&\quad -
\int_0^{\bar L}\left[1-g_\theta(F_\theta(l))\right] \frac{\partial R_\theta(l)}{\partial l}\,dl.
\end{align*}

\noindent For a deductible contract, we know that 
$\frac{\partial R_\theta(l)}{\partial l}=\mathbf 1_{\{l<d(\theta)\}}$, for a.e. $l\in[0,\bar L]$. 
Then the general premium expression reduces to the following
\begin{align*}
p_\theta
&=p_{\underline\theta}
+\int_0^{d(\underline\theta)}
\left[
1-g_{\underline\theta}
\bigl(F_{\underline\theta}(l)\bigr)
\right]\,dl- 
\int_{\underline\theta}^{\theta}\int_0^{d(s)} \left[\frac{\partial g_s}{\partial s}(F_s(l))+g_s'(F_s(l))\frac{\partial F_s(l)}{\partial s}\right]\,dl\,ds\\
&\quad -
\int_0^{d(\theta)}\left[1-g_\theta(F_\theta(l))\right]\,dl. 
\end{align*}
\qed

\bigskip
\subsection{Proof of Lemma \ref{re:numericalexampleassumption}}
Firstly, the maps $\theta \mapsto F_\theta(l)$ and $\theta \mapsto g_\theta(t)$ are differentiable for all $l \in [0,100]$ and $t \in [0,1]$ respectively. Furthermore, 
$
\left|\frac{\partial F_\theta(l)}{\partial\theta}\right|
\leq \frac{1}{e}$, for all $(\theta,l) \in [0,1]\times [0,100]$, 
and
$\left|\frac{\partial g_\theta(t)}{\partial\theta}\right|
\leq \frac{1}{e}$, for all $(\theta,t) \in [0,1]\times [0,1]$. Additionally, the map $t \mapsto g_\theta(t)$ is differentiable for all $\theta \in \Theta$ with
$0 \leq g_\theta^\prime (t)\leq 2$, for all $(\theta,t) \in [0,1]\times [0,1]$.

\medskip

We now verify that the map
$(\theta, t) \mapsto \frac{\partial g_\theta(t)}{\partial \theta}$ is jointly continuous. 
First, for any $t\in(0,1]$, the partial derivative of $g_\theta(t)$ with respect to $\theta$ is given by:
$
\frac{\partial g_\theta(t)}{\partial \theta} = t^{1+\theta} \, \ln(t). 
$
Additionally, since $ g_\theta(0)=0$ for all $\theta \in \Theta$, we have 
$\frac{\partial g_\theta(0)}{\partial \theta}=0.$
Hence, we can define
$$
h(\theta, t) = \frac{\partial g_\theta(t)}{\partial \theta} =
\begin{cases}
t^{1+\theta} \, \ln t & t\in(0,1], \\
0 & t=0. 
\end{cases}
$$

\noindent The function $h$ is jointly continuous on $[0,1]\times(0,1]$. Moreover, for every $\theta \in [0,1]$ and $t \in (0,1]$, we have $
t^{1+\theta} \leq t,$ which gives 
\begin{equation}
\label{eq:h_ineq}
|h(\theta, t)| = t^{1+\theta} |\ln t|\leq t |\ln t|, \ \text{ for all $\theta \in [0,1]$ and $t \in (0,1]$}. 
\end{equation}

\noindent Fix $\theta_0 \in [0,1]$. Since $t |\ln t| \to 0$ as $t \to 0^+$, for every $\varepsilon>0$, there exists $\delta>0$ such that if $0<t <\delta$, then  $t |\ln t| < \varepsilon.$
For any $\theta\in[0,1]$ and $t \in (0, \delta)$, it follows from \eqref{eq:h_ineq} that 
$$
|h(\theta, t) - h(\theta_0, 0)| = |h(\theta, t)| \leq t |\ln t| < \varepsilon.
$$

\noindent Additionally, it is easy to see that $|h(\theta, 0) - h(\theta_0, 0)| =0 <\varepsilon$. Since $\delta$ was chosen independently of $\theta$, this shows that 
$$
\underset{(\theta, t) \to (\theta_0, 0)}{\lim} h(\theta, t) =0 = h(\theta_0, 0), 
$$

\noindent which means that $h$ is continuous at every point $(\theta_0, 0)$, where $\theta_0 \in [0,1]$. Thus, $h$ is jointly continuous at every point of $[0,1]\times [0,1]$. Hence, Assumption \ref{ass:lipschitz1} is satisfied. Moreover, Assumption \ref{ass:lipschitz} is satisfied since
\begin{align*}
\left|\frac{\partial }{ \partial \theta } \, g_{\theta} \big(  F_{\theta} (l)  \big) \right|
&=\left|\frac{\partial  g_{\theta} }{ \partial \theta } \big(  F_{\theta} (l)  \big) 
+ g^{\prime}_{\theta}\big(  F_{\theta} (l)  \big)  \frac{\partial  F_{\theta}(l) }{ \partial \theta} \right| \\
&\leq \left|\frac{\partial  g_{\theta} }{ \partial \theta } \big(  F_{\theta} (l)  \big) \right| +  g^{\prime}_{\theta}\big(  F_{\theta} (l)  \big) \left| \frac{\partial  F_{\theta}(l) }{ \partial \theta} \right|
\leq \frac{1}{e} + \frac{2}{e} 
= \frac{3}{e}.
\end{align*}

\medskip

\noindent Assumption \ref{Ass:cdf_family} is satisfied since, for every $l\in(0,100]$, we have
$\frac{\partial F_\theta(l)}{\partial \theta}
= \left(\frac{l}{100}\right)^{1+\theta}\ln\left(\frac{l}{100}\right) \leq 0$.
Since $F_\theta(0)=0$ for every $\theta$, we also have $\frac{\partial F_\theta(0)}{\partial\theta}=0$.

\bigskip

Finally, Assumption \ref{ass:strictorder} is also satisfied, since for every $l \in (0, 100)$,
$0< F_\theta(l) < 1$, for all $\theta \in [0,1]$, and for all $t\in (0,1)$, the map $\theta\mapsto g_\theta(t)$ is strictly decreasing since
$\frac{\partial g_\theta(t)}{\partial\theta}
= t^{1+\theta} \,\ln t <0.$
\qed

\bigskip
\subsection{Proof of Proposition \ref{prop:proportional_rationiffsubmod}}
We know that 
$\frac{\partial R_\theta(l)}{\partial l} =1- a(\theta)$, for a.e. $l \in [0, \bar L]$.
Hence, we can immediately see that (1)$\iff$(2). That is, the collection of retention functions $\{R_\theta\}_{\theta \in \Theta}$ is submodular if and only if the map $\theta \mapsto a(\theta)$ is non-decreasing.

\bigskip

\noindent\underline{(2)$\implies$(3):} First, if $\{R_\theta\}_{\theta\in\Theta}$ is submodular then it follows from Proposition \ref{prop:cyclicmonotone} that the cyclical monotonicity condition is satisfied. Second, by Theorem \ref{th:ImplemIFF}, we conclude that the collection of proportional retention functions is implementable. 

\bigskip

\noindent\underline{(3)$\implies$(2):} Suppose that  $\{R_\theta\}_{\theta\in\Theta}$ is implementable. To show that it is submodular, it is enough to show that the map $\theta \mapsto a(\theta)$ is non-decreasing since (1)$\iff$(2). By Theorem \ref{th:ImplemIFF}, since $\{R_\theta\}_{\theta\in\Theta}$ is implementable, cyclical monotonicity is satisfied. That is, for every $\theta_1, \theta_2 \in \Theta$, 
$\theta_1<\theta_2$, we have
\begin{align*}
u_\cR(\theta_2, \theta_1) - u_\cR(\theta_1) + u_\cR(\theta_1, \theta_2) -u_\cR(\theta_2) \leq 0,  
\end{align*}

\medskip

\noindent where  $u_\cR(\theta_i, \theta_j) 
= -\left[1-a(\theta_j)\right] \displaystyle\int_0^{\bar L}\left[1-g_{\theta_i}(F_{\theta_i}(l))\right]\,dl$. Therefore, 
\begin{align*}
\left[a(\theta_2)-a(\theta_1)\right] \left[ \int_0^{\bar L}
\left[1-g_{\theta_2}(F_{\theta_2}(l))\right]\,dl -\int_0^{\bar L}
\left[1-g_{\theta_1}(F_{\theta_1}(l))\right]\,dl \right] \geq0. 
\end{align*}

\noindent We know from Remark \ref{re:strict_chain} that the map $\theta\mapsto\int_0^{\bar L}\left[1-g_\theta(F_\theta(l))\right]\,dl$ is strictly increasing. That is, for $\theta_1 < \theta_2$, 
$$
\int_0^{\bar L}
\left[1-g_{\theta_2}(F_{\theta_2}(l))\right]\,dl -\int_0^{\bar L}
\left[1-g_{\theta_1}(F_{\theta_1}(l))\right]\,dl>0. 
$$

\noindent Then $a(\theta_2) \geq a(\theta_1)$ which means that the map $\theta \mapsto a(\theta)$ is non-decreasing. Hence, the collection $\{R_\theta\}_{\theta\in\Theta}$ is submodular.
\qed

\bigskip
\subsection{Proof of Lemma \ref{prop:proportional_premia}}
Suppose that the collection of proportional retention functions
$\{R_\theta\}_{\theta\in\Theta}$ is implementable, then it follows from Definition \ref{def:PointMeasImplem} that there exists a premium schedule $p:\Theta \to \R_+$ such that the menu of contracts $(R_\theta, p_\theta)_{\theta \in \Theta}$ is incentive compatible. That is, for $\theta_1, \theta_2 \in \Theta$, $\theta_1<\theta_2$, we have 
$p_{\theta_2} -p_{\theta_1} \geq u_\cR(\theta_1, \theta_2) - u_\cR(\theta_1)$,
where
\begin{align*}
u_\cR(\theta_1, \theta_2) - u_\cR(\theta_1)
&= - \int_0^{\bar L}\left[ 1- g_{\theta_1}(F_{\theta_1}(l)) \right]\, \frac{\partial R_{\theta_2}(l)}{\partial l } \, dl +  \int_0^{\bar L}\left[ 1- g_{\theta_1}(F_{\theta_1}(l)) \right]\, \frac{\partial R_{\theta_1}(l)}{\partial l } \, dl \\
&= \int_0^{\bar L}\left[ 1- g_{\theta_1}(F_{\theta_1}(l)) \right]\, \left[\frac{\partial R_{\theta_1}(l)}{\partial l } -\frac{\partial R_{\theta_2}(l)}{\partial l } \right] \, dl. 
\end{align*}

\noindent Since $g_{\theta_1}$ is a distortion function, $ 1- g_{\theta_1}(F_{\theta_1}(l)) \geq 0$ for all $l \in [0, \bar L]$. Moreover, it follows from Proposition \ref{prop:proportional_rationiffsubmod} that the collection of retention functions $\{R_\theta\}_{\theta\in\Theta}$ is submodular. That is, 
$
\frac{\partial R_{\theta_1}(l)}{\partial l } -\frac{\partial R_{\theta_2}(l)}{\partial l }  \geq 0$, for a.e. $l \in [0, \bar L]$. Hence,
$p_{\theta_2} -p_{\theta_1} \geq u_\cR(\theta_1, \theta_2) - u_\cR(\theta_1) \geq 0,$
which means that $\theta \mapsto p_\theta$ is non-decreasing. Additionally, it follows from Proposition \ref{IC_characterization} that for each $\theta \in \Theta$, the corresponding premium satisfies
\begin{align*}
p_{\theta}
&=p_{\underline{\theta}}
+\int_0^{\bar L}
\left[1-g_{\underline{\theta}}\bigl(F_{\underline{\theta}}(l)\bigr)
\right]\frac{\partial R_{\underline{\theta}}(l)}{\partial l}\,dl -
\int_{\underline{\theta}}^{\theta}
\int_0^{\bar L}
\left[\frac{\partial g_s}{\partial s}(F_s(l))
+g'_s(F_s(l))\frac{\partial F_s(l)}{\partial s}
\right]\frac{\partial R_s(l)}{\partial l}\,dl\,ds\\
&\quad -
\int_0^{\bar L}\left[1-g_\theta(F_\theta(l))\right] \frac{\partial R_\theta(l)}{\partial l}\,dl.
\end{align*}

\noindent For proportional insurance, we know that 
$\frac{\partial R_\theta(l)}{\partial l}= 1-a(\theta)$, for $l\in[0,\bar L]$ and $\theta \in \Theta$. Then the general premium expression reduces to the following
\begin{align*}
p_{\theta}
&=p_{\underline{\theta}}
+\int_0^{\bar L}
\left[1-g_{\underline{\theta}}\bigl(F_{\underline{\theta}}(l)\bigr)\right](1-a(\underline{\theta}))\,dl -
\int_{\underline{\theta}}^{\theta}(1-a(s))
\int_0^{\bar L}\left[\frac{\partial g_s}{\partial s}(F_s(l))+g'_s(F_s(l))\frac{\partial F_s(l)}{\partial s}
\right]\,dl\,ds\\
&\quad -
\int_0^{\bar L}\left[1-g_\theta(F_\theta(l))\right](1-a(\theta))\,dl.
\end{align*}

\noindent Integrating the third term by parts yields
\begin{align*}
p_\theta
&=p_{\underline{\theta}}
+\int_0^{\bar L}\left[1-g_{\underline{\theta}}\bigl(F_{\underline{\theta}}(l)\bigr)\right](1-a(\underline{\theta}))\,dl
+(1-a(\theta)) \int_0^{\bar L}\left[1-g_\theta(F_\theta(l))\right]\,dl \\&\quad+\int_{\underline{\theta}}^\theta a^\prime(s) \int_0^{\bar L} \left[1- g_s(F_s(l)) \right]\,dl\,ds - (1-a(\theta)) \int_0^{\bar L}\left[1-g_\theta(F_\theta(l))\right]\,dl  \\ &\quad - \int_0^{\bar L}\left[1-g_{\underline{\theta}}\bigl(F_{\underline{\theta}}(l)\bigr)\right](1-a(\underline{\theta}))\,dl\\
&= p_{\underline\theta}+\int_{\underline\theta}^{\theta}\int_0^{\bar L}a^\prime (s)\left[1-g_s(F_s(l))\right]dl\,ds.
\end{align*}
\qed

\bigskip
\subsection{Proof of Proposition \ref{prop:policylimit_submodulariffimplementable}}
For policy-limit insurance, we know that 
$$
\frac{\partial R_\theta(l)}{\partial l}= \mathbf{1}_{ \{l>m(\theta)\} } =
\begin{cases}
0, & l<m(\theta),\\
1, & l>m(\theta), 
\end{cases}
\ \text{ for $\theta \in \Theta$ and for a.e. $l \in [0, \bar L]$}.
$$

\bigskip

\noindent\underline{(1)$\implies$(2):} Suppose that $\theta \mapsto m(\theta)$ is non-decreasing, that is, for $\theta_1, \theta_2 \in \Theta$, $\theta_1< \theta_2$, we have $m(\theta_1) \leq m(\theta_2)$. Then, 
$\{l\in[0,\bar L]:l>m(\theta_2)\}
\subseteq
\{l\in[0,\bar L]:l>m(\theta_1)\},$ 
which implies that $ \mathbf{1}_{\{l>m(\theta_2)\}}\leq \mathbf{1}_{\{l>m(\theta_1)\}}$, for a.e. $l \in [0, \bar L]$. Hence, the collection of retention functions $\{R_\theta\}_{\theta \in \Theta}$ is submodular since $\frac{\partial R_{\theta_2}(l)}{\partial l} \leq \frac{\partial R_{\theta_1}(l)}{\partial l}$, for a.e.\ $l \in [0, \bar L]$. 

\bigskip

\noindent\underline{(2)$\implies$(1):} Suppose that $\{R_\theta\}_{\theta \in \Theta}$ is submodular and let us show that the map $\theta \mapsto m(\theta)$ is non-decreasing. For the sake of contradiction, suppose that $\theta \mapsto m(\theta)$ is not non-decreasing. Then there exist $\theta_1, \theta_2 \in \Theta$, $\theta_1<\theta_2$ such that $m(\theta_1) > m(\theta_2)$. We choose any $l \in (m(\theta_2), m(\theta_1))$. Then 
$\frac{\partial R_{\theta_1}(l)}{\partial l} =0$ and $\frac{\partial R_{\theta_2}(l)}{\partial l} =1$. 
That is, on the interval $\left(m(\theta_2), m(\theta_1)\right)$ that has a positive Lebesgue measure, we have
$\frac{\partial R_{\theta_1}(l)}{\partial l} < \frac{\partial R_{\theta_2}(l)}{\partial l},$ 
contradicting the submodularity of $\{R_\theta\}_{\theta \in \Theta}$. Hence $\theta \mapsto m(\theta)$ is non-decreasing.  

\bigskip

\noindent\underline{(2)$\implies$(3):} Suppose that $\{R_\theta\}_{\theta\in\Theta}$ is submodular, then it follows from Proposition \ref{prop:cyclicmonotone} that the cyclical monotonicity condition is satisfied. Moreover, by Theorem \ref{th:ImplemIFF}, we conclude that the collection of  retention functions is implementable. 

\bigskip

\noindent\underline{(3)$\implies$(2):} Suppose that $\{R_\theta\}_{\theta\in\Theta}$ is implementable. To show submodularity of $\{R_\theta\}_{\theta\in\Theta}$, it suffices to show that the map $\theta \mapsto m(\theta)$ is non-decreasing since (1)$\iff$(2). For the sake of contradiction, suppose that $\theta \mapsto m(\theta)$ is not non-decreasing, then there exist $\theta_1, \theta_2 \in \Theta$, $\theta_1<\theta_2$ such that $m(\theta_1)>m(\theta_2)$. 
We know that $\frac{\partial R_\theta(l)}{\partial l} = \mathbf{1}_{\{l>m(\theta)\}}$, \ for $l \in [0, \bar L]$ and $\theta \in \Theta$. 
By Theorem \ref{th:ImplemIFF}, since $\{R_\theta\}_{\theta\in\Theta}$ is implementable, cyclical monotonicity is satisfied. That is, for every $\theta_1, \theta_2 \in \Theta$, 
$\theta_1<\theta_2$, we have
\begin{align*}
u_\cR(\theta_2, \theta_1) - u_\cR(\theta_1) + u_\cR(\theta_1, \theta_2) -u_\cR(\theta_2) \leq 0, 
\end{align*}

\noindent where 
\begin{align*}
u_\cR(\theta_i, \theta_j)
&= - \int_0^{\bar L} \left[ 1- g_{\theta_i}(F_{\theta_i}(l)) \right] \mathbf{1}_{\{l>m(\theta_j)\}} \,dl \\
&= -\int_{m(\theta_j)}^{\bar L } \left[ 1- g_{\theta_i}(F_{\theta_i}(l)) \right] \,dl \\
&=m(\theta_j) -\bar L +\int_{m(\theta_j)}^{\bar L }g_{\theta_i}(F_{\theta_i}(l)) \,dl. 
\end{align*}

\noindent Then, 
\begin{align*}
-\int_{m(\theta_1)}^{\bar L } \left[ g_{\theta_1}(F_{\theta_1}(l)) - g_{\theta_2}(F_{\theta_2}(l)) \right] \,dl + \int_{m(\theta_2)}^{\bar L } \left[ g_{\theta_1}(F_{\theta_1}(l)) - g_{\theta_2}(F_{\theta_2}(l)) \right] \,dl \leq 0.  
\end{align*}

\medskip

\noindent Since $m(\theta_1)>m(\theta_2)$ by assumption, splitting the second integral of the above expression at $m(\theta_1)$ yields $\displaystyle\int_{m(\theta_2)}^{m (\theta_1) } \left[ g_{\theta_1}(F_{\theta_1}(l)) - g_{\theta_2}(F_{\theta_2}(l)) \right] \,dl \leq 0$. On the other hand, for $\theta_1<\theta_2$, Remark \ref{re:strict_chain} implies that $g_{\theta_1}(F_{\theta_1}(l))
-g_{\theta_2}(F_{\theta_2}(l)) > 0$ for all $l \in (0, \bar L)$. Moreover, since $(m(\theta_2), m(\theta_1))$ has positive Lebesgue measure, it follows that
$$
\int_{m(\theta_2)}^{m(\theta_1)}
\left[g_{\theta_1}(F_{\theta_1}(l))
-g_{\theta_2}(F_{\theta_2}(l))\right]dl > 0. 
$$

\noindent This contradicts the preceding inequality. Hence, the map $\theta \mapsto m(\theta)$ is non-decreasing, and therefore $\{R_\theta\}_{\theta \in \Theta}$ is submodular by (1)$\iff$(2).\qed

\bigskip
\subsection{Proof of Lemma \ref{prop:policylimit_premia}}
Suppose that the collection of retention functions
$\{R_\theta\}_{\theta\in\Theta}$ is implementable, then it follows from Definition \ref{def:PointMeasImplem} that there exists a premium schedule $p:\Theta \to \R_+$ such that the menu of contracts $(R_\theta, p_\theta)_{\theta \in \Theta}$ is incentive compatible. That is, for $\theta_1, \theta_2 \in \Theta$, $\theta_1<\theta_2$, we have
$p_{\theta_2} -p_{\theta_1} \geq u_\cR(\theta_1, \theta_2) - u_\cR(\theta_1),$
where
\begin{align*}
u_\cR(\theta_1, \theta_2) - u_\cR(\theta_1)
&= - \int_0^{\bar L}\left[ 1- g_{\theta_1}(F_{\theta_1}(l)) \right]\, \frac{\partial R_{\theta_2}(l)}{\partial l } \, dl +  \int_0^{\bar L}\left[ 1- g_{\theta_1}(F_{\theta_1}(l)) \right]\, \frac{\partial R_{\theta_1}(l)}{\partial l } \, dl \\
&= \int_0^{\bar L}\left[ 1- g_{\theta_1}(F_{\theta_1}(l)) \right]\, \left[\frac{\partial R_{\theta_1}(l)}{\partial l } -\frac{\partial R_{\theta_2}(l)}{\partial l } \right] \, dl. 
\end{align*}

\medskip

\noindent Since $g_{\theta_1}$ is a distortion function, we have $ 1- g_{\theta_1}(F_{\theta_1}(l)) \geq 0$ for all $l \in [0, \bar L]$. Moreover, it follows from Proposition \ref{prop:policylimit_submodulariffimplementable} that the collection of policy limit retention functions $\{R_\theta\}_{\theta\in\Theta}$ is submodular. That is, 
$\frac{\partial R_{\theta_1}(l)}{\partial l } -\frac{\partial R_{\theta_2}(l)}{\partial l }  \geq 0$, for a.e.\ $l \in [0, \bar L]$.
Hence, we obtain $p_{\theta_2} -p_{\theta_1} \geq u_\cR(\theta_1, \theta_2) - u_\cR(\theta_1) \geq 0,$
which means that $\theta \mapsto p_\theta$ is non-decreasing. Additionally, it follows from Proposition \ref{IC_characterization} that for each $\theta \in \Theta$, the corresponding premium satisfies
\begin{align*}
p_{\theta}
&=p_{\underline{\theta}}
+\int_0^{\bar L}
\left[1-g_{\underline{\theta}}\bigl(F_{\underline{\theta}}(l)\bigr)
\right]\frac{\partial R_{\underline{\theta}}(l)}{\partial l}\,dl -
\int_{\underline{\theta}}^{\theta}
\int_0^{\bar L}
\left[\frac{\partial g_s}{\partial s}(F_s(l))
+g'_s(F_s(l))\frac{\partial F_s(l)}{\partial s}
\right]\frac{\partial R_s(l)}{\partial l}\,dl\,ds\\
&\quad -
\int_0^{\bar L}\left[1-g_\theta(F_\theta(l))\right] \frac{\partial R_\theta(l)}{\partial l}\,dl.
\end{align*}

\medskip

For policy limit insurance, we know that 
$R_\theta(l) = l - \min \{l, m(\theta)\} = ( l - m(\theta))^+$, for all $\theta \in \Theta$ and $l \in [0, \bar L]$. The corresponding marginal retention is given by
$\frac{\partial R_\theta(l)}{\partial  l}= \mathbf{1}_{\{l>m(\theta)\}}$, for $\theta \in \Theta$ and for a.e.\ $l \in [0, \bar L]$. Then the general premium expression reduces to the following
\begin{align*}
p_\theta
&= p_{\underline{\theta}}
+\int_{m(\underline{\theta})}^{\bar L}
\left[1-g_{\underline{\theta}}\bigl(F_{\underline{\theta}}(l)\bigr)
\right]\,dl -
\int_{\underline{\theta}}^{\theta}
\int_{m(s)}^{\bar L}
\left[\frac{\partial g_s}{\partial s}(F_s(l))
+g'_s(F_s(l))\frac{\partial F_s(l)}{\partial s}
\right]\,dl\,ds -
\int_{m(\theta)}^{\bar L}\left[1-g_\theta(F_\theta(l))\right] \,dl.
\end{align*}

\medskip

Now, for all $s \in \Theta$, let
$$b(s):=\int_{m(s)}^{\bar L} \left[ 1- g_s(F_s(l)) \right] \,dl.$$ 

\noindent We show the absolute continuity of $b$ and identify its derivative. For any $s, z \in \Theta$, we know that 
\begin{align*}
b(z)- b(s) 
&= \int_{m(z)}^{\bar L} \left[ 1- g_z(F_z(l))\right]\,dl - \int_{m(s)}^{\bar L} \left[ 1- g_s(F_s(l))\right]\,dl 
\end{align*}

\noindent Adding and subtracting the integral $\displaystyle\int_{m(z)}^{\bar L} \left[ 1- g_s(F_s(l))\right]\,dl $ gives
\begin{align*}
b(z)- b(s) &= \int_{m(z)}^{\bar L} \left[ g_s(F_s(l))- g_z(F_z(l))\right]\,dl + \int_{m(z)}^{\bar L} \left[ 1- g_s(F_s(l))\right]\,dl  - \int_{m(s)}^{\bar L} \left[ 1- g_s(F_s(l))\right]\,dl. 
\end{align*}

\medskip

\noindent We know from Assumption \ref{ass:lipschitz} that $ |\frac{\partial}{\partial \theta} g_\theta(F_\theta(l))|\leq K$, for all $\theta \in \Theta$ and $l \in [0, \bar L]$. Thus it follows from the mean value theorem that 
$\left| g_z(F_z(l)) - g_s(F_s(l))\right|\leq K |z-s|$, for $ s,z \in \Theta$ and for all $l \in [0, \bar L]$. 
Since $0 \leq g_\theta(F_\theta(l)) \leq 1 $ for all $\theta \in \Theta$ and $l \in [0, \bar L]$, the following inequality holds for all $s, z\in \Theta$
\begin{align*}
\left|b(z) -b(s) \right|
&\leq K(\bar L-m(z)) |z-s| + \int_{\min\{m(s), m(z)\}}^{\max\{m(s), m(z)\}}\left| 1- g_s(F_s(l))\right| \,dl\\
&\leq K\bar L |z-s| +|m(z) - m(s)|.
\end{align*}

\medskip

\noindent Fix $\varepsilon>0$. Since $m$ is absolutely continuous, \cite[Definition 7.17]{rudin1987real}, there exists $\delta_m>0$ such that for every finite collection of non-overlapping intervals $[s_i, z_i]\subseteq \Theta$, $s_i \leq z_i$, if $\sum_{i=1}^N (z_i - s_i) < \delta_m$, we have $\sum_{i=1}^N|m(z_i) - m(s_i)| < \frac{\varepsilon}{2}$. 
Let 
$$
\delta := \min \left \{\delta_m, \frac{\varepsilon}{2 (K\bar L +1)}\right\} > 0. 
$$
\noindent For any such collection of non-overlapping intervals with $\sum_{i=1}^N (z_i -s_i) < \delta$, 
\begin{align*}
\sum_{i=1}^N \left|b(z_i) -b(s_i) \right|
&\leq K \bar L \sum_{i=1}^N (z_i-s_i) + \sum_{i=1}^N|m(z_i) -m(s_i)| < \frac{\varepsilon}{2} + \frac{\varepsilon}{2} = \varepsilon,
\end{align*}

\noindent which implies that $b$ is absolutely continuous. Thus by the Fundamental Theorem of Calculus for absolutely continuous functions, the derivative of $b$ exists, it is finite a.e.\ on $(\underline{\theta}, \bar \theta)$, and satisfies
$b(\theta) - b(\underline{\theta}) = \int_{\underline{\theta}}^\theta b^\prime(s) \, ds$, for every $\theta \in \Theta$.

\medskip

For every $s \in \Theta$, the cumulative distribution function $F_s$ is right-continuous, and $g_s$ is continuous. Consequently, the map $l \mapsto g_s(F_s(l))$ is right-continuous. We consider the following set:
$$
E:= \left\{ s \in (\underline{\theta}, \bar \theta): \ b \text{ and } m  \text{ are differentiable at $s$ with finite derivatives}  \right\}.
$$

\medskip

\noindent Since $b$ and $m$ are absolutely continuous, they are both differentiable a.e. Therefore, the set $E$ has full Lebesgue measure in $(\underline\theta, \bar\theta)$.

\medskip

Fix $s \in E$, and consider $h>0$ with $s+h \in \Theta$. Since $m$ is non-decreasing, the quantity $\Delta_h := m(s+h) - m(s)$ is non-negative. Continuity and differentiability of $m$ at $s$ give $\Delta_h \to 0$, and $\frac{\Delta_h}{h}\to m^\prime(s) \in [0,+\infty)$ as $h \downarrow 0$. Moreover, 
$$
\frac{b(s+h)-b(s)}{h}= - \int_{m(s)}^{\bar L}\frac{g_{s+h}(F_{s+h}(l)) - g_s(F_s(l))}{h}\, dl - \frac{1}{h}\int_{m(s)}^{m(s+h)}\left[ 1- g_{s+h}(F_{s+h}(l))\right]\,dl. 
$$

\medskip

\noindent We evaluate the two limits separately. For the first term, by the Dominated Convergence Theorem, we have 
$$
\underset{h \downarrow 0 }{\lim} \int_{m(s)}^{\bar L}\frac{g_{s+h}(F_{s+h}(l)) - g_s(F_s(l))}{h}\, dl=\int_{m(s)}^{\bar L}\frac{\partial }{\partial s} g_s(F_s(l)) \, dl. 
$$

\noindent Moreover, we have
\begin{equation}\label{eq:deltah}
\left|  \frac{1}{h}\int_{m(s)}^{m(s+h)}\left[  g_{s+h}(F_{s+h}(l))- g_s(F_s(l))\right]\,dl \right| \leq \frac{1}{h}\int_{m(s)}^{m(s+h)} K h \, dl = K \Delta_h \to 0, \ \text{ as } h \to 0. 
\end{equation}

\medskip

\noindent Hence, 
\begin{align*}
\left|\frac{1}{h} \int_{m(s)}^{m(s+h)} \left[1- g_s(F_s(l)) \right]\, dl - \frac{\Delta_h}{h}  \bigl[1- g_s\big(F_s(m(s))\big) \bigr]\right|
&\leq \frac{\Delta_h}{h} \underset{l \in [m(s), m(s+h)]}{\sup} \left| g_s(F_s(l)) -g_s\big(F_s(m(s))\big) \right|. 
\end{align*}

\medskip

\noindent The supremum tends to zero, and the first factor $\frac{\Delta_h}{h}$ is bounded for sufficiently small $h$. Thus it follows that
\begin{align*}
\underset{h \downarrow 0 }{\lim} \, \frac{1}{h} \int_{m(s)}^{m(s+h)} \left[1- g_s(F_s(l)) \right]\, dl= m^\prime(s)  \left[1- g_s(F_s(m(s))) \right]. 
\end{align*}

\medskip

\noindent Combining this limit with \eqref{eq:deltah} yields
$$
\underset{h \downarrow 0 }{\lim} \, \frac{1}{h} \int_{m(s)}^{m(s+h)} \left[1- g_{s+h}(F_{s+h}(l)) \right]\, dl= m^\prime(s)  \left[1- g_s(F_s(m(s))) \right]. 
$$

\noindent Hence, we obtain
$$
b^\prime(s)= \int_{m(s)}^{\bar L} 
-\left[\frac{\partial g_s}{\partial s}(F_s(l))
+g_s'(F_s(l))\frac{\partial F_s(l)}{\partial s}\right] \,dl - m^\prime(s)\left[ 1- g_s\big(F_s(m(s)) \big)\right], \ \text{ for a.e. $s \in \Theta$} .
$$

\medskip

\noindent Or equivalently, 
$$
\int_{m(s)}^{\bar L} 
-\left[\frac{\partial g_s}{\partial s}(F_s(l))
+g_s'(F_s(l))\frac{\partial F_s(l)}{\partial s}\right] \,dl
=b^\prime(s)+m^\prime(s)\left[ 1- g_s\big(F_s(m(s)) \big)\right], \text{ for a.e. $s \in \Theta$} .
$$

\medskip

\noindent Integrating from $\underline{\theta}$ to $\theta$ yields
\begin{align*}
\int_{\underline{\theta}}^\theta\int_{m(s)}^{\bar L} 
-\left[
\frac{\partial g_s}{\partial s}(F_s(l))
+g_s'(F_s(l))\frac{\partial F_s(l)}{\partial s}\right] \,dl\,ds
&= b(\theta) - b(\underline{\theta}) +\int_{\underline{\theta}}^\theta m^\prime(s)\left[ 1- g_s\big(F_s(m(s)) \big)\right]\, ds\\
&=  \int_{m(\theta)}^{\bar L} \left[ 1- g_\theta(F_\theta(l)) \right] \,dl -\int_{m(\underline{\theta})}^{\bar L} \left[ 1- g_{\underline{\theta}}(F_{\underline{\theta}}(l)) \right] \,dl\\
&\qquad + \int_{\underline{\theta}}^\theta m^\prime(s)\left[ 1- g_s\big(F_s(m(s)) \big)\right]\, ds. 
\end{align*}

\medskip

\noindent Substituting the above expression into the premium formula, we obtain
\begin{align*}
p_\theta
&= p_{\underline{\theta}} 
+ \int_{\underline{\theta}}^\theta m^\prime(s)\left[ 1- g_s\big(F_s(m(s)) \big)\right]\, ds.
\end{align*}
\qed 

\bigskip
\subsection{Proof of Proposition \ref{prop:rationiffsub_capped}}
Firstly, we know that for capped deductible insurance, the marginal retention is given by $\frac{\partial R_\theta(l)}{\partial l}= \mathbf{1}_{\{l <d(\theta)\}} + \mathbf{1}_{ \{l> d(\theta) +m(\theta) \}}$, for all $\theta \in \Theta$ and for a.e. $l \in [0, \bar L]$. Suppose that the map $\theta\mapsto d(\theta)$ is non-increasing and that the map $\theta\mapsto d(\theta)+m(\theta)$ is non-decreasing. Then for any $\theta_1, \theta_2 \in \Theta$, with $\theta_1<\theta_2$, we have 
$$
\{l\in[0,\bar L]:l<d(\theta_2)\}
\subseteq
\{l\in[0,\bar L]:l<d(\theta_1)\}, 
$$
and 
$$
\{l\in[0,\bar L]:l>d(\theta_2)+m(\theta_2)\}
\subseteq
\{l\in[0,\bar L]:l>d(\theta_1)+m(\theta_1)\}.
$$

\medskip

\noindent That is, 
$\mathbf{1}_{ \{l<d(\theta_2) \} } \leq \mathbf{1}_{ \{l<d(\theta_1)\}}$ and $\mathbf{1}_{ \{l>d(\theta_2) +m(\theta_2)\} } \leq \mathbf{1}_{ \{l > d(\theta_1) +m(\theta_1)\}}$, for a.e.\ $l \in [0, \bar L]$, 
implying that the collection of retention functions $\{R_\theta\}_{\theta \in \Theta}$ is submodular since 
$\frac{\partial R_{\theta_2}(l)}{\partial l} \leq \frac{\partial R_{\theta_1}(l)}{\partial l}$, for a.e.\ $l \in [0, \bar L]$. 

\medskip

Conversely, suppose that the collection of retention functions $\{R_\theta\}_{\theta \in \Theta}$ is submodular. For capped deductible insurance, the marginal retention is given by
$\frac{\partial R_\theta(l)}{\partial l}= \mathbf{1}_{\{l <d(\theta)\}} + \mathbf{1}_{ \{l> d(\theta) +m(\theta) \}}$, for all $\theta \in \Theta$ and for a.e.\ $l \in [0, \bar L]$. That is, 
$$
\frac{\partial R_\theta(l)}{\partial l}
=
\begin{cases}
1, & l<d(\theta),\\
0, & d(\theta)<l<d(\theta)+m(\theta),\\
1, & l>d(\theta)+m(\theta).
\end{cases}
\ \text{for all $\theta \in \Theta$ and for a.e. $l \in [0, \bar L]$}. 
$$

\medskip

\noindent Fix $\theta_1, \theta_2 \in \Theta$, with $\theta_1\leq\theta_2$. If $\theta_1 =\theta_2$, then $d(\theta_1) = d(\theta_2)$ and $d(\theta_1) +m(\theta_1) = d(\theta_2) + m(\theta_2)$. Suppose now that $\theta_1<\theta_2$, and let 
$J_i := \left( d(\theta_i), d(\theta_i)+  m(\theta_i) \right)$, for $i \in \{1, 2\}$. 
Since $m(\theta_i)>0$ for $i \in \{1, 2\}$, each interval $J_i$ has positive length. Moreover, the marginal retention can be rewritten as
$\frac{\partial R_{\theta_i}(l)}{\partial l}= 1 - \mathbf{1}_{J_i}(l)$, for a.e.\ $l \in [0, \bar L]$ and $i \in \{1,2\}$. 
Since $\{R_\theta\}_{\theta \in \Theta}$ is submodular, it follows that $\mathbf{1}_{J_1}(l)\leq \mathbf{1}_{J_2}(l)$, for a.e.\ $l \in [0, \bar L]$.
Consequently, $J_1\setminus J_2$ has Lebesgue measure zero.

\medskip

Next, we show that $d(\theta_2) \leq d(\theta_1)$. Suppose for the sake of contradiction that $d(\theta_1)< d(\theta_2)$, then 
$$
\Big(d(\theta_1), \min\{d(\theta_2),d(\theta_1)+ m(\theta_1)\}\Big) \subseteq J_1 \setminus J_2.
$$

\noindent This interval has positive length since both $d(\theta_2)$ and $d(\theta_1) + m(\theta_1)$ are strictly larger than $d(\theta_1)$ contradicting the fact that $J_1 \setminus J_2$ has Lebesgue measure zero. Thus, $d(\theta_2) \leq d(\theta_1)$, implying that the map $\theta \mapsto d(\theta)$ is non-increasing. 

\medskip

We now show that $d(\theta_2) + m(\theta_2) \geq d(\theta_1) + m(\theta_1)$. For the sake of contradiction, suppose that $d(\theta_2) + m(\theta_2) < d(\theta_1) + m(\theta_1)$, then 
$$
\Big( \max \{ d(\theta_1), d(\theta_2) + m(\theta_2)\},d(\theta_1) + m(\theta_1) \Big)\subseteq J_1 \setminus J_2.
$$

\noindent This interval also has positive length since both $d(\theta_1)$ and $d(\theta_2) + m(\theta_2)$ are strictly smaller than $d(\theta_1) + m(\theta_1)$ contradicting the fact that $J_1 \setminus J_2$ has Lebesgue measure zero. Thus, 
$d(\theta_2) + m(\theta_2) \geq d(\theta_1) + m(\theta_1)$, implying that the map $\theta \mapsto d(\theta) + m(\theta)$ is non-decreasing.
\qed

\bigskip
\subsection{Proof of Lemma \ref{prop:capped_premium}}
Suppose that $\theta\mapsto d(\theta)$ is non-increasing and that
$\theta\mapsto d(\theta)+m(\theta)$ is non-decreasing. It follows from Proposition \ref{prop:rationiffsub_capped} that the collection of retention functions  $\{R_\theta\}_{\theta \in \Theta}$ is submodular and hence implementable by Corollary \ref{prop:capped_implementable}. We know from Definition \ref{def:PointMeasImplem} that there exists a premium schedule $p:\Theta \to \R_+$ such that the menu of contracts $(R_\theta, p_\theta)_{\theta \in \Theta}$ is incentive compatible. That is, for $\theta_1, \theta_2 \in \Theta$, with $\theta_1<\theta_2$, we have
$p_{\theta_2} -p_{\theta_1} \geq u_\cR(\theta_1, \theta_2) - u_\cR(\theta_1),$
where
\begin{align*}
u_\cR(\theta_1, \theta_2) - u_\cR(\theta_1)
&= - \int_0^{\bar L}\left[ 1- g_{\theta_1}(F_{\theta_1}(l)) \right]\, \frac{\partial R_{\theta_2}(l)}{\partial l } \, dl +  \int_0^{\bar L}\left[ 1- g_{\theta_1}(F_{\theta_1}(l)) \right]\, \frac{\partial R_{\theta_1}(l)}{\partial l } \, dl \\
&= \int_0^{\bar L}\left[ 1- g_{\theta_1}(F_{\theta_1}(l)) \right]\, \left[\frac{\partial R_{\theta_1}(l)}{\partial l } -\frac{\partial R_{\theta_2}(l)}{\partial l } \right] \, dl \geq 0, 
\end{align*}

\medskip

\noindent since $ 1- g_{\theta_1}(F_{\theta_1}(l)) \geq 0$ for all $l \in [0, \bar L]$. Hence, $\theta \mapsto p_\theta$ is non-decreasing.

\medskip

Since $\{R_\theta\}_{\theta \in \Theta}$ is implementable, it follows from Proposition \ref{IC_characterization}, that the corresponding premium satisfies
\begin{align*}
p_{\theta} = 
p_{\underline {\theta} } \,\, &+\int_0^{ \bar L}  \left[ 1 - g_{ \underline{ \theta }} \big( F_{\underline{ \theta }}(l)  \big)   \right] \,\, \frac{ \partial R_{\underline{\theta}}(l) }{ \partial l } \, dl 
-
\int_{ \underline{\theta}} ^ { \theta} \int_0 ^{\bar L}  \left[ \frac{\partial g_s}{ \partial s} (F_s(l)) + g'_s(F_s(l)) \frac{\partial F_s(l)}{\partial s} \right] 
\frac{\partial R_s(l)}{\partial l } \, dl \, ds \\
&\quad  -
\int_0^{ \bar L }  \left[ 1 - g_{\theta} \big( F_{\theta}(l)  \big)   \right] \,\frac{ \partial R_{\theta}(l) }{ \partial l } \,dl.  
\end{align*} 

\medskip

\noindent The capped deductible marginal retention function is given by 
$\frac{\partial R_\theta(l)}{\partial l}= \mathbf{1}_{\{l <d(\theta)\}} + \mathbf{1}_{ \{l> d(\theta) +m(\theta) \}}$, for all $\theta \in \Theta$ and for a.e. $l \in [0, \bar L]$. 
Thus the premium's expression becomes
\begin{align*}
p_\theta
&= p_{\underline\theta}
+\int_0^{d(\underline\theta)}
\left[1-g_{\underline\theta}(F_{\underline\theta}(l))\right]\,dl
+\int_{d(\underline\theta)+m(\underline\theta)}^{\bar L}
\left[1-g_{\underline\theta}(F_{\underline\theta}(l))\right]\,dl \\
&-\int_{\underline\theta}^{\theta}
\int_0^{d(s)}
\left[\frac{\partial g_s}{\partial s}(F_s(l))
+g^\prime_s(F_s(l))\frac{\partial F_s(l)}{\partial s}
\right]\,dl \,ds 
-\int_{\underline\theta}^{\theta}
\int_{d(s)+m(s)}^{\bar L}
\left[\frac{\partial g_s}{\partial s}(F_s(l))
+g^\prime_s(F_s(l))\frac{\partial F_s(l)}{\partial s}
\right]\,dl\,ds\\
&-\int_0^{d(\theta)}\left[1-g_\theta(F_\theta(l))\right]\,dl
-\int_{d(\theta)+m(\theta)}^{\bar L} \left[1-g_\theta(F_\theta(l))\right]\,dl. 
\end{align*}

\medskip

For all $s \in \Theta$, let 
$$
C(s):=\int_{d(s)+m(s)}^{\bar L} \left[1-g_s(F_s(l))\right]\,dl, \ \text{ and } \
D(s) := \displaystyle\int_{0}^{d(s)} \left[ 1- g_s(F_s(l)) \right] \,dl.
$$

\medskip

\noindent Similarly to the proof of Lemma \ref{prop:policylimit_premia}, we can show that 
\begin{equation}\label{eq:D}
|D(z)-D(s)| \leq K\bar L|z-s|+|d(z)-d(s)|, 
\end{equation}
and
\begin{equation}\label{eq:C}
|C(z) - C(s)| \leq K \bar L |z-s| + \left| d(z) +m(z) - d(s) - m(s)\right|. 
\end{equation}

\medskip

\noindent Since $d$ and $m$ are absolutely continuous, their sum $d+m$ is also absolutely continuous. Summing \eqref{eq:D} and \eqref{eq:C} separately, over finite collections of non-overlapping intervals yields absolute continuity of $D$ and $C$ \cite[Definition 7.17]{rudin1987real}. Thus, both functions $D$ and $C$ are differentiable almost everywhere, and satisfy
$D(\theta)-D(\underline\theta)=\int_{\underline\theta}^{\theta}D'(s)\,ds$
and
$
C(\theta)-C(\underline\theta)=\int_{\underline\theta}^{\theta}C'(s)\,ds$.

\medskip

Following the same steps of the proof of Lemma \ref{prop:policylimit_premia}, since $d+m$ is non-decreasing in $s\in \Theta$, we can show that, for almost every $s \in \Theta$, 
\begin{align*}
C'(s)&=-\int_{d(s)+m(s)}^{\bar L}
\left[\frac{\partial g_s}{\partial s}(F_s(l))
+g_s'(F_s(l))\frac{\partial F_s(l)}{\partial s}\right]\,dl -
\left(d(s)+m(s) \right)^\prime\big[1-g_s\big(F_s(d(s)+m(s))\big)
\big]. 
\end{align*}

\medskip

\noindent To identify the derivative of $D$, we first consider the following set:
$$
E_D:= \left\{ s\in (\underline{\theta}, \bar \theta): d \text{ and } D \text{ are differentiable at $s$ with finite derivatives.} \right\}. 
$$

\medskip

\noindent Fix $s \in E_D$. For $h>0$ sufficiently small that $s-h \in \Theta$, let $\Delta_h^d:= d(s-h) - d(s)$.
Since $d$ is non-increasing, $\Delta^d_h \geq 0$. Differentiability of $d$ gives $\Delta_h^d \to 0$, and $\frac{\Delta_h^d}{h}\to -d^\prime(s)$. Moreover, we know that 
\begin{align*}
\frac{D(s)-D(s-h)}{h}
&=-\int_0^{d(s)}
\frac{g_s(F_s(l))-g_{s-h}(F_{s-h}(l))}{h}\,dl -
\frac{1}{h}\int_{d(s)}^{d(s-h)}
\left[1-g_{s-h}(F_{s-h}(l))\right]\,dl.
\end{align*}

\noindent By the dominated convergence theorem, 
$$
\underset{h\downarrow0}{\lim }\int_0^{d(s)}
\frac{g_s(F_s(l))-g_{s-h}(F_{s-h}(l))}{h}\,dl  = \int_0^{d(s)}\frac{\partial }{\partial s} g_s(F_s(l))\, dl. 
$$

\noindent On the other hand, 
\begin{align*}
&\left| \frac{1}{h}
\int_{d(s)}^{d(s-h)}
\left[ g_{s-h}(F_{s-h}(l))-g_s(F_s(l))\right]\,dl
\right|
\leq
\frac{1}{h}\int_{d(s)}^{d(s-h)}Kh\,dl
=K\Delta_h^d \to 0 ,\ \text{ as $h \downarrow 0$} .
\end{align*}

\noindent Hence, 
\begin{align*}
\left|
\frac{1}{h} \int_{d(s)}^{d(s-h)}[1-g_s(F_s(l))]\,dl
-\frac{\Delta_h^d}{h}[1-g_s(F_s(d(s)))]\right|
\leq \frac{\Delta_h^d}{h}
\underset{l\in[d(s),d(s-h)]}{\sup}
\Big|g_s(F_s(l))-g_s\big(F_s(d(s))\big)\Big|.
\end{align*}

\medskip

\noindent The cumulative distribution function $F_s$ is right-continuous, and the distortion function $g_s$ is continuous, hence $l \mapsto g_s(F_s(l))$ is right-continuous. The supremum tends to zero, meanwhile $\frac{\Delta_h^d}{h}\to - d^\prime(s)$ for sufficiently small $h>0$. Thus, 
\begin{align*}
\underset{h\downarrow0}{\lim}
\frac{1}{h}
\int_{d(s)}^{d(s-h)}
[1-g_s(F_s(l))]\,dl
=-d'(s)[1-g_s(F_s(d(s)))], 
\end{align*}
and
\begin{align*}
\underset{h\downarrow0}{\lim}
\frac{1}{h}
\int_{d(s)}^{d(s-h)}
[1-g_{s-h}(F_{s-h}(l))]\,dl
=-d'(s)[1-g_s(F_s(d(s)))]. 
\end{align*}

\medskip

\noindent Hence, the derivative of $D$ for a.e.\ $s \in \Theta$ is given by 
$$
D^\prime(s)=
\int_0^{d(s)}
-\left[\frac{\partial g_s}{\partial s}(F_s(l))+g_s'(F_s(l))\frac{\partial F_s(l)}{\partial s}
\right]\,dl +d'(s)\left[ 1- g_s\big( F_s(d(s))\big) \right].  
$$

\medskip

\noindent We can equivalently rewrite the derivatives of $D$ and $C$ respectively as follows: 
\begin{align*}
-\int_0^{d(s)}
\left[
\frac{\partial g_s}{\partial s}(F_s(l))
+g_s'(F_s(l))
\frac{\partial F_s(l)}{\partial s}
\right]\,dl
=D'(s)-d'(s)\left[1-g_s(F_s(d(s))) \right], 
\end{align*}
and
\begin{align*}
-\int_{d(s)+m(s)}^{\bar L}
\left[\frac{\partial g_s}{\partial s}(F_s(l))
+g_s'(F_s(l))\frac{\partial F_s(l)}{\partial s}\right]\,dl
&= C'(s) + \left(d(s)+m(s) \right)^\prime\left[1-g_s\big(F_s(d(s)+m(s))\big)
\right]. 
\end{align*}

\medskip

\noindent Integrating the above two expressions from $\underline{\theta}$ to $\theta$ respectively, yields
\begin{align*}
-\int_{\underline{\theta}}^\theta\int_0^{d(s)}
\left[\frac{\partial g_s}{\partial s}(F_s(l))
+g_s'(F_s(l))
\frac{\partial F_s(l)}{\partial s}
\right]\,dl\,ds
&=D(\theta) - D(\underline{\theta})- \int_{\underline{\theta}} ^\theta d'(s)
\left[1-g_s(F_s(d(s)))
\right]\, ds\\
&= \int_0^{d(\theta)} \left[ 1- g_\theta(F_\theta(l))\right]\,dl - \int_0^{d(\underline{\theta})} \left[ 1- g_{\underline{\theta}}(F_{\underline{\theta}}(l))\right]\,dl\\
&\qquad - \int_{\underline{\theta}} ^\theta d'(s)
\left[1-g_s(F_s(d(s)))
\right]\, ds, 
\end{align*}
and
\begin{align*}
-\int_{\underline\theta}^{\theta}\int_{d(s)+m(s)}^{\bar L}
\left[\frac{\partial g_s}{\partial s}(F_s(l))
+g_s'(F_s(l))\frac{\partial F_s(l)}{\partial s}\right]\,dl\,ds
&=\int_{\underline\theta}^{\theta}\left(d(s)+m(s) \right)^\prime \left[1-g_s\left(F_s\big( d(s)+m(s) \big)\right)
\right]\,ds \\ &\quad + C(\theta) - C(\underline{\theta}),
\\
&=\int_{\underline\theta}^{\theta}
\left[d'(s)+m'(s)\right]
\left[1-g_s \big( F_s(d(s)+m(s)) \big) \right]\,ds\\
&\quad +
\int_{d(\theta)+m(\theta)}^{\bar L}
\left[1-g_\theta(F_\theta(l))\right]\,dl\\
&\quad -
\int_{d(\underline\theta)+m(\underline\theta)}^{\bar L}\left[1-g_{\underline\theta}
(F_{\underline\theta}(l))\right]\,dl.
\end{align*}

\medskip

\noindent Substituting these expressions in the premium yields
\begin{align*}
p_\theta
&=p_{\underline\theta}
-\int_{\underline\theta}^{\theta}d^\prime(s)\left[1-g_s(F_s(d(s)))\right]\,ds
+\int_{\underline\theta}^{\theta}\left[d^\prime(s)+m^\prime(s)\right]\left[1-g_s(F_s(d(s)+m(s)))\right]\,ds.
\end{align*}
\qed


\vspace{0.6cm}
\bibliographystyle{ecta}
\bibliography{biblio}
\vspace{0.6cm}

\end{document}